\documentclass[traditabstract]{aa}
\usepackage{txfonts}
\usepackage{graphicx}
\usepackage{caption}
\usepackage{subcaption}
\usepackage{natbib}
\bibpunct{(}{)}{;}{a}{}{,} 
\usepackage{epsfig}
\usepackage[pdftex]{hyperref}   
\hypersetup{bookmarks=true,colorlinks=true,linkcolor=black,citecolor=blue,urlcolor=blue}

\begin{document}

\title{Planets Transiting Non-Eclipsing Binaries}

\author{David V. Martin\inst{1}
\and Amaury H.~M.~J. Triaud\inst{2,3}
}

\offprints{David.Martin@unige.ch}

\institute{Observatoire Astronomique de l'Universit\'e de Gen\`eve, Chemin des Maillettes 51, CH-1290 Sauverny, Switzerland
\and Kavli Institute for Astrophysics \& Space Research, Massachusetts Institute of Technology, Cambridge, MA 02139, USA
\and Fellow of the Swiss National Science Foundation 
}

\date{Received date / accepted date}
\authorrunning{Martin D.\,V. \& Triaud A.\,H.\,M.\,J.}
\titlerunning{Planets Transiting Non-Eclipsing Binaries}

\abstract{The  majority of binary stars do not eclipse. Current searches for transiting circumbinary planets concentrate on eclipsing binaries, and are therefore restricted to a small fraction of potential  hosts. We investigate the concept of finding planets transiting non-eclipsing binaries, whose geometry would require mutually inclined planes.

Using an N-body code we explore how the number and sequence of transits vary as functions of observing time and orbital parameters. The concept is then generalised thanks to a suite of simulated circumbinary systems. Binaries are constructed from radial-velocity surveys of the solar neighbourhood. They are then populated with orbiting gas giants, drawn from a range of distributions. The binary population is shown to be compatible with the {\it Kepler} eclipsing binary catalogue, indicating that the properties of binaries may be as universal as the initial mass function. 

These synthetic systems produce transiting circumbinary planets occurring on both eclipsing and non-eclipsing binaries. Simulated planets transiting eclipsing binaries are compared with published {\it Kepler} detections. We obtain 1) that planets transiting non-eclipsing binaries probably exist in the {\it Kepler} data, 2) that observational biases alone cannot account for the observed over-density of circumbinary planets near the stability limit, implying a physical pile-up, and  3) that the distributions of gas giants orbiting single and binary stars are likely different.

Estimating the frequency of circumbinary planets is degenerate with the spread in mutual inclination. Only a minimum occurrence rate can be produced, which we find to be compatible with 9\%. 
Searching for inclined circumbinary planets may significantly increase the population of known objects and will test our conclusions. Their existence, or absence, will reveal the true occurrence rate and help develop circumbinary planet formation theories.
 }

\maketitle


Over the past two decades, hundreds of exoplanets have been discovered \citep{Mayor:1995uq,Wright:2011fj,Schneider:2011lr}\footnote{Two exoplanet compilations can 
be reached online at the following addresses: \href{http://exoplanet.eu}{exoplanet.eu} and \href{http://exoplanets.org}{exoplanets.org}.}. Yet, only few have been detected as part of binary systems despite the fact that a large fraction of systems in the Galaxy are composed of multiple stars \citep{Duquennoy:1991kx,Tokovinin:2006la,Raghavan:2010lr}. It would be
interesting to know if planet formation --and their subsequent orbital evolution (e.g. \citealt{Lin:1996yq,Rasio:1996ly,Baruteau:2013lr,Davies:2013qy})-- happens in a similar manner around single stars 
compared to around multiple stars. Any difference might offer crucial clues about which formation and evolution scenarios are dominant.

In some cases, planets orbit just one component of the binary, called an s-type orbit, which was highlighted in the recent discovery of \object{Alpha Centauri B}b \citep{Dumusque:2012lr}. In this paper we only consider planets orbiting the centre of mass of the binary, also called a circumbinary or p-type orbit. 
\citet{Schneider:1994lr} proposed such planets could be discovered using the transit method and, in the course of its photometric survey, NASA's {\it Kepler} telescope discovered a handful of circumbinary systems, all of which are around eclipsing binaries and nearly coplanar.

Because most binaries do not eclipse, present searches for circumbinary planets have been restricted to a very small fraction of the binary population. In this paper we explore the topic of circumbinary planets transiting non-eclipsing binaries. Whilst the exact number is not presently known, the complete binary population within the {\it Kepler} field invariably numbers in the tens of thousands. 
This is a significant fraction of all observed systems within the {\it Kepler} catalog and a large amount of potential planet hosts.

Our study is produced in the context of the {\it Kepler} telescope. The high quality of its photometric data and the fact that it concluded its observations of the same field for four years provide a controlled sample within which our conclusions can be quickly tested. Our work  nonetheless remains easily applicable to past surveys, for example {\it CoRoT}, and applicable to future surveys, such as {\it TESS}, {\it NGTS}, {\it PLATO} and {\it Kepler's} extended mission {\it K2}.

This paper is organised in the following way: We first summarise the current circumbinary discoveries (Sec.~\ref{sec:circumbinary_discoveries}). We then move onto the concept of misaligned circumbinary systems. This is first motivated in Section~\ref{sec:evidence}, by rapidly reviewing observational and theoretical evidence for mutually inclined planes. Then, in Section~\ref{sec:concept} we introduce the concept of planets transiting non-eclipsing binaries. In Section~\ref{sec:observational_consequences} we explore some observational consequences.

In Section~\ref{sec:distributions} we detail our method for constructing a synthetic distribution of circumbinary systems, drawing from a realistic distribution of binary stars, taken from the {\it Kepler} and radial-velocity surveys. We then draw circumbinary planets using a number of possible distributions in mutual inclination and in orbital separation. Based on these distributions, we make predictions on the number of exoplanets transiting both eclipsing and non-eclipsing binaries (Sec.~\ref{sec:results}), before comparing these results with those found already in the {\it Kepler} data (Sec.~\ref{sec:comparisons_with_kepler}). Finally, in Section \ref{sec:discuss} we discuss our results and ways of finding these planets using current and future telescopes, before concluding.

\section{Current circumbinary discoveries}\label{sec:circumbinary_discoveries}

The {\it Kepler} telescope has, so far, discovered seven circumbinary systems, with one having multiple planets (Kepler-47; \citealt{Orosz:2012lr}).  Aside from {\it Kepler}, there have been circumbinary planets discovered by the radial-velocity method \citep{Correia:2005lr} and through eclipse timing variations (e.g. \citealt{Beuermann:2010fk}) although some doubts exist about some systems \citep{Mustill:2013lr}. 

The {\it Kepler} circumbinary planets were discovered by seeking repeated, quasi-periodic photometric dimmings, caused by the planet transiting one of the stars (departure from strictly periodic transits --transit timing variations-- is caused by the binary motion). The requirement of semi-regular transits biased these searches to coplanar systems: inclined planets will generally produce transits that are either singular or irregularly spaced, something which we demonstrate throughout the paper.

Whilst the number of detected planets is currently low, two trends are emerging in the orbital distribution of circumbinary planets:

1) All the innermost planets reside close to the stability limit, as defined by \citet{Holman:1999lr}, with planets closer to the binary being dynamically unstable (see also \citealt{Dvorak:1986fk,Dvorak:1989vn}). The location of that limit is a function of the binary mass ratio and of the planet and binary eccentricities. It can be approximated by $P_{\rm p}\sim4.5~P_{\rm bin}$. There is a weak dependence on the mutual orbital inclination, with inclined systems generally remaining stable at shorter orbits \citep{Pilat-Lohinger:2003qy,Doolin:2011lr,Morais:2012qy}.

The origin of this accumulation remains an open question. Inward, disc-driven orbital migration is possible (e.g. \citealt{Pierens:2013kx,Kley:2014rt}) and could preferentially place planets near the stability limit. Alternatively this can result from an observing bias since, for a given binary, our probability is highest to detect the shortest possible planetary orbit. A recent commentary on circumbinary detections by \citet{Welsh:2013lr} suggested that both explanations were possible. We tested both hypotheses, with results in Section \ref{sec:planet_distribution} and discussion in Section~\ref{sec:formation}.

2) There are no claims of planets transiting very short period binaries ($<$ 5 days), despite more than half of the eclipsing systems in the {\it Kepler} catalog having periods in this range. The cause for this dearth of planets is presently unknown.\\

{\it Kepler}'s circumbinary systems' main orbital parameters are summarised in Table~\ref{tab:KeplerDiscoveries}, with $P_{\rm bin}$ standing for the binary period, and $e_{\rm bin}$ for its eccentricity. $P_{\rm p}$ and $e_{\rm p}$ are the planets' period and eccentricity. $P_{\rm crit}$ is the shortest, stable planetary orbit using the criterion from \citet{Holman:1999lr}.
\begin{table}
\caption{Circumbinary planets detected by {\it Kepler}.} 
\centering 
\begin{tabular}{l | c | c | c | c | c } 
\hline\hline 
Name & $P_{\rm bin}$ (days) & $e_{\rm bin}$ & $P_{\rm p}$ (days) & $e_{\rm p}$ & $P_{\rm p}/P_{\rm crit}$\\ 
[0.5ex] 
\hline 
\object{Kepler-16} & 40.1 & 0.16 & 228.8 & 0.01 & 1.14\\
\object{Kepler-34} & 28.0 & 0.52 & 288.8 & 0.18 & 1.21\\
\object{Kepler-35} & 20.7 & 0.14 & 131.4 & 0.04 & 1.24\\
\object{Kepler-38} & 18.8 & 0.10 & 106.0 & 0.07 & 1.42\\
\object{Kepler-47}b & 7.4 & 0.02 & 49.5 & 0.04 & 1.77\\
\object{Kepler-47}d & 7.4 & 0.02 & 187.3 & - & -\\
\object{Kepler-47}c & 7.4 & 0.02 & 303.1 & $<$ 0.41 & 10.8\\
\object{Kepler-64} & 20.0 & 0.21 & 138.5 & 0.07 & 1.29\\
\object{Kepler-413} & 10.1 & 0.04 & 66.3 & 0.12 & 1.60\\
\hline 
\multicolumn{6}{l}{\footnotesize{{\bf Note:} Kepler-47d awaits publication and has no $e_{\rm p}$ measurement.}}\\
\multicolumn{6}{l}{\footnotesize{{\bf Refs:} \citet{Doyle:2011vn,Welsh:2012lr,Orosz:2012yq,Orosz:2012lr}}}\\
\multicolumn{6}{l}{\footnotesize{\citet{Schwamb:2013kx,Kostov:2013lr,Kostov:2014qy}}.}\\

\end{tabular}
\label{tab:KeplerDiscoveries}
\end{table}

\section{Observational and theoretical evidence for mutually inclined planes}\label{sec:evidence}

Observations of circumbinary debris  discs show they are largely found coplanar (\citealt{Watson:2011mz,Kennedy:2012lr,Greaves:2014vn}), but some misaligned discs exist (e.g. 99\,Herculis; \citealt{Kennedy:2012zr}). 

The detections and monitoring of {\it winking} binaries, (\object{KH\,15D}, \citealt{Winn:2004lr}; \object{WL~4}, \citealt{Plavchan:2008fk}; \object{YLW~16A}, \citealt{Plavchan:2013lr}) indicate that they are composed of a binary surrounded by an optically thick and inclined protoplanetary disc. The disc is warped and precesses, due to distance-dependent torques exerted by the binary \citep{Chiang:2004fk}. The precession modulates the amount of light obscured by the disc and causes the {\it winking} phenomenon \citep{Winn:2006qy}. Some binary stars possess rotation spins inclined with respect to their orbital spin (e.g. \citealt{Albrecht:2009fy,Zhou:2013fk,Albrecht:2014lr}) and may reflect the same processes that lead to inclined discs.

Proto-binaries formed by fragmentation must have an initial separation larger than 10 AU, with typical separations larger than 100 AU \citep{Bate:2012rt}. Migration of both components is therefore required to form closer binaries, which can occur under certain disc conditions for an accreting binary \citep{Artymowicz:1996fk}. \citet{Foucart:2013ys} found that during mass accretion, angular momentum gets redistributed leading the binary and disc planes to align. Their corollary is that relatively short period binaries (sub-AU) will only harbour coplanar circumbinary planets, because they would likely form after the binary migration and thus, after the disc has realigned.

An alternative result was proposed by \citet{Lodato:2013uq}, who realised that the warped disc can reach a misaligned steady-state and subsequently precess as a right body, as it has been assumed by other studies dealing with similar situations (e.g. \citealt{Terquem:1999pd}). To maintain equilibrium with the binary torques, the disc needs to be 
more inclined the further away from the binary, which is observationally confirmed \citep{Capelo:2012cr}. 

\begin{figure*}  
\begin{center}  
	\begin{subfigure}[b]{0.23\textwidth}
		\caption{Transit of a single star}
		\includegraphics[width=\textwidth]{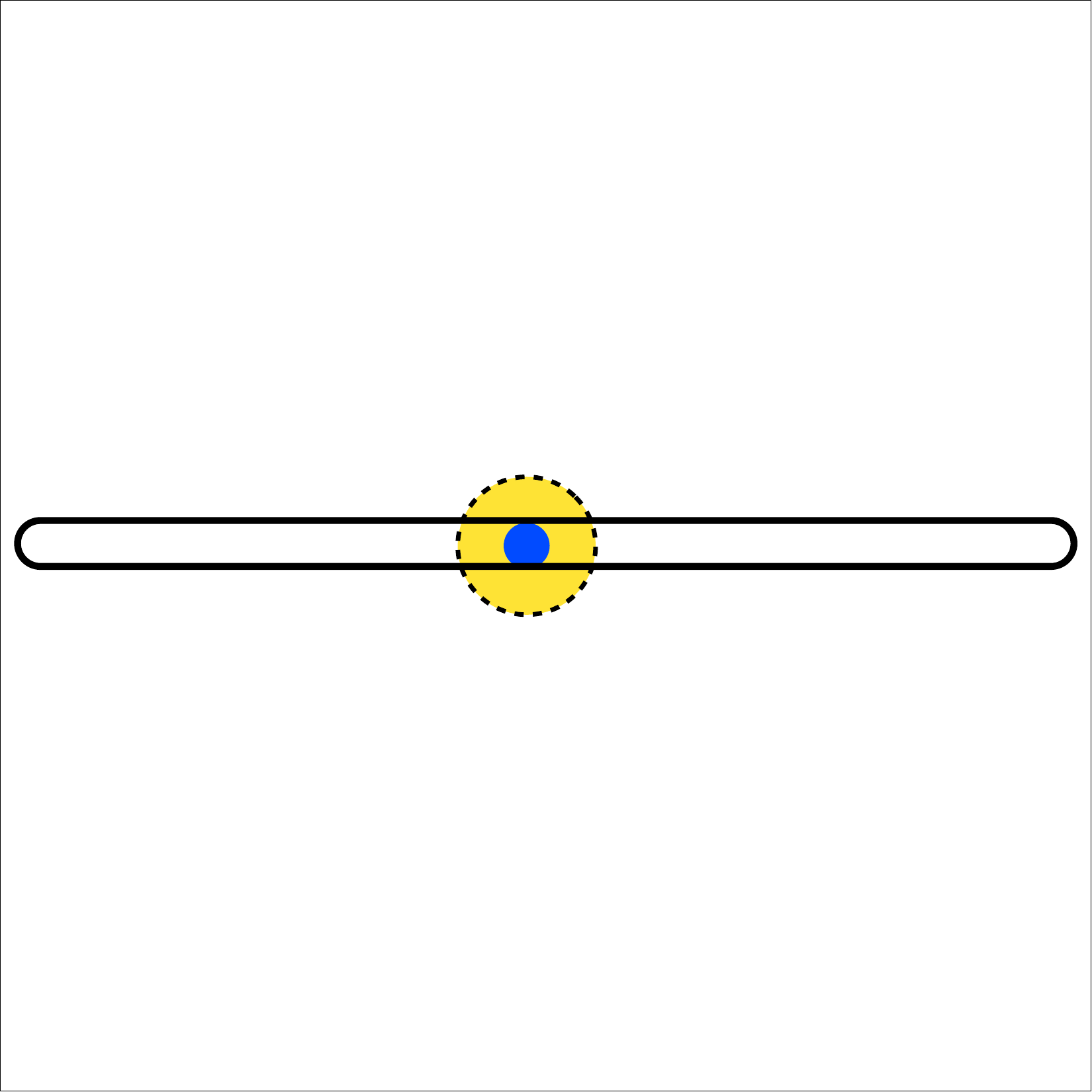}  
		\label{fig:Geometry_a}  
	\end{subfigure}
	\hspace{0.1cm}
	\begin{subfigure}[b]{0.23\textwidth}
		\caption{Transit of an eclipsing binary by a coplanar planet}
		\includegraphics[width=\textwidth]{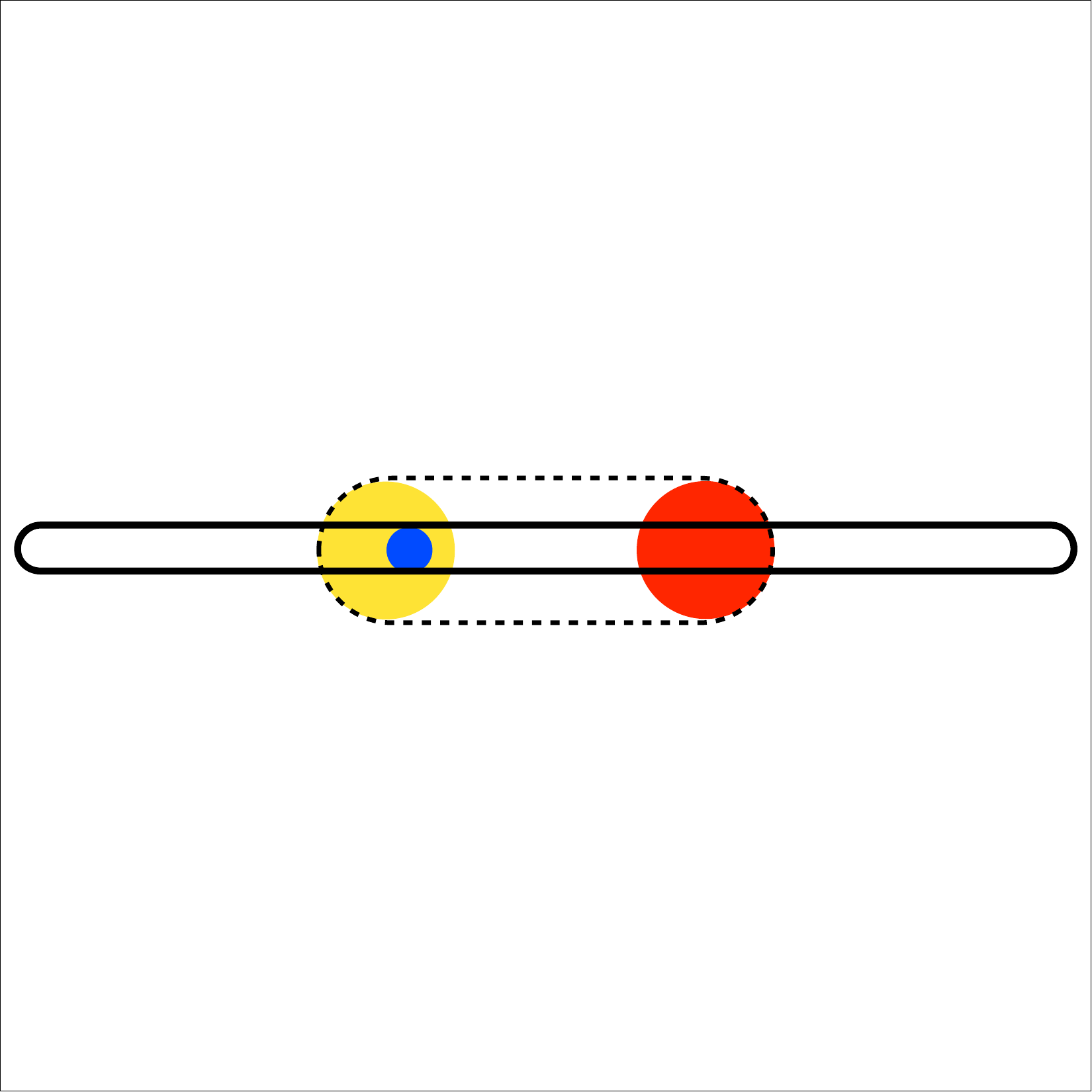}  
		\label{fig:Geometry_b}  
	\end{subfigure}
	\hspace{0.1cm}	
	\begin{subfigure}[b]{0.23\textwidth}
		\caption{Transit of an eclipsing binary by a misaligned planet}
		\includegraphics[width=\textwidth]{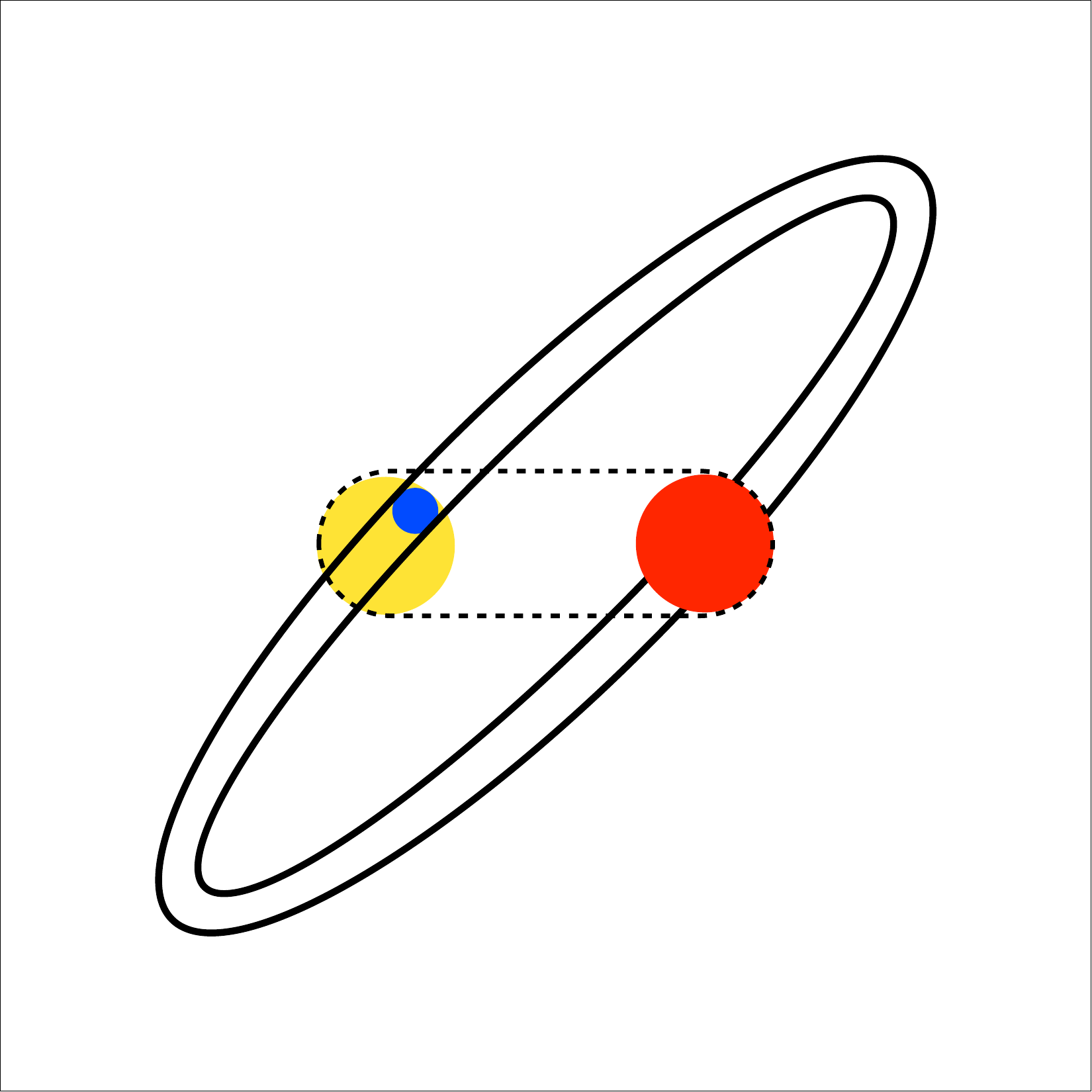}  
		\label{fig:Geometry_c}  
	\end{subfigure}
	\hspace{0.1cm}
	\begin{subfigure}[b]{0.23\textwidth}
		\caption{Transit of a non-eclipsing binary by a misaligned planet}
		\includegraphics[width=\textwidth]{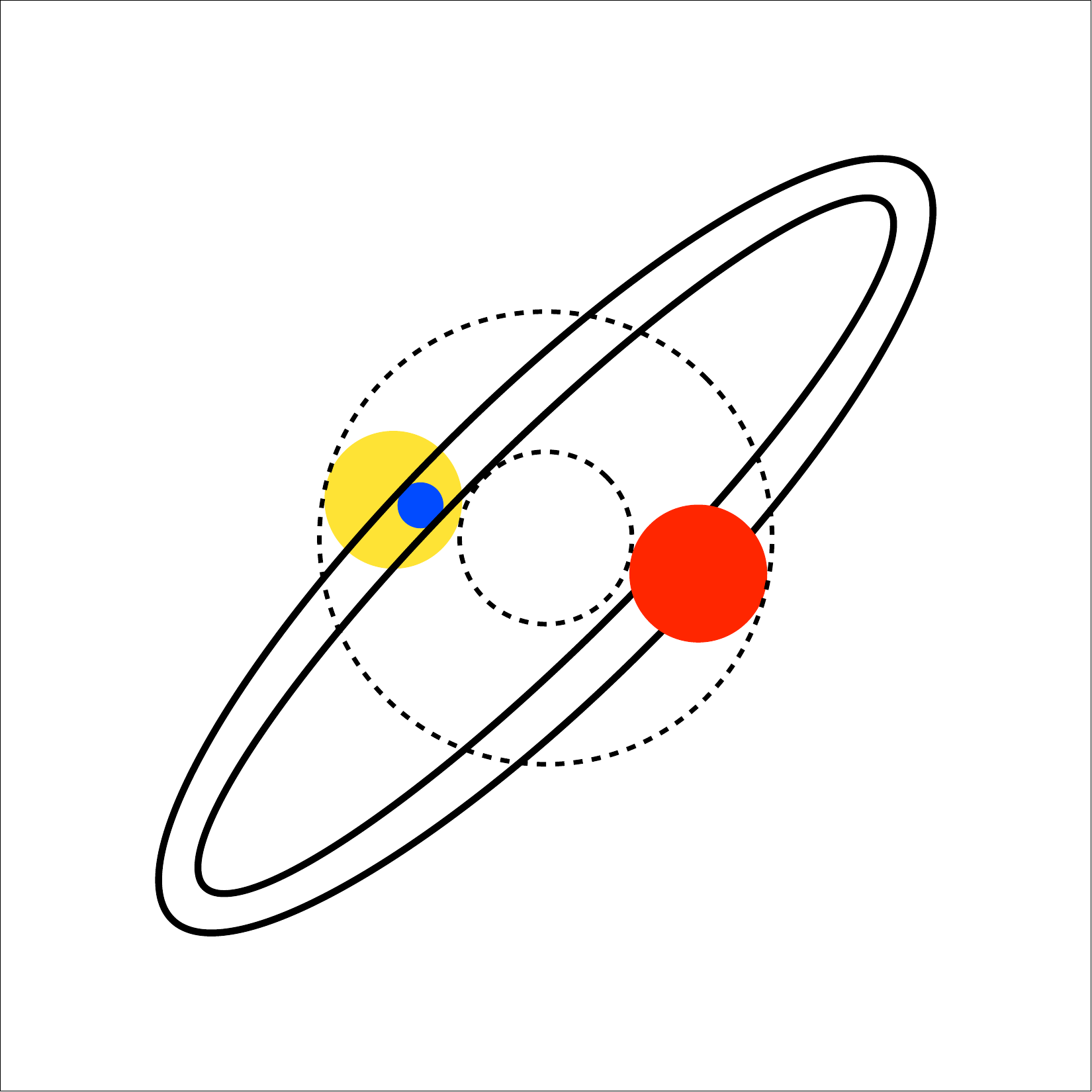}  
		\label{fig:Geometry_d}  
	\end{subfigure}
	\caption{Transit configurations of a planet that moves within the solid lines, around single and binary stars that move within the dashed lines.}
\label{fig:OrbitGeometry}  
\end{center}  
\end{figure*}


Earlier considerations aside, it is likely that the ultimate orientation of the disc does not completely determine the inclinations of planetary orbits, as epitomised by the existence of inclined and even retrograde {\it hot Jupiters} \citep{Hebrard:2008mz,Winn:2009lr,Triaud:2010fr,Schlaufman:2010fk,Triaud:2011qy,Brown:2012lr,Albrecht:2012lp}. Even if circumbinary planets can ultimately only form in coplanar discs, concluding that planets will also be coplanar bears no regards to post-formation dynamical processes, such as planet-planet scattering. These processes likely exist in circumbinary systems just as they do in single star systems, where scattering is an important mechanism to explain eccentric and misaligned planets \citep{Rasio:1996ly,Adams:2003wd,Juric:2008qy,Chatterjee:2008uq,Nagasawa:2008gf,Matsumura:2010ve,Matsumura:2010ul}. These processes may be enhanced in circumbinary planetary systems. An inner binary induces large eccentricity variations in the orbits of its planets \citep{Leung:2013fk}, potentially leading to significant planet-planet scattering.
Finally, multi-body dynamics can lead to a wide variety of configurations and we cannot exclude that more exotic scenarios such as an exchange of planets between components of a multiple stellar system would not provide a population of non-coplanar orbits \citep{Moeckel:2012lr,Kratter:2012lr}. 

\section{Thinking out of the plane}\label{sec:concept}

\begin{figure}  
\begin{center}  
\includegraphics[width=0.4\textwidth]{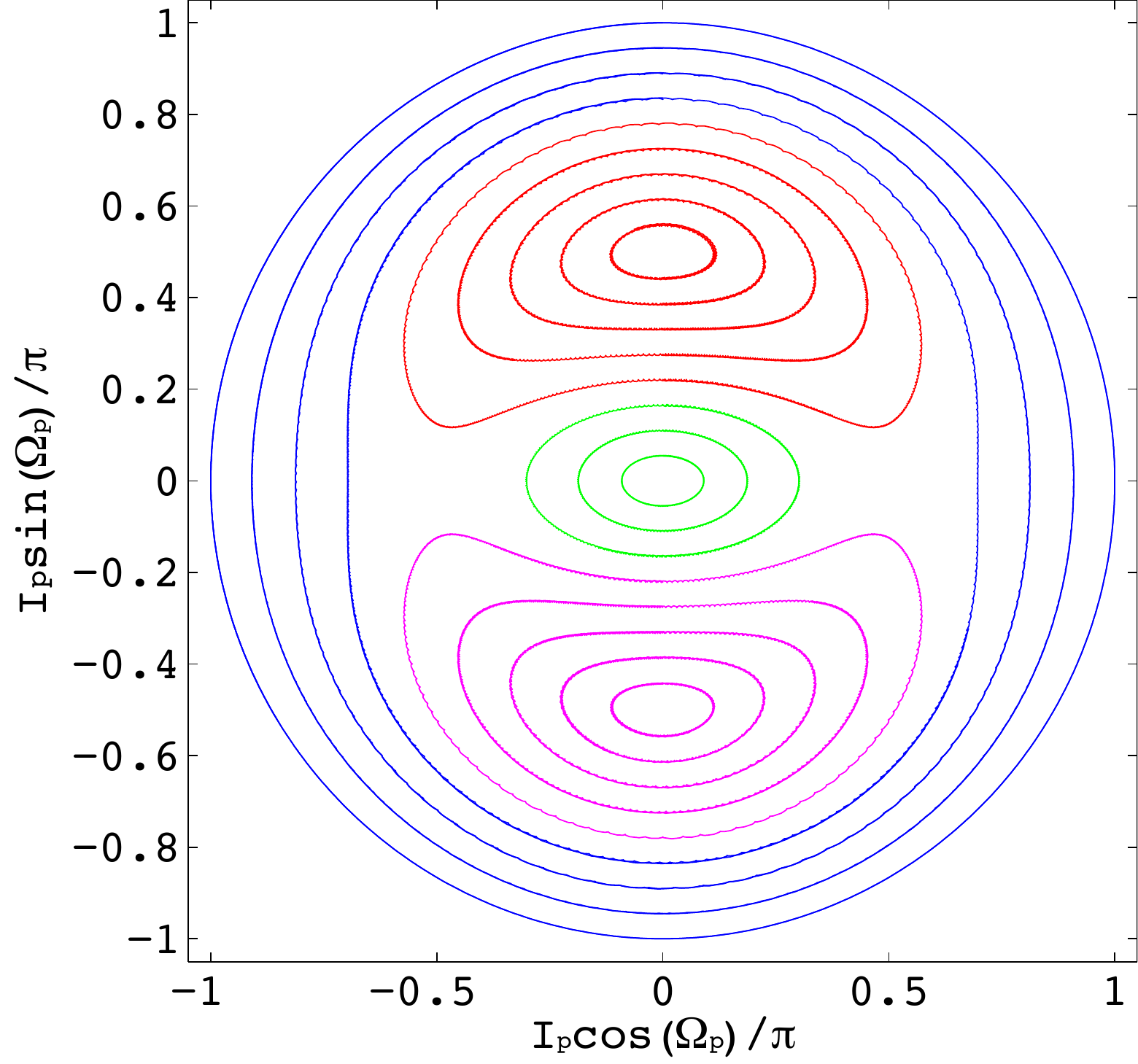}  
\caption{The ($I_{\rm p}\cos(\Omega_{\rm p})$, $I_{\rm p}\sin(\Omega_{\rm p})$) surface of section for an example circumbinary system and over a range of initial $I_{\rm p}$ and $\Omega_{\rm p}$ obtained using our N-body code. The different colours refer to four different types of precession: green orbits are prograde with clockwise precession, blue orbits are retrograde with anticlockwise precession, red orbits are in an island of libration with anticlockwise precession and magenta orbits are in a second island of libration with anticlockwise precession.
}\label{fig:DBTest}  
\end{center}  
\end{figure}

The concept of a planet transiting a single star is well understood and many have now developed a sort of {\it intuitive} knowledge of this technique. Some of these concepts do not directly translate to the case of circumbinary planets, particularly when there is a large mutual inclination between the binary and planet orbits. The purpose of this section is to build a similar intuition for misaligned circumbinary planets.

\subsection{Numerical method}\label{sec:numericalmethods}

All of the numerical simulations in this paper were conducted using an N-body code that integrates Newton's equations of motion with a fourth-order Runge-Kutta algorithm. This algorithm generates a slight energy loss over time, that is dependent on the size of the fixed integration time-step. Despite this limitation, the fourth-order Runge-Kutta algorithm is sufficient for our purpose: we only required short integrations (up to 200 years) where energy loss is insignificant. As a matter of caution, the energy integral was computed inside each simulation. Its variation was monitored and we made sure it stayed within an acceptable level (generally less than $\sim 10^{-7}\%$ over a standard 4-year integration). The effects of spin, tides and relativity were assumed negligible over these timescales, and consequently were not included in the model.

\subsection{Transitability}\label{sec:geometry}

\begin{figure*}  
\begin{center}  
\includegraphics[width=0.95\textwidth]{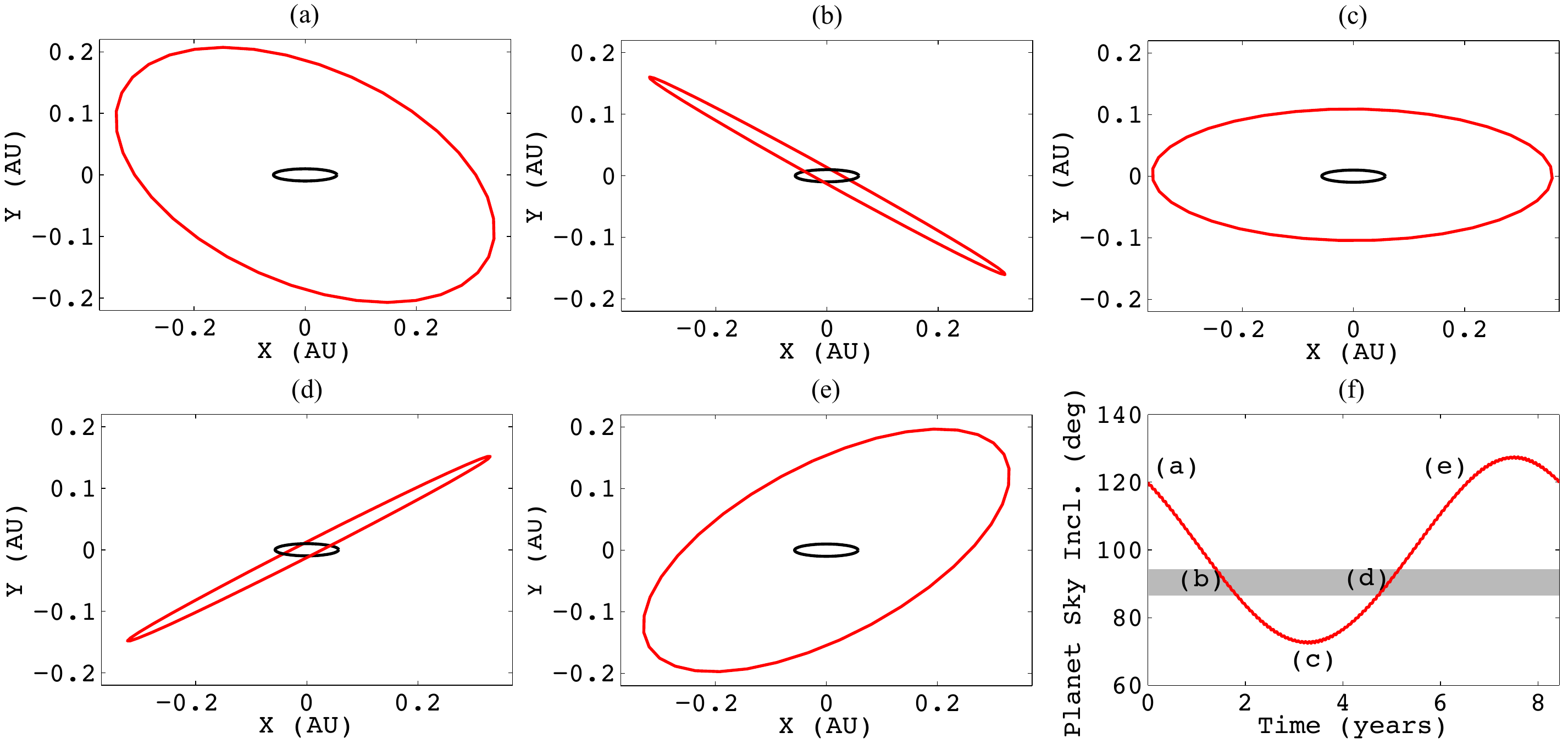}  
\caption{Orbital precession of the osculating Keplerian orbit of a 55 day circumbinary planet (outer, red) 
inclined by 30$^\circ$ with respect to a 10 day equal-mass binary (inner, black). The different panels
show different epochs of the precessing period. Those times are labelled and displayed in panel (f) where we draw the planet's precessing orbital inclination with respect to the plane of the sky as a function of time. The grey band illustrates the range of sky inclinations for which the planet has a chance to transit. The overlap of the red line with the grey band corresponds to epochs of transitability. 
}\label{fig:Precession_Illustration}
\end{center}  
\end{figure*} 

To a given observer, a planet orbiting a single star will transit if its orbital inclination is very close to perpendicular to the plane of the sky. The probability that a planet will transit is a simple function of only the {\it scaled radius} $R_\star/a$, where $R_\star$ is the stellar radius, $a$ is the separation between the planet and the star and we neglect the radius of the planet \citep{Seager:2003qy,Winn:2010lz}. The scaled radius corresponds to a circular area projected on the sky. This probability is purely geometric and is static over time; planets either always or never transit\footnote{An exception to this rule has been presented by \citet{Barnes:2013lv} and concerns close-in planets occupying an inclined orbit about a rotationally deformed star.}. The inclination of the planet with respect to the stellar spin axis does not affect the transit probability. This geometry is depicted in Fig.~\ref{fig:Geometry_a}.

A single star presents an axisymmetric geometry while a binary does not. This implies that estimating the transit probability of circumbinary planets is more complex; one must consider the inclination of the planet with respect to both the plane of the sky (the sky inclination) and the plane of the binary (the mutual inclination). In Fig. \ref{fig:OrbitGeometry} we show the four possible transit arrangements for single and binary stars.

In case (b) the planet will transit both stars every orbit. In cases (c) and (d) the planet and binary orbits overlap but transits will not occur at every period, due to the relative motion of the bodies. Any given transit requires a chance alignment between the star and the planet
This leads us to define the following term:\\

\noindent {\it Transitability: An orbital configuration where the planet and binary orbits intersect on the sky, such that transits are possible but not guaranteed at every planetary orbit.}\\

 The concept of transitability leads to a description of each system and about when to expect and search for transits. The term is used at several occasions in the paper.
 Finding planets require transits, and not just the fact that two orbits are crossing. For instance, both stellar radii have to be taken into account  (the planet radius is generally negligible). Finding the fraction of systems that experience a transit is what we will now describe.

\subsection{Three classes of transiting circumbinary planets}\label{sec:types_of_transits}


Throughout the paper we categorise transiting circumbinary planets according to the following three categories:

\begin{itemize}
\item consecutive transits on eclipsing binaries (EBs consecutive);
\item infrequent transits on eclipsing binaries (EBs sparse);
\item any transits on non-eclipsing binaries (NEBs).
\end{itemize}

Planets consecutively transiting eclipsing binaries are defined as presenting at least three transits on the primary star, each separated by approximately one planetary period. This particular class encompasses all circumbinary planets currently found by {\it Kepler}, and has a strong bias towards coplanar planets. 

The remaining planets transiting eclipsing binaries are labelled as ``sparse". These planets are either misaligned, such that the transit sequence is irregular, or coplanar but on a sufficiently long period that they have not been caught transiting three times within a given observing timeframe.

Planets transiting non-eclipsing binaries are accounted for if a planet transits either star at least once. This is the most relaxed criterion possible. No distinction is made between planets exhibiting consecutive or infrequent transits on non-eclipsing binaries, although consecutive transits will be rare. Confirmation of a planet transiting a non-eclipsing binary will require follow-up observations, like for  most other transiting systems. Our focus will remain solely on gas giants whose transit depths are sufficient to be noticed in one event.


\subsection{The dynamic nature of circumbinary orbits}\label{sec:dyn_int}

Circumbinary orbits are not static. Perturbations from the binary cause the orbital elements\footnote{In this paper we use osculating Jacobi orbital elements.} of the planet to change on quick timescales (typically years for the close binaries we considered), whilst maintaining a stable orbit. This has been calculated analytically \citep{Farago:2010fj,Leung:2013fk} and numerically \citep{Doolin:2011lr}. It has also been demonstrated observationally in the {\it Kepler} circumbinary discoveries. The variation which effects most significantly the transit probabilities is a precession of the longitude of the ascending node, $\Omega_{\rm p}$, and the mutual inclination, $I_{\rm p}$.

This effect is illustrated in Fig.~\ref{fig:DBTest}, for an example circumbinary system with binary masses 0.67 and 0.33 $M_\odot$, period $P_{\rm bin}$ = 5 days, and eccentricity $e_{\rm bin}$ = 0.5. The planet was given a period $P_{\rm p}$ = 37.5 days, eccentricity $e_{\rm p}$ = 0. A range of starting values for $I_{\rm p}$ and $\Omega_{\rm p}$ were used, each one corresponding to a different trace in Fig.~\ref{fig:DBTest}. On the timescale of years the planet mass has a negligible effect on the dynamics, and consequently was inserted as a test particle, like in  \citet{Doolin:2011lr}. The integration was conducted over a full precession period. Our results match Fig. 2 of \citet{Doolin:2011lr}.

\begin{figure*}  
\begin{center}  
	\begin{subfigure}[b]{0.4\textwidth}
		\caption{Binary sky inclination of 80$^\circ$}
		\includegraphics[width=\textwidth]{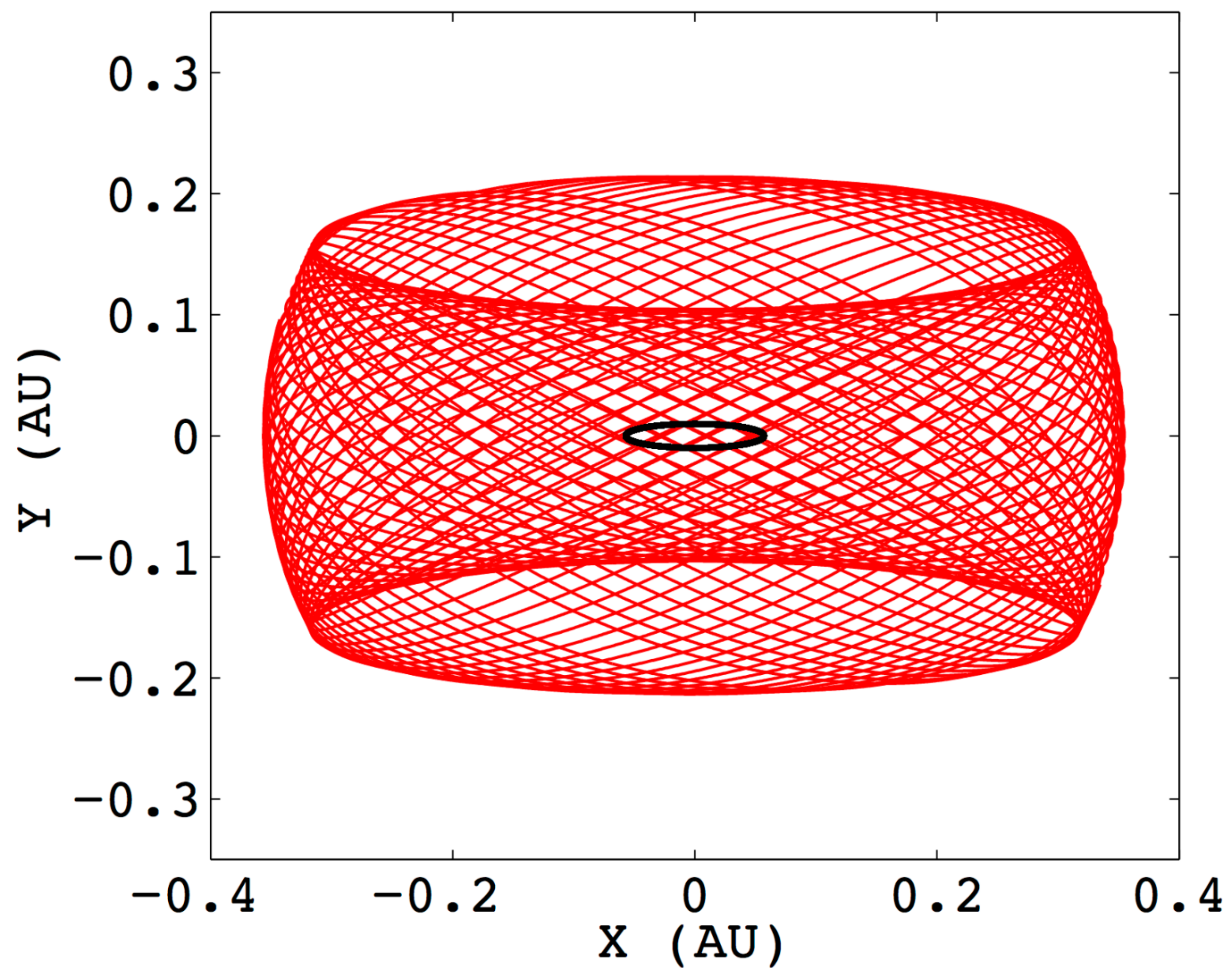}  
		\label{fig:TransitabilityExample_a}  
	\end{subfigure}
	\hspace{1cm}
	\begin{subfigure}[b]{0.4\textwidth}
		\caption{Binary sky inclination of 50$^\circ$}
		\includegraphics[width=\textwidth]{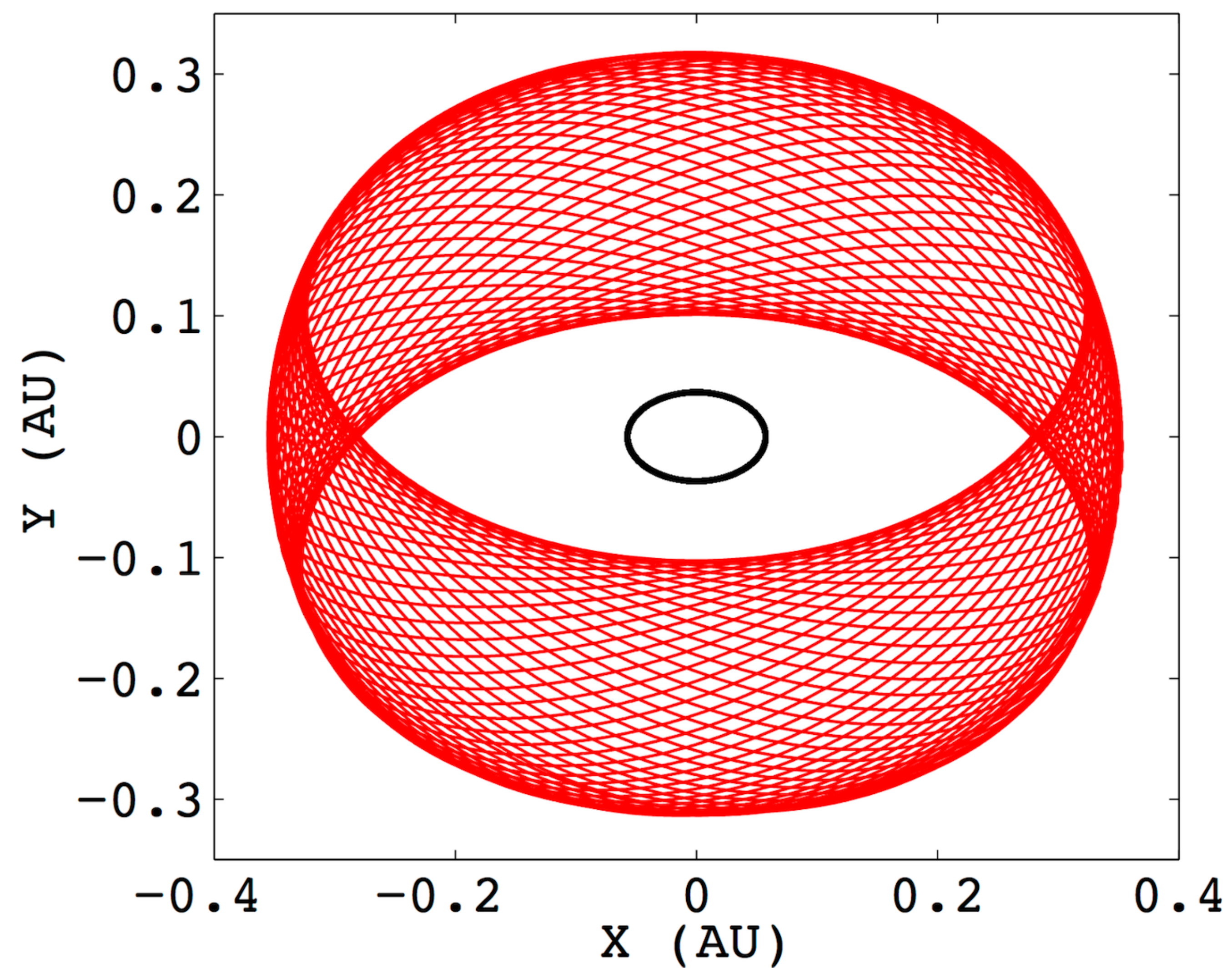}  
		\label{fig:TransitabilityExample_b}  
	\end{subfigure}
	\caption{a) Integrated path of the planet simulated in Fig. \ref{fig:Precession_Illustration} on its orbit (outer, red) around a binary (inner, black) that is inclined by 80$^\circ$ to the plane of the sky. b) same system
tilted by 50$^\circ$ on the plane of the sky.
}\label{fig:TransitabilityExample}  
\end{center}  
\end{figure*} 

In addition to varying the alignment of the planetary orbit with respect to the binary orbit, precession changes the inclination of the planet's orbital plane on the sky. This can cause the orbit to come in and out of transitability over time. To emphasise the importance of this effect, we simulated a new circumbinary system composed of a planet (outer red orbit) on a 55 day orbit inclined by 30$^\circ$ with respect to a binary  of two $1~M_\odot$ stars (inner black orbit) with a period of 10 days (Fig.~\ref{fig:Precession_Illustration}). This system is similar to the existing {\it Kepler} circumbinary systems, except with a large mutual inclination. Five snapshots of the osculating Keplerian period are shown in Fig.~\ref{fig:Precession_Illustration}a to e. These are labelled in Fig.~\ref{fig:Precession_Illustration}f, which displays the variation of the planet's orbital inclination on the sky, over time. The orbit of the planet precesses completely in slightly more than eight years. It enters intervals of transitability when the sky inclination is between 94$^\circ$ and 86$^\circ$, which corresponds to $\sim$10\% of the precession period.

The existing {\it Kepler} circumbinary systems have slight but non-negligible mutual inclinations. This means that precession causes them to exit transitability for extended periods of time. Kepler-16b, for example, will start missing the primary star in 2018, after first stopping to transit the secondary in 2014. Primary and secondary transits are calculated to return approximately in 2042 and 2049, respectively. A quicker effect is observed in Kepler-413, whose 4.1$^\circ$ mutual inclination is the largest found so far \citep{Kostov:2014qy}. 

The alternance between being in  and  out of transitability, implies that there exist several systems whose parameters may resemble very closely the current {\it Kepler} detection, that remained out of transitability for the duration of the survey. Any attempt to estimate occurrence rates from transiting circumbinary planets must therefore account for the probability that a system is in transitability and the time it spends in that configuration.


It is important to note that not every misaligned circumbinary planet will pass through a region of transitability. In Fig.~\ref{fig:TransitabilityExample_a} we present the same planet that we simulated in Fig.~\ref{fig:Precession_Illustration}, but showing its trace over an entire precession period for a binary plane inclined on the sky by 80$^\circ$. This graph also shows an inherent property of an inclined circumbinary planet: its probability to exhibit transitability, in other words, the probability that its orbit crosses the binary's,  is linked to the scaled extent of the planet's orbit on the sky. This is much larger than the scaled radius of the star in an equivalent circumstellar case.
Fig.~\ref{fig:TransitabilityExample_b} represents the exact same system with the difference that it is tilted by 50$^\circ$ with respect to the plane of the sky. The planet no-longer crosses any region of transitability, and will never transit. 

One could wonder whether a planet can exhibit permanent transitability on a non-eclipsing binary. We find that this is possible but only in an extreme scenario of perfectly perpendicular planet and binary orbits, seen by an observer where $I_{\rm bin}=0^{\circ}$ and $I_{\rm p}=90^{\circ}$. One might also consider a resonant system where $P_{\rm p}=NP_{\rm bin}$ for some integer $N$, and the orbits are perfectly aligned such that the planet transits on every passing of the binary. Due to precession and other short-term variations in the orbital elements (e.g. \citealt{Leung:2013fk}), such a configuration is not sustainable.

\section{Observational consequences}\label{sec:observational_consequences}

It is important to understand how an inclined planet population would affect the rate of detectable transiting planets. In this section, hypothetical misaligned circumbinary systems are used for illustration. With Newtonian physics being scalable, the results were easily generalised in Section~\ref{sec:distributions} and applied to {\it Kepler} to allow their empirical verification. 


\subsection{Transit timing}\label{sec:chromosomes}

A single planet transiting a single star does so with strict periodicity. In the presence of multiple planets, each contributes a ``wobble" to the star, which consequently causes each planet to transit with slight aperiodicity \citep{Miralda-Escude:2002uq,Agol:2005qy,Holman:2005fk}. An additional effect comes from planet-planet interactions that add an extra perturbation on their orbits. The resulting transit timing variations (TTVs) are generally on the order of seconds or minutes, and have become detectable with the advent of the {\it Kepler} telescope (e.g. \citealt{Holman:2010lr}).
In the context of circumbinary planets, the main contribution to the TTVs is not from dynamical interactions but from simple geometry. 

In the case of a coplanar planet transiting an eclipsing binary,  both stars move significantly about their centre of mass. The position of a star at the time of transit varies on the order of twice the binary semimajor axis. This induces a geometric TTV effect $\approx(P_{\rm p}P_{\rm bin}^{2})^{1/3}/(2\pi)$, which is generally on the order of days \citep{Armstrong:2013rt}. Perturbations on a planet's orbit due to planet-binary interactions or with another planet still take place, but are of smaller amplitude.

In the case of planet inclined with respect to the binary plane, its precessing orbit (Fig.~\ref{fig:Precession_Illustration}) adds an extra complication to the estimation of transit timings. One must consider three important periods: 
\begin{itemize}
\item the planet period: transits can only occur at approximately integer multiples of the planet period; 
\item the binary period: it affects how much the transit times can deviate from integer multiples;
\item the precession period: it determines the frequency and length of transitability (Fig. \ref{fig:Precession_Illustration}f).
\end{itemize}

\begin{figure}  
\begin{center}  
\includegraphics[width=0.35\textwidth]{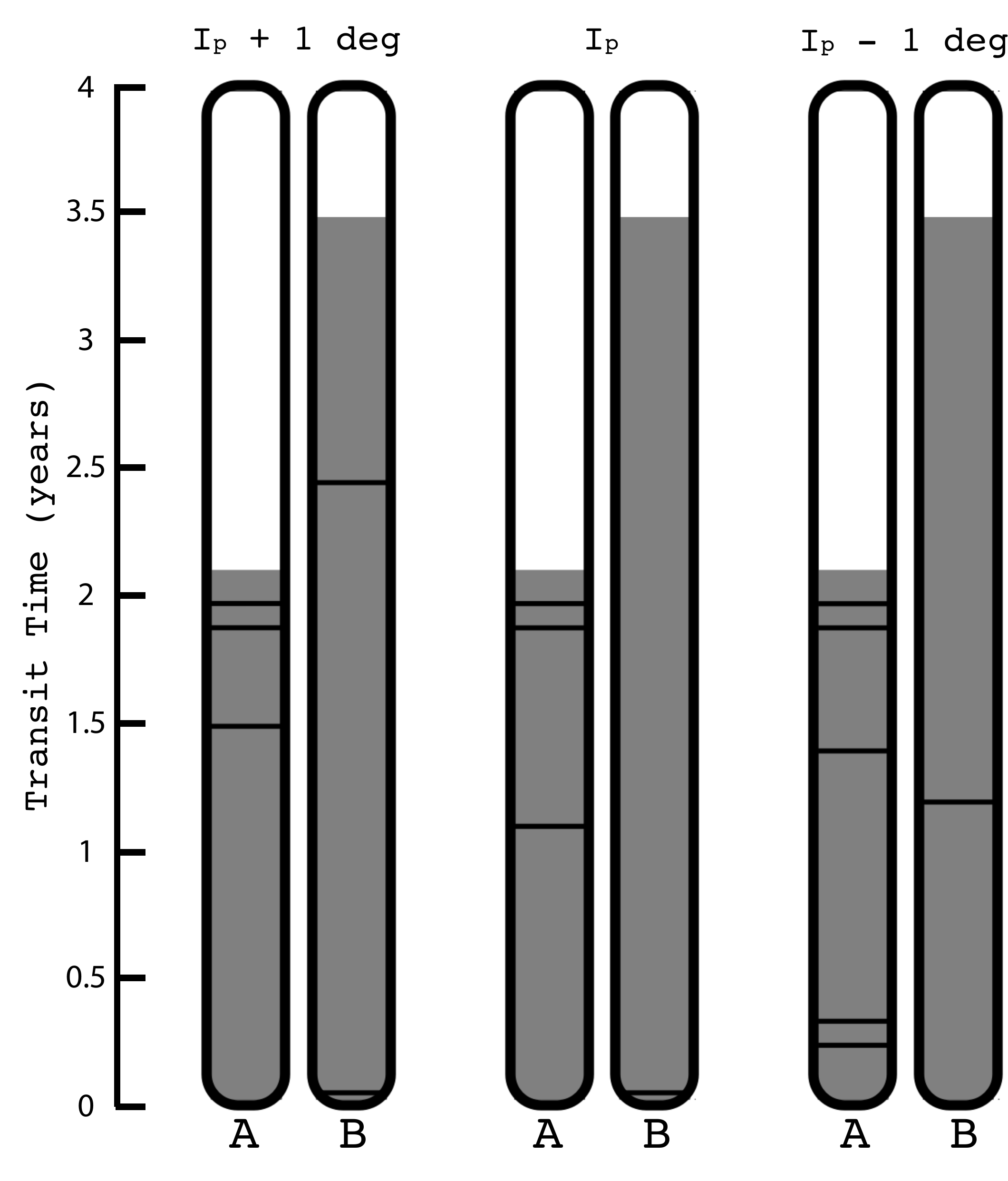}  
\caption{Transit times (horizontal black lines) within regions of transitability (grey regions) on the primary (A) and secondary (B) stars of a non-eclipsing binary over 4 years. To the left and the right, the mutual inclination $I_{\rm p}$ was altered by $+1^\circ$ and $-1^\circ$ compared to the central pair.}
\label{fig:Chromosomes}
\end{center}  
\end{figure}

In Fig.~\ref{fig:Chromosomes} we illustrate the transit timing of a circumbinary planet transiting a non-eclipsing binary, extracted from one of our synthetic populations described in Section~\ref{sec:distributions}. The system is composed of a 35.1 day planet orbiting a 7.1 day binary with masses 1.06 and 0.44 $M_{\odot}$. The orbit of the planet, with respect to the binary, is nearly polar: $I_{\rm p} = 102^{\circ}$. Transits, represented by horizontal black lines, occur within regions of transitability (grey area) on the primary (A) and secondary (B) stars. The time spent in transitability for the secondary star is longer because, being less massive, it has a wider orbit.

The difference between the central system and the systems on either side, is only a $1^{\circ}$ change in the mutual inclination. Despite such a small difference, the transit sequences vary significantly. This is due to the relative motion between one of the stars and the planet, who move in different directions on the sky. Any slight changes to the orbital parameters can cause the planet to be early or late enough such that it misses a transit, for instance the transit on the secondary star near 2.5 years; it is present on the system at the left hand side but missing in the middle and on the right. 

At other times, the same change does not cause particular transits to be missed: there is pair of transits on the primary, near 2 years, that is present in all three variations of the system. Whilst the transits still occur on these particular orbits, the exact mid-transit times and durations vary by hours, although this is not discernible on the plot. 

By being so sensitive to the orbital parameters, one can obtain tight constraints on a system's parameters, provided that there are enough transits. Inversely, it implies that a high precision is required to predict future transits accurately.

\subsection{ Impact of the observing timespan}\label{sec:transit_prob}

\begin{figure*}  
\begin{center}  
	\begin{subfigure}[b]{0.33\textwidth}
		\caption{Uniform 0$^\circ$ to 15$^\circ$}
		\includegraphics[width=\textwidth]{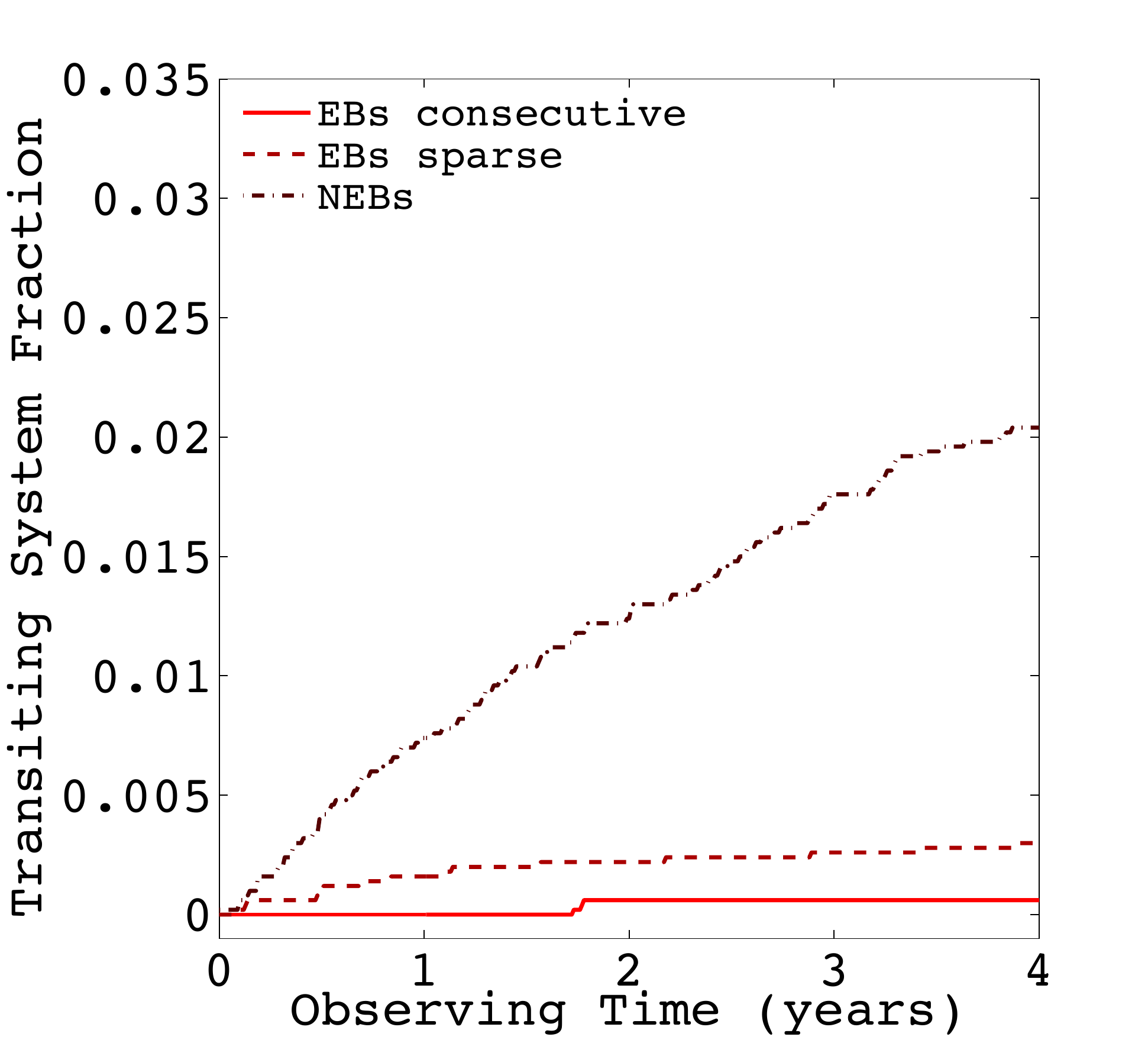}  
		\label{fig:ObservingTime_Zoom_15deg}  
	\end{subfigure}
	\begin{subfigure}[b]{0.33\textwidth}
		\caption{Uniform 0$^\circ$ to 30$^\circ$}
		\includegraphics[width=\textwidth]{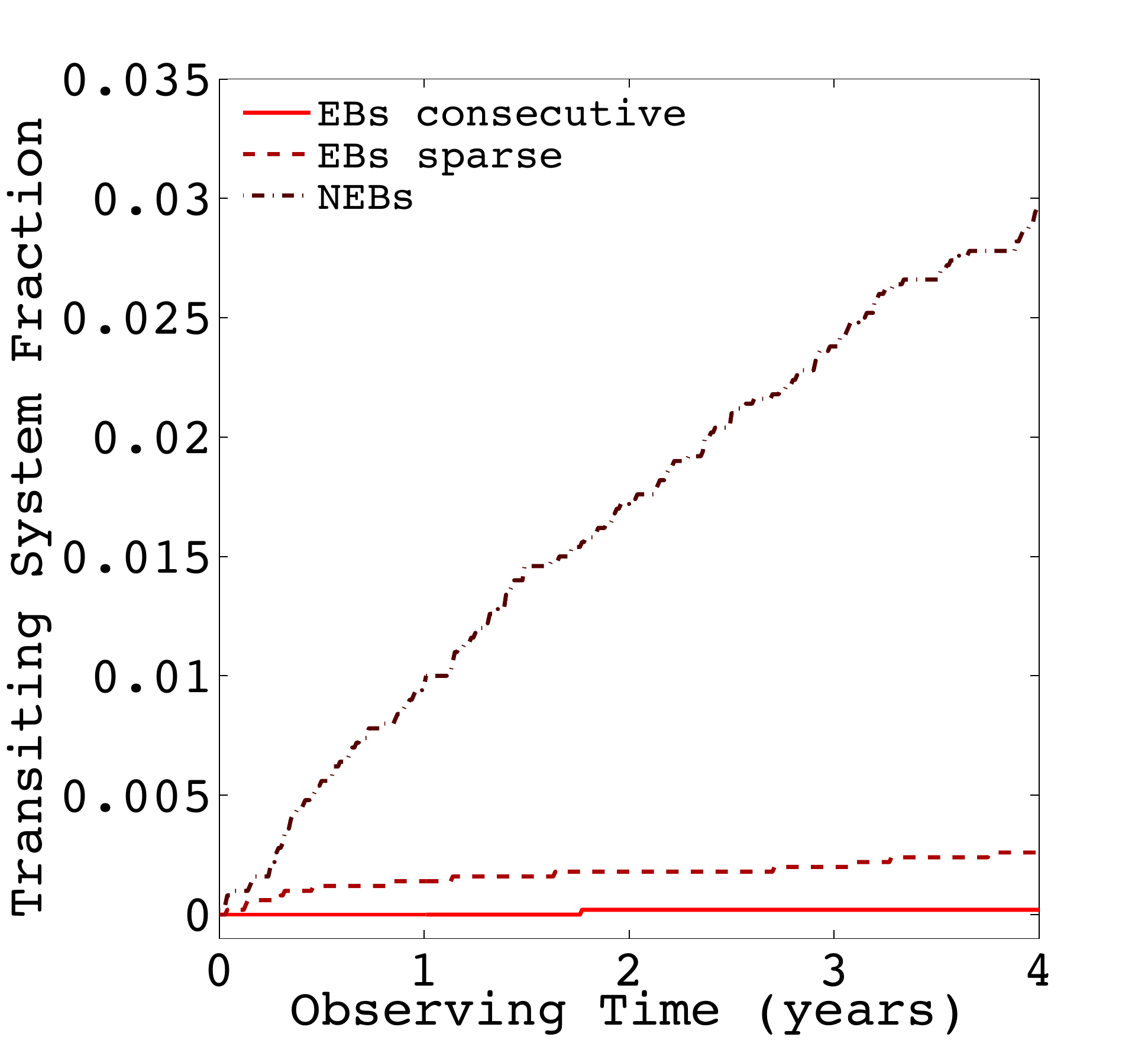}  
		\label{fig:ObservingTime_Zoom_30deg}  
	\end{subfigure}
	\begin{subfigure}[b]{0.33\textwidth}
		\caption{Uniform 0$^\circ$ to 45$^\circ$}
		\includegraphics[width=\textwidth]{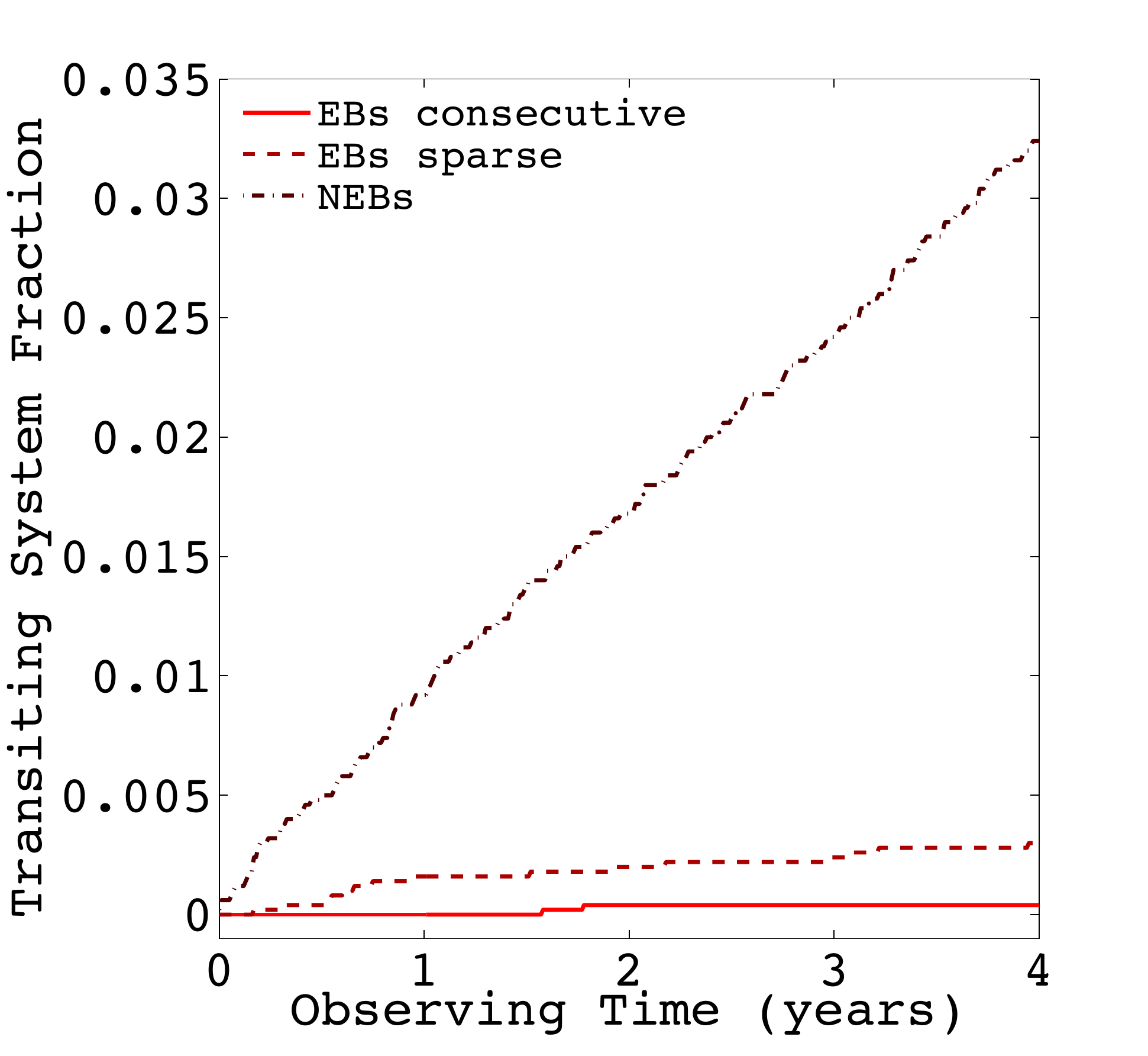}  
		\label{fig:ObservingTime_Zoom_45deg}  
	\end{subfigure}
	\caption{Integrated probability of transit as a function of time for a sample of simulated Kepler-16-like system ($P_{\rm p}$ = 228 d) having a range of mutual inclinations as labeled on individual panels. Those plots are limited to the length of a typical {\it Kepler} timeseries.
}\label{fig:ObservingTime_Zoom}  
\end{center}  
\end{figure*} 

\begin{figure*}  
\begin{center}  
	\begin{subfigure}[b]{0.33\textwidth}
		\caption{Uniform 0$^\circ$ to 15$^\circ$}
		\includegraphics[width=\textwidth]{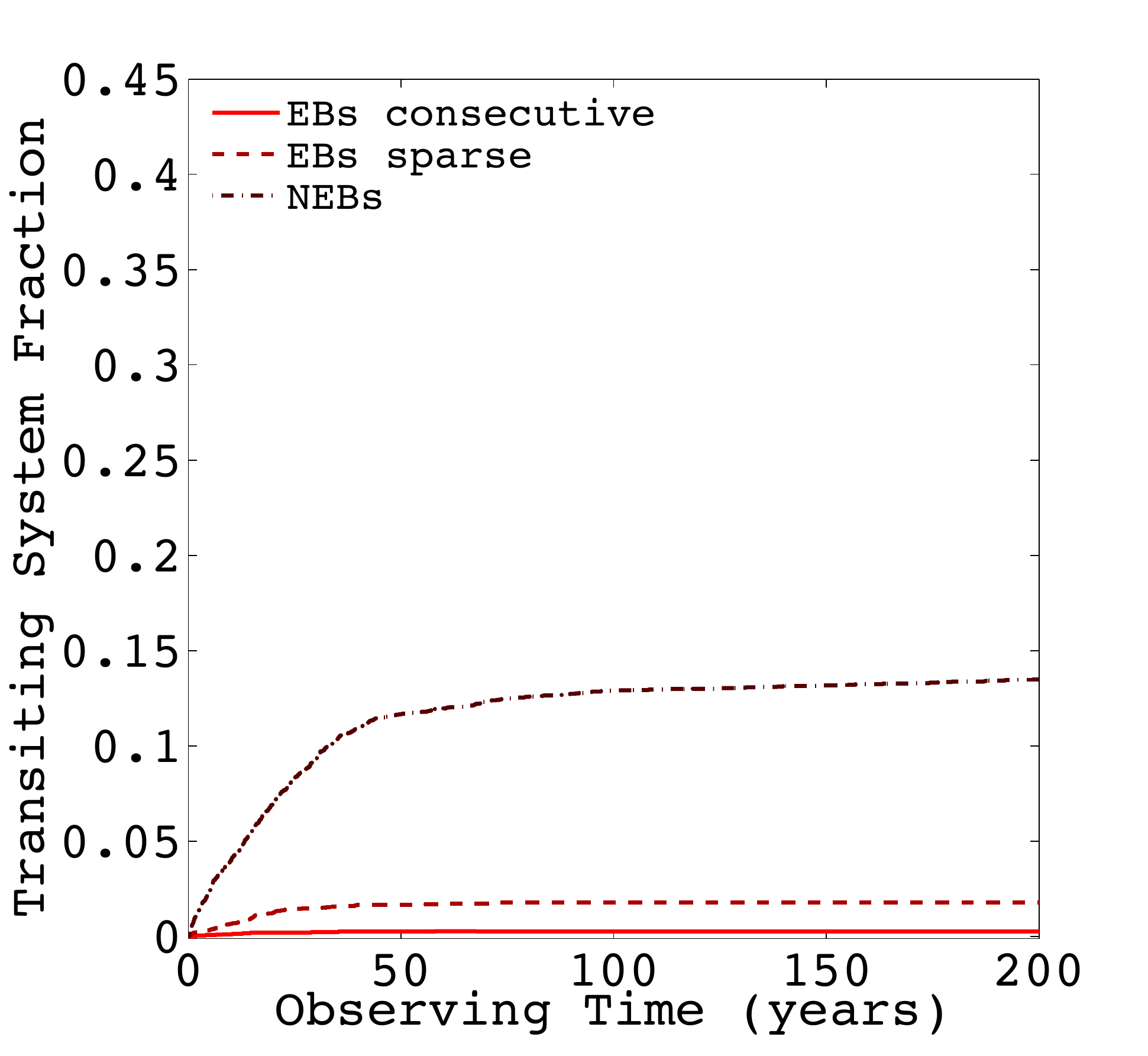}  
		\label{fig:ObservingTime_15deg}  
	\end{subfigure}
	\begin{subfigure}[b]{0.33\textwidth}
		\caption{Uniform 0$^\circ$ to 30$^\circ$}
		\includegraphics[width=\textwidth]{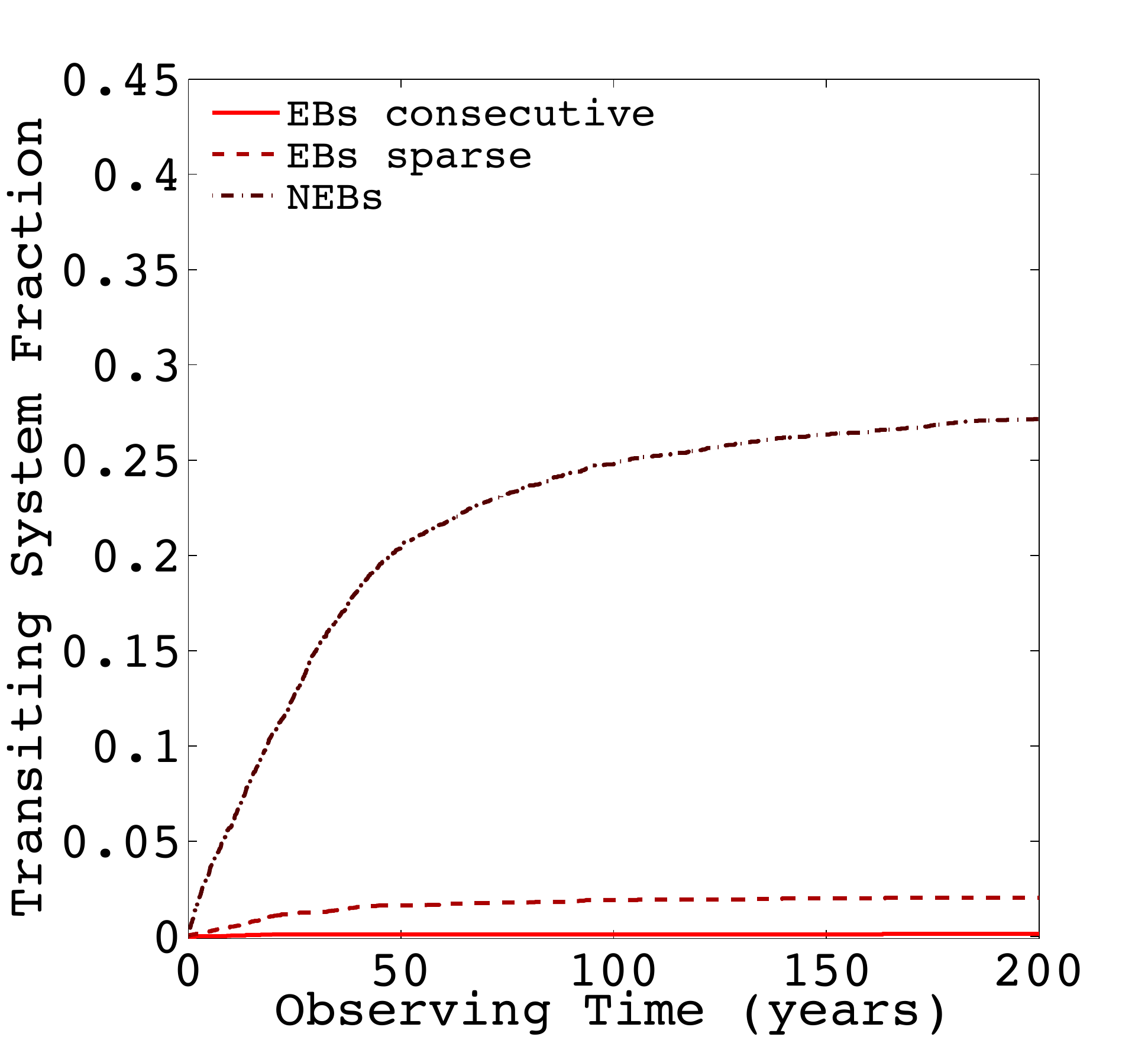}  
		\label{fig:ObservingTime_30deg}  
	\end{subfigure}
	\begin{subfigure}[b]{0.33\textwidth}
		\caption{Uniform 0$^\circ$ to 45$^\circ$}
		\includegraphics[width=\textwidth]{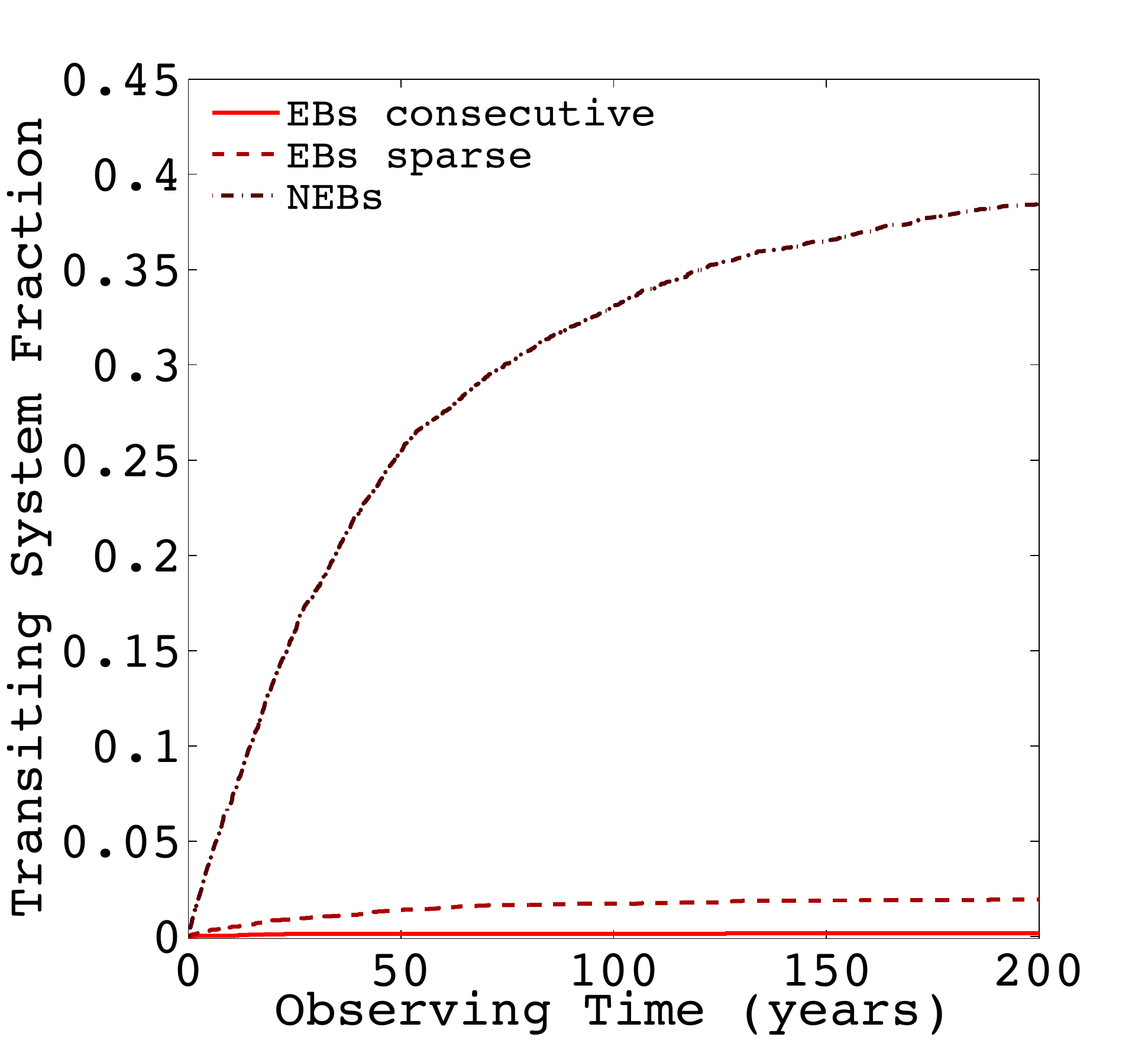}  
		\label{fig:ObservingTime_45deg}  
	\end{subfigure}
	\caption{The same as on Fig. \ref{fig:ObservingTime_Zoom} but extended to 200 years.
}\label{fig:ObservingTime}  
\end{center}  
\end{figure*}

Unlike the transit probabilities around single stars, which are static, the probability of observing a transit of a misaligned circumbinary system increases with time. There are two principal contributions:

\begin{itemize}
\item For circumbinary systems exhibiting transitability, any given period has a transit probability between 0 and 1. Longer observations mean that the planet covers more orbits, and hence has more opportunities to transit.
\item The state of transitability is time-dependent, due to precession (Fig. \ref{fig:Precession_Illustration}). Longer observations yield more time spent in transitability, and consequently a greater chance of catching a transit.
\end{itemize}


To demonstrate how transit probabilities evolve with time, and explore the link with mutual inclination, we took the Kepler-16 system as a template. We reproduced its binary and planet masses, radii, periods and eccentricities. The angles within the orbital planes, $\omega_{\rm bin}$, $M_{\rm bin}$, $\omega_{\rm p}$ and $M_{\rm p}$, where $M$ denotes the initial mean anomaly, were randomised between 0$^\circ$ and 360$^\circ$. For the inclinations we drew $I_{\rm p}$ from three uniform distributions: 0$^\circ$ and 15$^\circ$, 0$^\circ$ and 30$^\circ$, and 0$^\circ$ and 45$^\circ$. The nodal angle $\Omega_{\rm p}$ was randomly drawn between 0$^\circ$ and 360$^\circ$, since observations could coincide with any point on the precession period. The 3D orientation of the circumbinary system on the sky, with respect to a given observer, was randomised by applying a uniform 3D rotation algorithm by \citet{Arvo:1992lr}.

\begin{figure*}[t]  
\begin{center}  
	\begin{subfigure}[b]{0.4\textwidth}
		\caption{$P_{\rm bin}= 41 {\rm d}, P_{\rm p}= 228 {\rm d}$}
		\includegraphics[width=\textwidth]{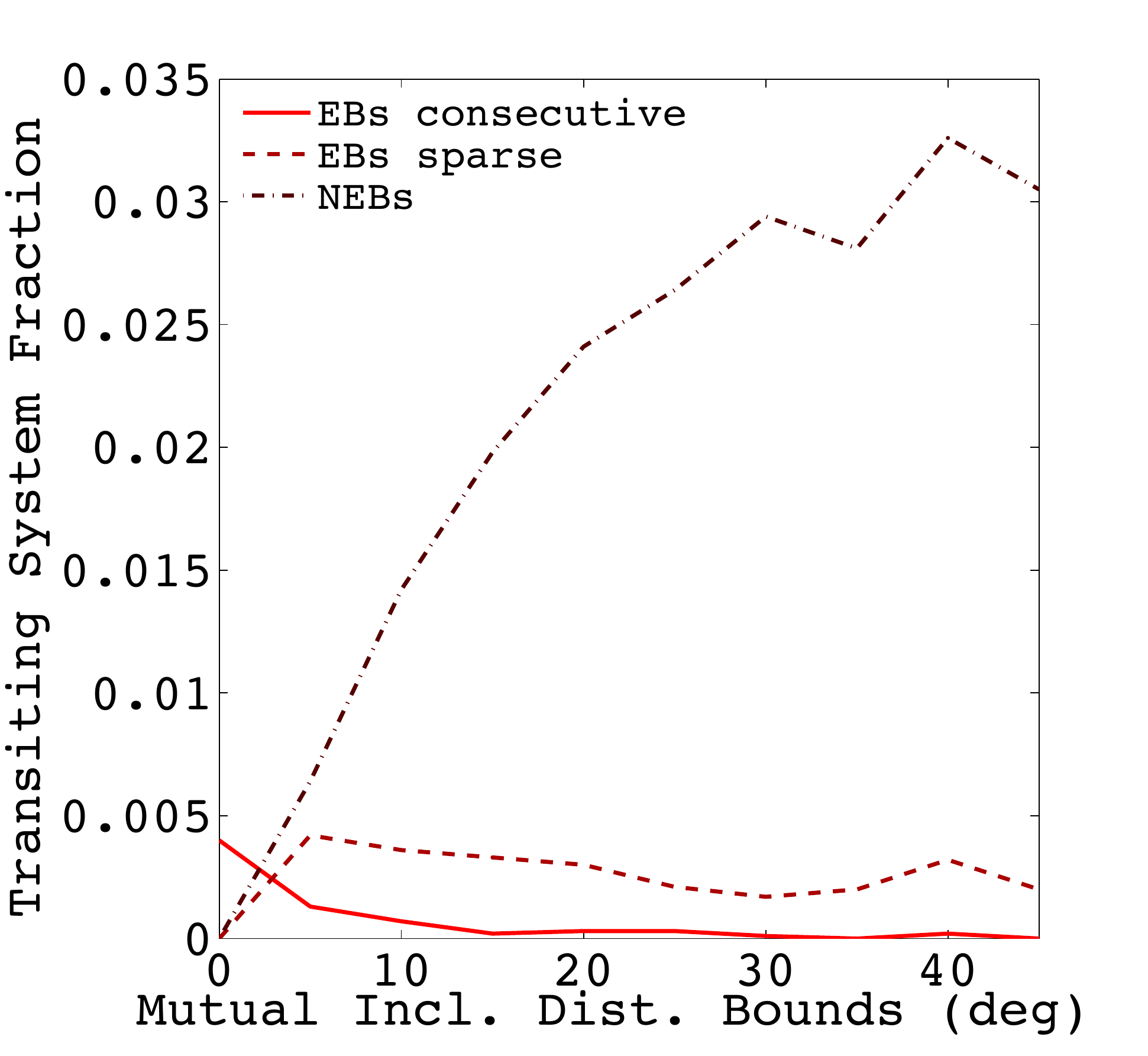}  
		\label{fig:MutualIncl_a}  
	\end{subfigure}
	\begin{subfigure}[b]{0.4\textwidth}
		\caption{$P_{\rm bin}= 5.1 {\rm d}, P_{\rm p}= 28.5 {\rm d}$}
		\includegraphics[width=\textwidth]{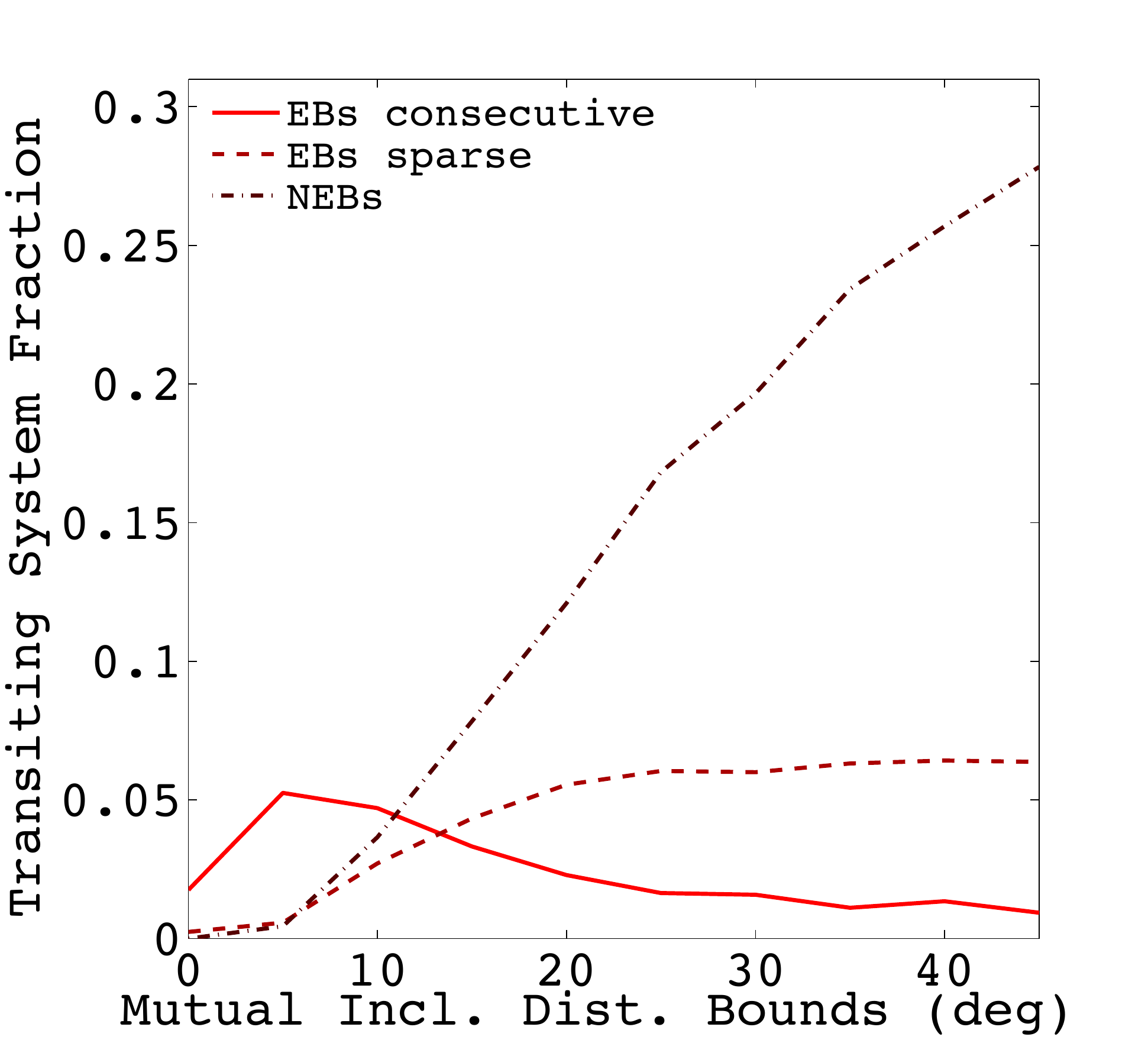}  
		\label{fig:MutualIncl_b}  
	\end{subfigure}	\caption{Simulated fractions of transiting systems for a Kepler-16-like system transiting eclipsing and non-eclipsing binaries, as a function of the mutual inclination distribution. The periods are reduced by a factor of 8 in b). Note that the two vertical axes are different.}
	 \label{fig:MutualIncl}
	 \end{center}  
\end{figure*} 

We simulated 10\,000 randomly drawn systems over 4 years, and recorded stellar eclipses and planet transits. Circumbinary systems were then discard as non-transiting or classified as transiting in one of the three categories described in Section~\ref{sec:types_of_transits}. In Fig.~\ref{fig:ObservingTime_Zoom} we plot the fraction of systems in each of the three classes as a function of time.

The fraction in the category of ``EBs consecutive" remains at zero until the planet has completed at least two orbits (1.25 years), due to the requirement of three consecutive primary transits. The small jump corresponds to the small fraction of  systems transiting eclipsing binaries. Once three orbits (1.87 years) have been completed, there are no additional detections. 

The fraction for the other two categories can increase from zero immediately since we set a requirement of a single transit. Three trends are immediately recognisable:
\begin{itemize}
\item There is a higher fraction of systems in the ``EBs sparse" class, than in the ``EBs consecutive" class. This is because a planet inclined with respect to the binary plane precesses, leading more systems to enter into transitability.
\item The fraction of systems whose planet transits an ``NEB" is larger than in the ``EBs sparse" category. This is due to a property that pervades this entire paper: whilst the probability of detecting a transit in any given non-eclipsing system is low, non-eclipsing systems make up the large majority of all binaries (98\% in this particular example).
\item The total fraction of systems whose planet transits an ``NEB" increases with the spread in orbital inclination. This indicates that binary systems at higher orbital inclination on the sky are reached, increasing the number of considered systems (more details in Sec.~\ref{subsec:incl}).
\end{itemize}



Out of interest we extended the simulations of Fig. \ref{fig:ObservingTime_Zoom} to a hypothetical 200 year continuous observing run. The results, provided in Fig. \ref{fig:ObservingTime}, indicate that the fraction of transiting systems asymptotes. After such a long observing time, all systems exhibiting transitability have had at least one transit occur: it becomes unlikely that a planet would pass the binary orbit so many times and avoid transiting. The fraction of planets transiting eclipsing binaries is barely discernible from zero on this scale.

The asymptote is the total fraction of misaligned circumbinary systems that could ever be observed to transit, at some remote point in time. It is a significant departure from traditional considerations about transit probabilities around single stars, particularly for planets with periods greater than 200 days. This confirms the interest of defining the concept of {\it transitability}: the probability that a given system will go through transitability is static. It is a property linked to a system's orbital and physical parameters whose only remaining unknown is the inclination of the system on the sky (see Fig.~\ref{fig:TransitabilityExample}), like the probability of transit is, in the case of a planet revolving around a single star.

\subsection{The effects of orbital parameters}\label{sec:orbit_params}

\begin{figure}  
\begin{center}  
\includegraphics[width=0.45\textwidth]{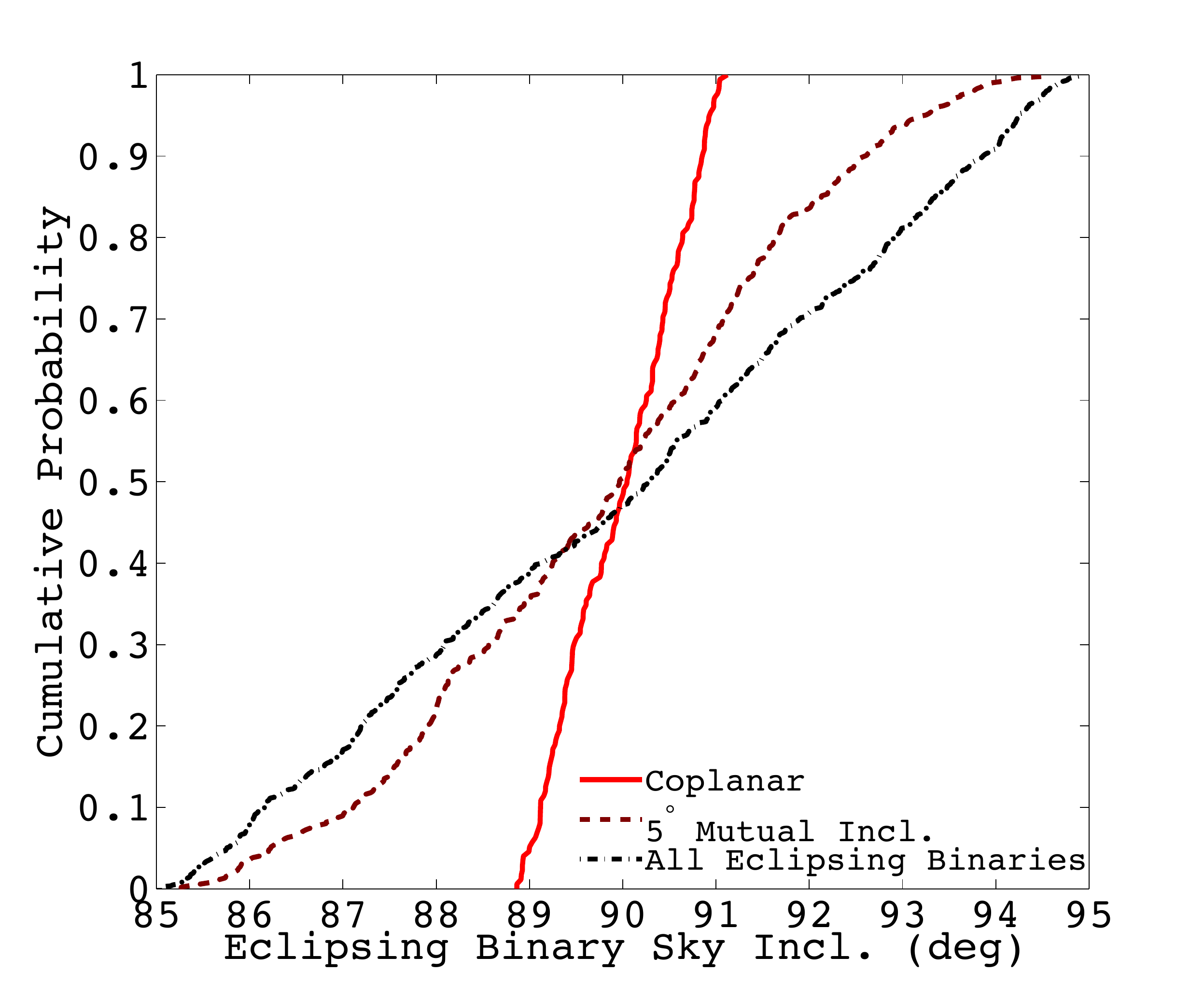}  
\caption{Cumulative probabilities of the sky inclination of eclipsing binaries. All eclipsing binaries are compared to the sky inclinations of eclipsing systems with coplanar transiting planets, and with two distributions of misaligned systems. 
}\label{fig:InclinationSpread} 
\end{center}  
\end{figure}

The current section will explore how different parameters change the dynamical and geometric properties, and ultimately affect the likelihood of transit. To demonstrate this we took the same template as in Section~\ref{sec:transit_prob}: a Kepler-16 like system. Individual parameters were then altered in specific tests.

\subsubsection{Mutual inclination and orbital periods}\label{subsec:incl}

The parameters with the most impact on transit probabilities are the mutual inclination between the planetary and binary planes, and their respective orbital periods. In this section we quantify their effects on the expected frequency of transiting systems. 

First, for a Kepler-16 analog, we simulated a suite of uniformly drawn mutual inclination distributions whose upper bounds ranged from 0$^\circ$ (strictly coplanar) to 45$^\circ$, with 5$^\circ$ increments. The results, depicted in Fig. \ref{fig:MutualIncl_a}, show the fraction of each of the three types of transiting systems. The simulations were then re-run with the binary and planet periods reduced by a factor of eight (Fig.~\ref{fig:MutualIncl_b}). Despite some kinks in the plots that reflect statistical noise, our simulations are sufficiently sampled to illustrate some key trends.

Fig. \ref{fig:MutualIncl_a} reveals that the consecutive-transit population is maximised in the coplanar case. Its percentage is similar to its equivalent single-star system: 0.43\%. For the other two categories there are, as expected, no transits.
Moving away from the strictly coplanar case, the frequency of planets transiting consecutively eclipsing binaries decreases and reaches a low plateau. If a planet has a significant mutual inclination, it can still transit eclipsing binaries but they will more likely be in the sparse category. Except in the coplanar case, the majority of transiting systems are around non-eclipsing binaries. Like in Section~\ref{sec:transit_prob}, this is due to the large majority of binaries being non-eclipsing. Aside from some statistical noise between 30$^\circ$ and 40$^\circ$, the amount of transiting systems increases with mutual inclination spread, which is consistent with Section~\ref{sec:transit_prob}.

As one would expect, shortening the orbital periods produces an increase in the fraction of transiting systems (Fig.~\ref{fig:MutualIncl_b}, note the higher vertical scale).
An additional trend has also appeared. In Fig.~\ref{fig:MutualIncl_a} consecutive transits of eclipsing binaries are maximised for completely coplanar systems. For shorter period systems in Fig.~\ref{fig:MutualIncl_b}. the peak instead corresponds to a spread in mutual inclinations. To determine the cause of this, consider Fig. \ref{fig:InclinationSpread}, where we have plotted the cumulative distribution of the sky inclination of the binaries being transited, taken from the simulation in Fig.~\ref{fig:MutualIncl_b} ($P_{\rm bin} = 5.1$ days). For comparison, we have also included the cumulative distribution of all eclipsing binaries in our simulation, irrespective of transits (Fig.~\ref{fig:InclinationSpread}, black dash-dotted line). The binaries eclipse if their sky inclination is between 85$^\circ$ to 95$^\circ$, and the inclination distribution is approximately uniform. When planets are all coplanar, however, only eclipsing binaries with inclinations between 89$^\circ$ and 91$^\circ$ can be transited consecutively (Fig.~\ref{fig:InclinationSpread}, red solid line). This is only approximately 20\% of all eclipsing systems. When the mutual inclination spread is increased to 5$^\circ$, the planet is able consecutively transit across eclipsing binaries of all inclinations on the sky. Increasing the mutual inclination spread too much causes too many planets to be too misaligned to transit consecutively.

A spread in mutual inclination can produce transits on the full spectrum of possible eclipsing binaries. A similar effect happens on the probability of a second planet transiting in a system where one is already known to be \citep{Gillon:2011lr,Figueira:2012lr}.

\begin{figure*}  
\begin{center}  
	\begin{subfigure}[b]{0.4\textwidth}
		\caption{$P_{\rm bin}= 41 {\rm d}, P_{\rm p}= 228 {\rm d}$}
		\includegraphics[width=\textwidth]{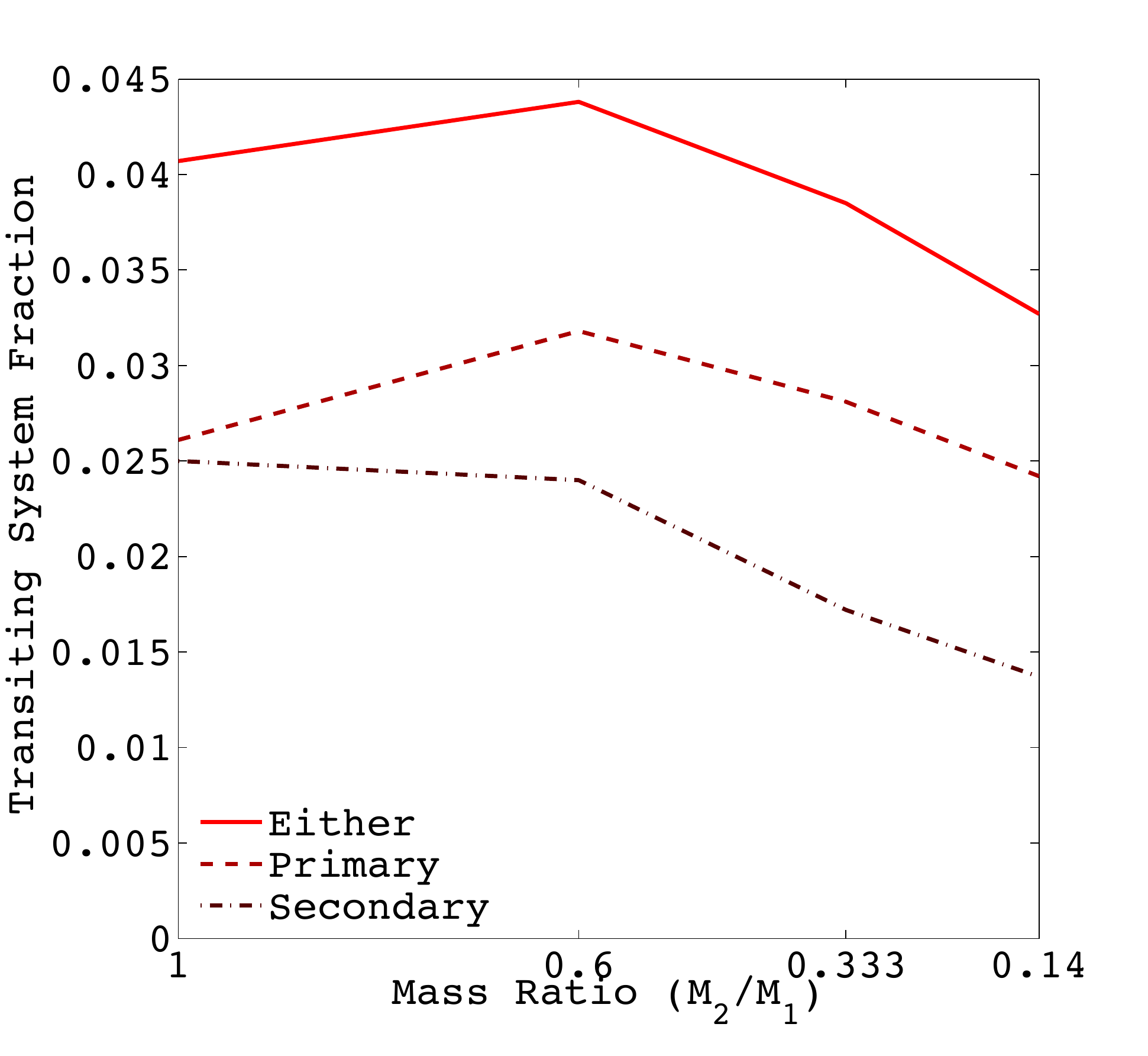}  
		\label{fig:MassRatioTest_a}  
	\end{subfigure}
	\hspace{1cm}
	\begin{subfigure}[b]{0.4\textwidth}
		\caption{$P_{\rm bin}= 5.1 {\rm d}, P_{\rm p}= 28.5 {\rm d}$}
		\includegraphics[width=\textwidth]{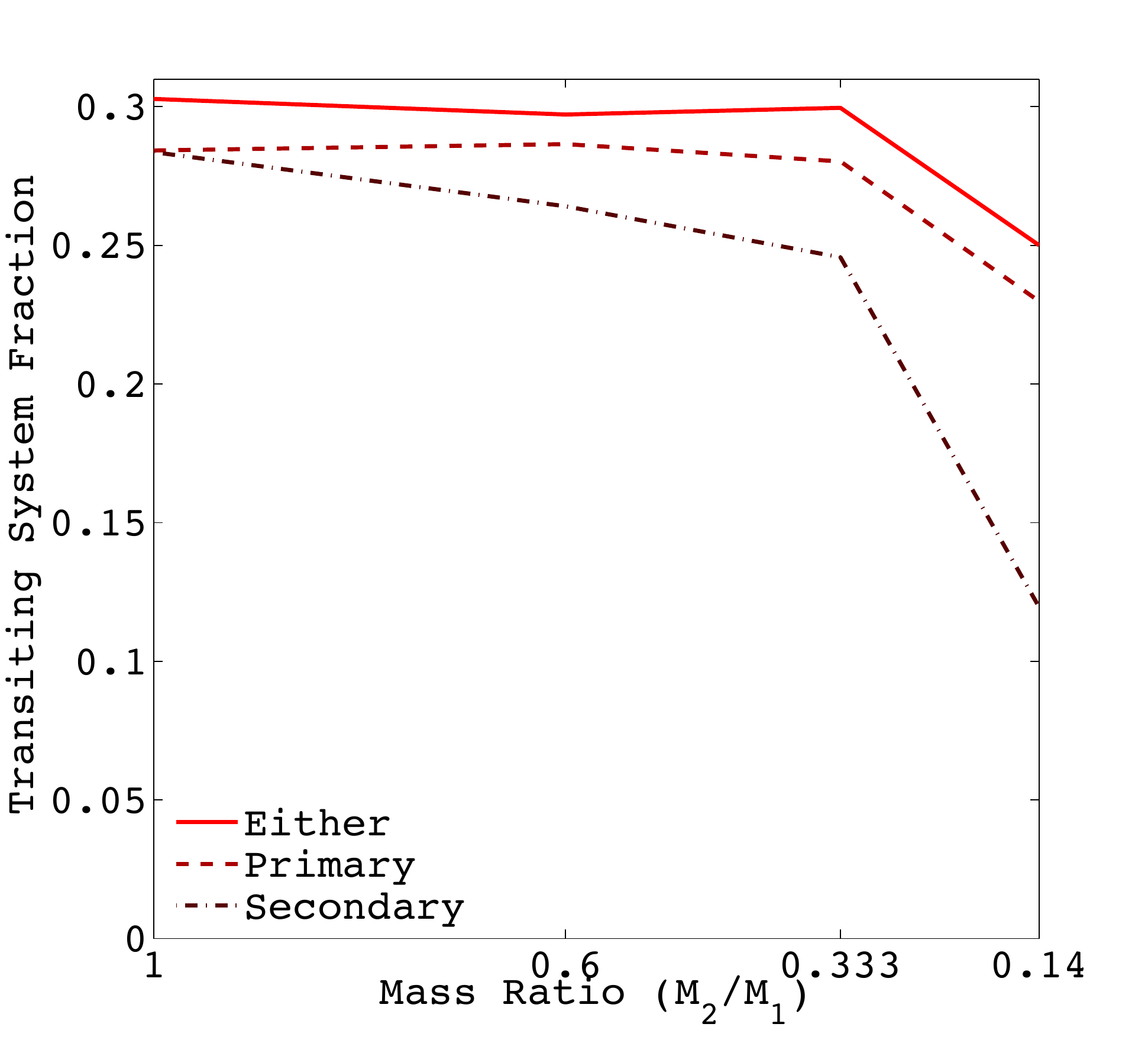}  
		\label{fig:MassRatioTest_b}  
	\end{subfigure}
	\caption{Effect of a change in mass ratio --keeping total mass constant-- on transit probabilities over either star; calculated for four years and for a mutual inclination distribution uniform between 0$^\circ$ and 30$^\circ$.}
	 \label{fig:MassRatioTest}
	 \end{center}  
\end{figure*}

\subsubsection{Binary mass ratio}\label{sec:MassRatio}

To test the effect of the binary mass ratio we constructed four test systems, again based on Kepler-16, with primary masses of 1, 1.25, 1.5 and 1.75~$M_{\odot}$. The secondary masses were assigned such that the total mass was kept constant at 2~$M_{\odot}$. A radius was calculated for each star using the mass-radius relation \citep{Kippenhahn:1994fj}:
\begin{equation}\label{eq:massradius}
R/R_{\odot} = M^{\xi}/M_{\odot},
\end{equation}
where $\xi=0.57$ for $M \geq1M_{\odot}$ and $\xi=0.8$ for $M<1M_{\odot}$. \\

The mutual inclination was drawn from a uniform distribution between 0$^\circ$ and 30$^\circ$. For each test system, 10\,000 simulations were run for four years. Transits on both stars in non-eclipsing binaries were counted. The results are shown in Fig.~\ref{fig:MassRatioTest_a}. This process was repeated with binary and planet periods reduced by a factor of 8 (Fig.~\ref{fig:MassRatioTest_b}).

For an equal mass binary the chance of transiting either star is the same, assuming that the radii are the same (Eq. \ref{eq:massradius}). By varying the mass ratio away from unity, two effects are observed. Transits have an increased occurrence on the primary compared to the secondary simply because the primary star has a larger radius. In Fig.~\ref{fig:MassRatioTest_b} a sharp decrease in transits occurs between mass ratios of 0.33 and 0.14. The precession period increases with decreasing mass ratio \citep{Doolin:2011lr} such that the precession period of the lower mass ratio systems becomes longer than the four years of observations that set the boundary of our simulation. Some systems do not have enough time to move into a window of transitability and consequently, the fraction of transiting systems drops.

Knowing the rate at which planets precess in and out of view is vital for calculating their abundance. The binary mass ratio is a key component to estimating the occurrence rates of transiting circumbinary planets for surveys limited in time. 

\subsubsection{Other trends}

There are too many degrees of freedom in a circumbinary system to analyse all of them in depth. The most obvious that we have not investigated in detail is the impact of eccentricity of both the binary and planet. This parameter affects orbital precession \citep{Farago:2010fj,Doolin:2011lr} and stability \citep{Dvorak:1986fk,Dvorak:1989vn,Holman:1999lr}. Its effects are however included by construction in the population synthesis that will be described in Section~\ref{sec:distributions}.

\subsection{Varied transit shapes}

\begin{figure*}  
\begin{center}  
	\begin{subfigure}[b]{0.28\textwidth}
		\caption{EB consecutive}
		\includegraphics[width=\textwidth]{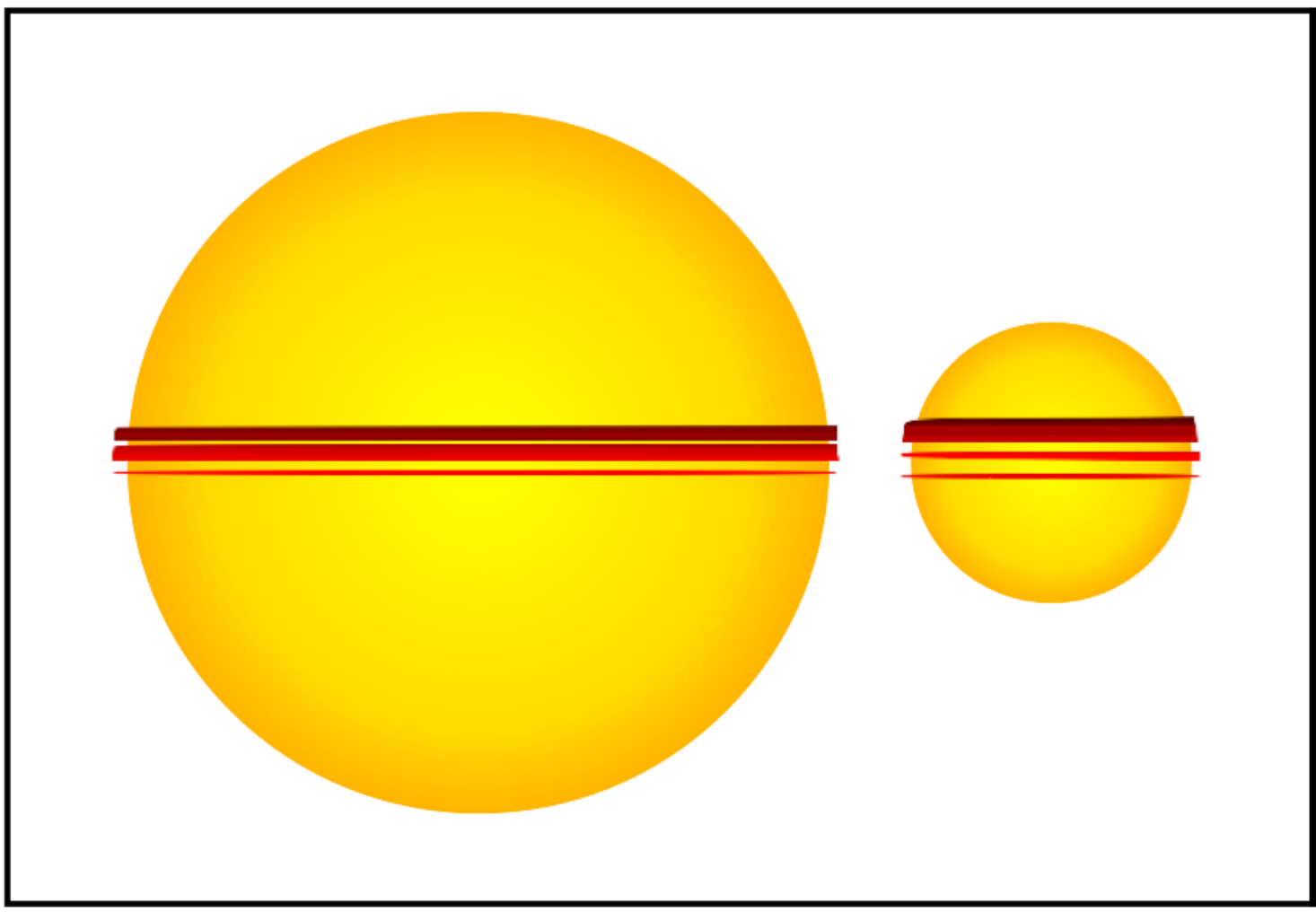}  
		\label{fig:StellarTomography_a}  
	\end{subfigure}
	\hspace{0.1cm}
	\begin{subfigure}[b]{0.28\textwidth}
		\caption{EB sparse}
		\includegraphics[width=\textwidth]{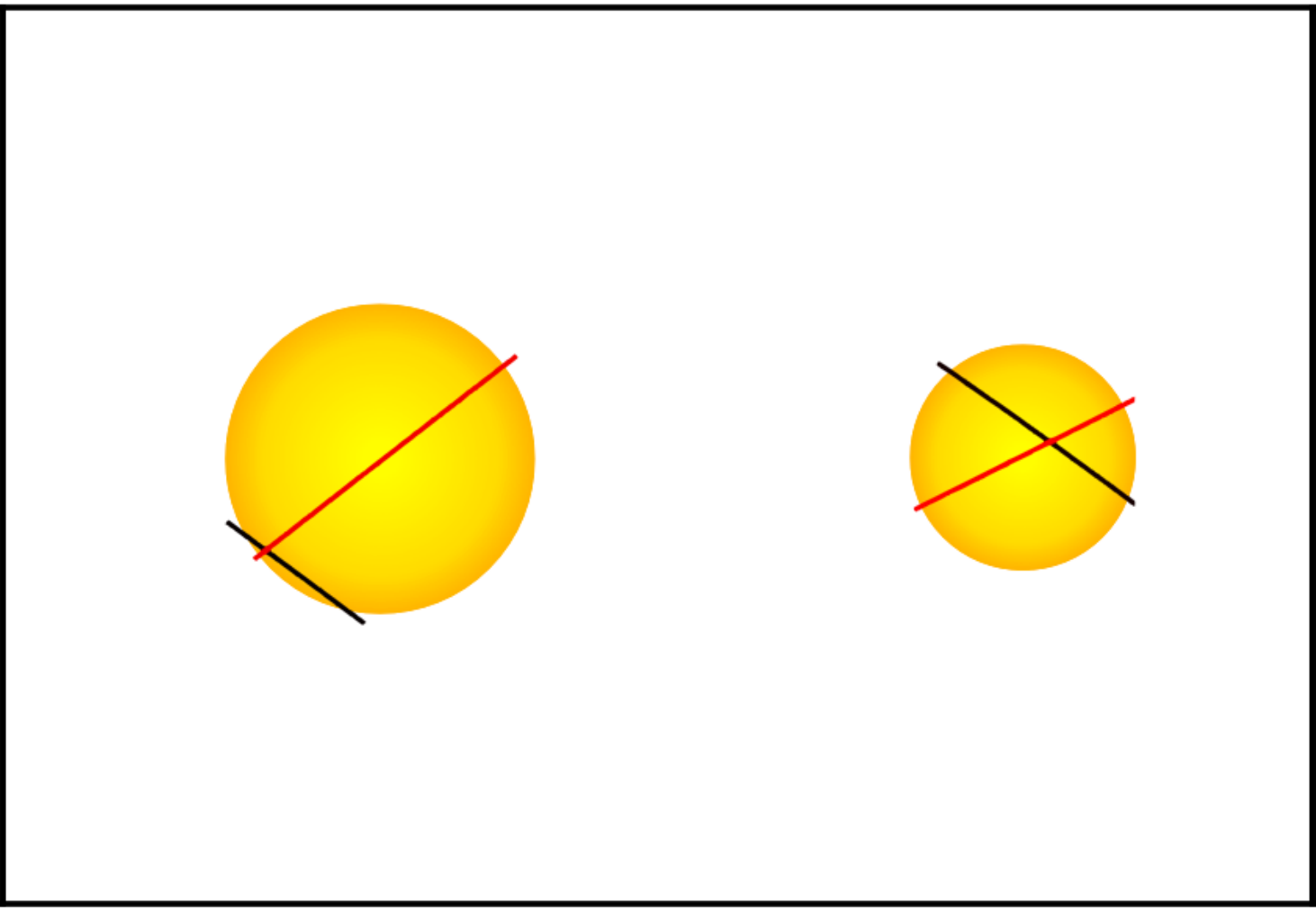}  
		\label{fig:StellarTomography_b}  
	\end{subfigure}
	\hspace{0.1cm}
	\begin{subfigure}[b]{0.28\textwidth}
		\caption{NEB}
		\includegraphics[width=\textwidth]{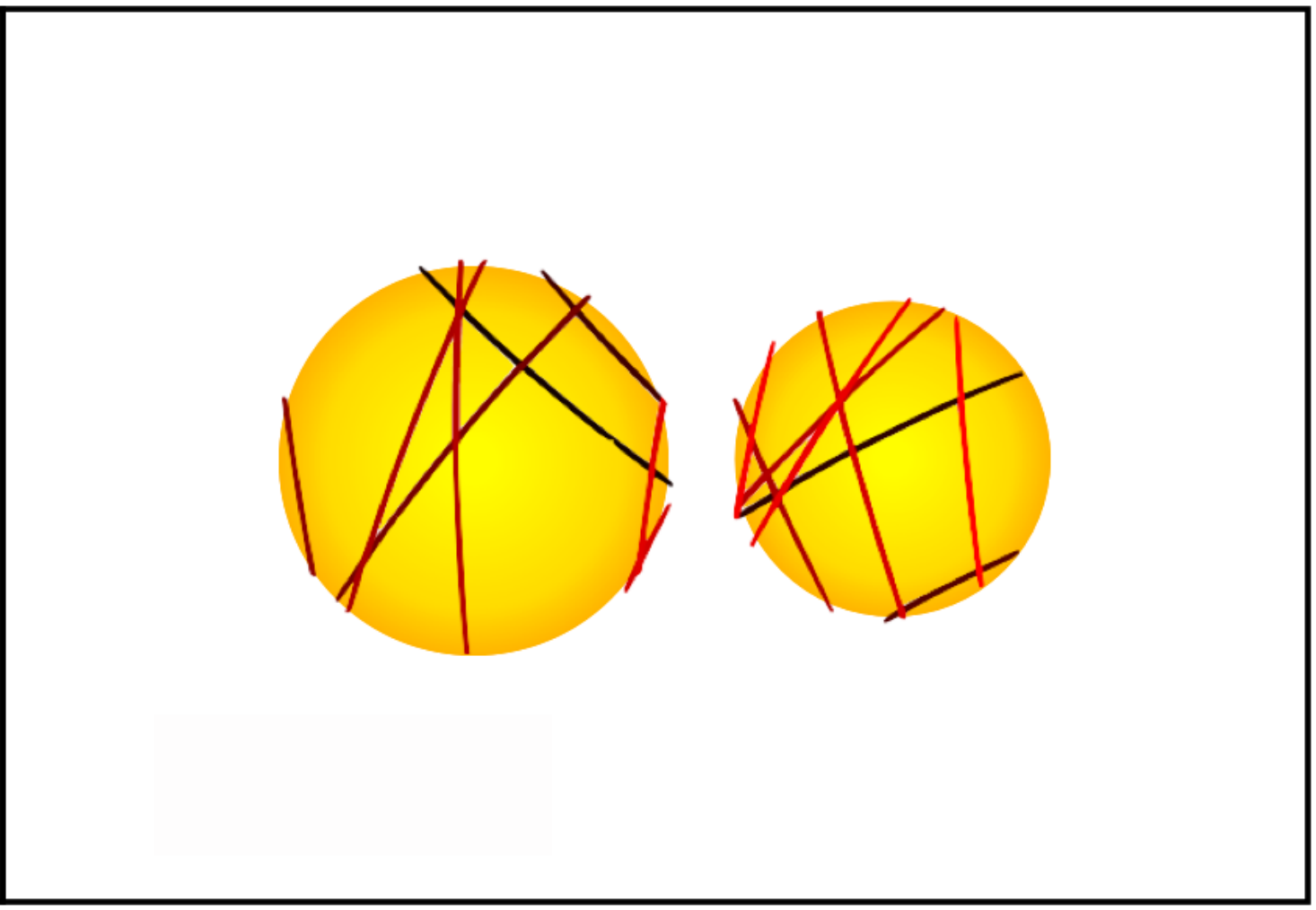}  
		\label{fig:StellarTomography_c}  
	\end{subfigure}
	\hspace{0.1cm}
	\begin{subfigure}[b]{0.07\textwidth}
		\includegraphics[width=\textwidth]{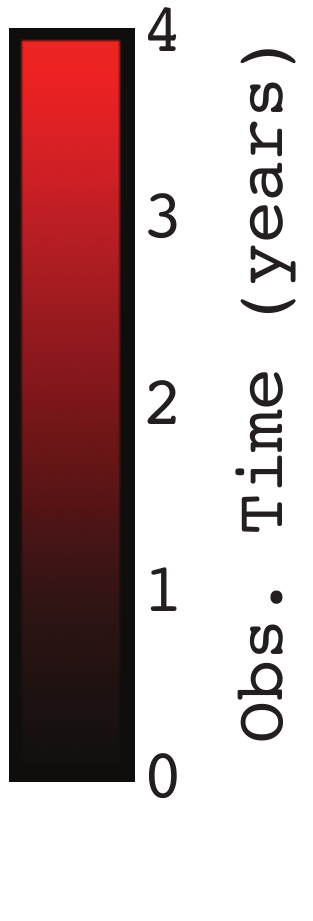}  
	\end{subfigure}
	\caption{Transit chords across both stars for: a) consecutive transits on an eclipsing binary, b) irregular transits over an eclipsing binary and c) a non-eclipsing binary. Colours are defined as a function of time, going from black to red. Stellar sizes are all shown to scale, but not the separations. The circumbinary systems were extracted from our populations described in Section~\ref{sec:distributions}. }
	 \label{fig:StellarTomography}
	 \end{center}  
\end{figure*}

The transit chords of a coplanar planet across an eclipsing binary are constant, and so are its two transit depths. The transit width can vary significantly between consecutive transits, since it is a function of the planet-star relative velocity. A small mutual inclination will make the transit chord vary from transit to transit, due to precession. An example is shown in Fig.~\ref{fig:StellarTomography_a}, where colour corresponds to the transit time over a 4 year simulation (starting black and finishing red).

If there is a significant mutual inclination, the planet is likely to fall into the ``sparse" transiter category. In addition to many transits being missed, the transits that do occur can have significantly different chords across the star, as exhibited in Fig.~\ref{fig:StellarTomography_b}. Consequently, there can be significant variations in the transit depth, width and asymmetry.

Moving to the case of non-eclipsing binaries, the only possible transiting planets are misaligned ones. Highly variable transit shapes are expected. An example is given in Fig.~\ref{fig:StellarTomography_c} is for a highly inclined system. Slight curvature is visible in some of the transit chords. The complexity in the transit shapes could make detecting these systems more difficult.

Chords spanning an entire surface, at different angles and for several phases of the stellar rotation would allow to construct tomographic maps and collect a detailed knowledge of the shape of the stars. This would be particularly interesting for studying tidal deformation in close binaries.

\section{Creating synthetic distributions}\label{sec:distributions}

In the previous section we demonstrated how different orbital parameters affect the fraction and type of transiting systems. We now extend our simulations from a small set of illustrative examples to a complete synthetic population of binary stars (Sec.~\ref{sec:binarydist}) and orbiting planets (Sec.~\ref{sec:planetdist}). This population forms the basis for Section~\ref{sec:results}, where we predict the number of transiting circumbinary planets in the {\it Kepler} data, around both eclipsing and non-eclipsing binaries. The methods of forming the population are general enough such that they could be applied to any telescope.


\subsection{The binary distribution}\label{sec:binarydist}
	
\begin{figure*}  
	\begin{center}  
	\centering
	\begin{subfigure}[b]{0.31\textwidth}
		\caption{}
		\includegraphics[width=\textwidth]{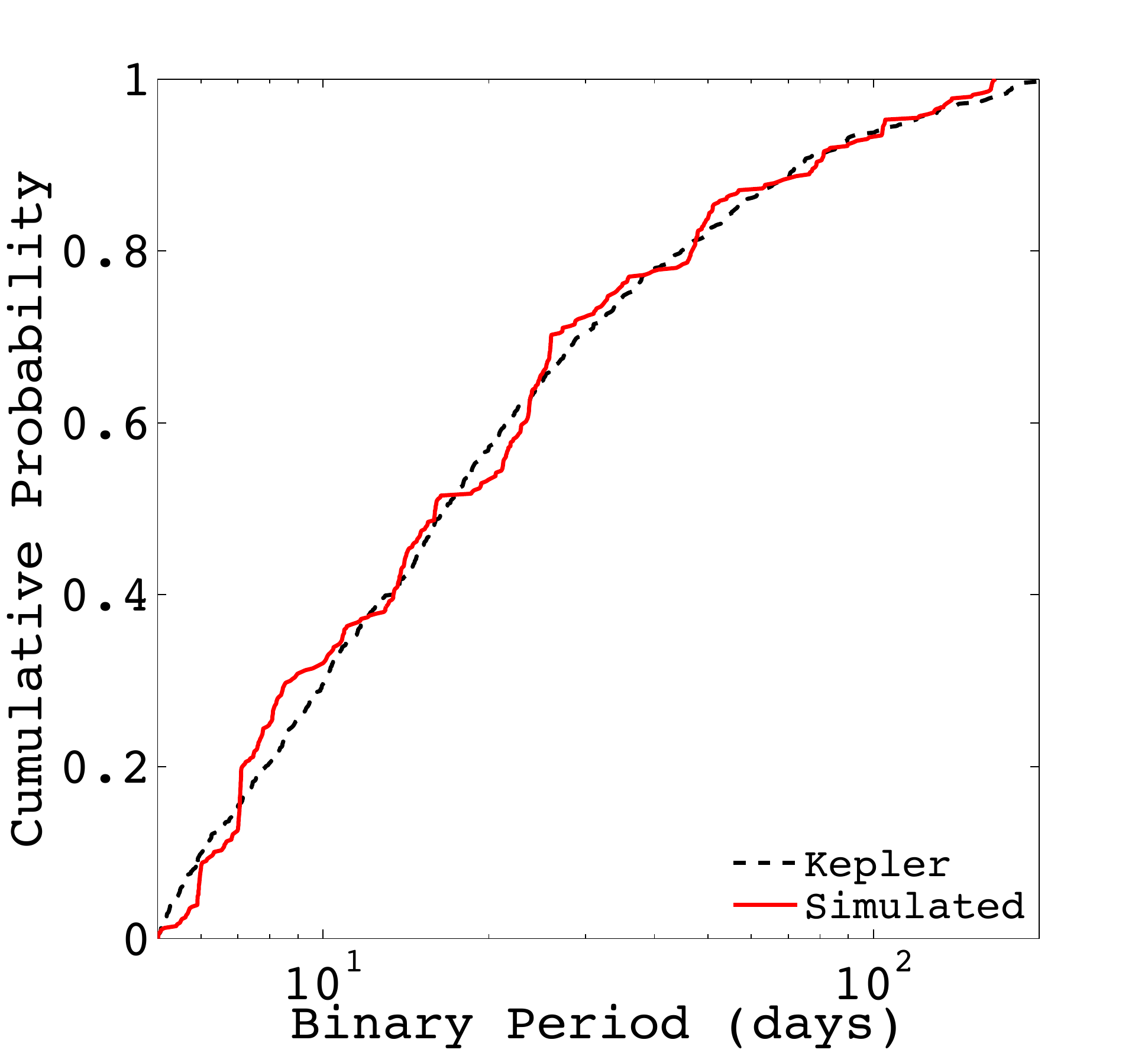}  
		\label{fig:BinaryTests_Period}  
	\end{subfigure}
	\begin{subfigure}[b]{0.31\textwidth}
		\caption{}
		\includegraphics[width=\textwidth]{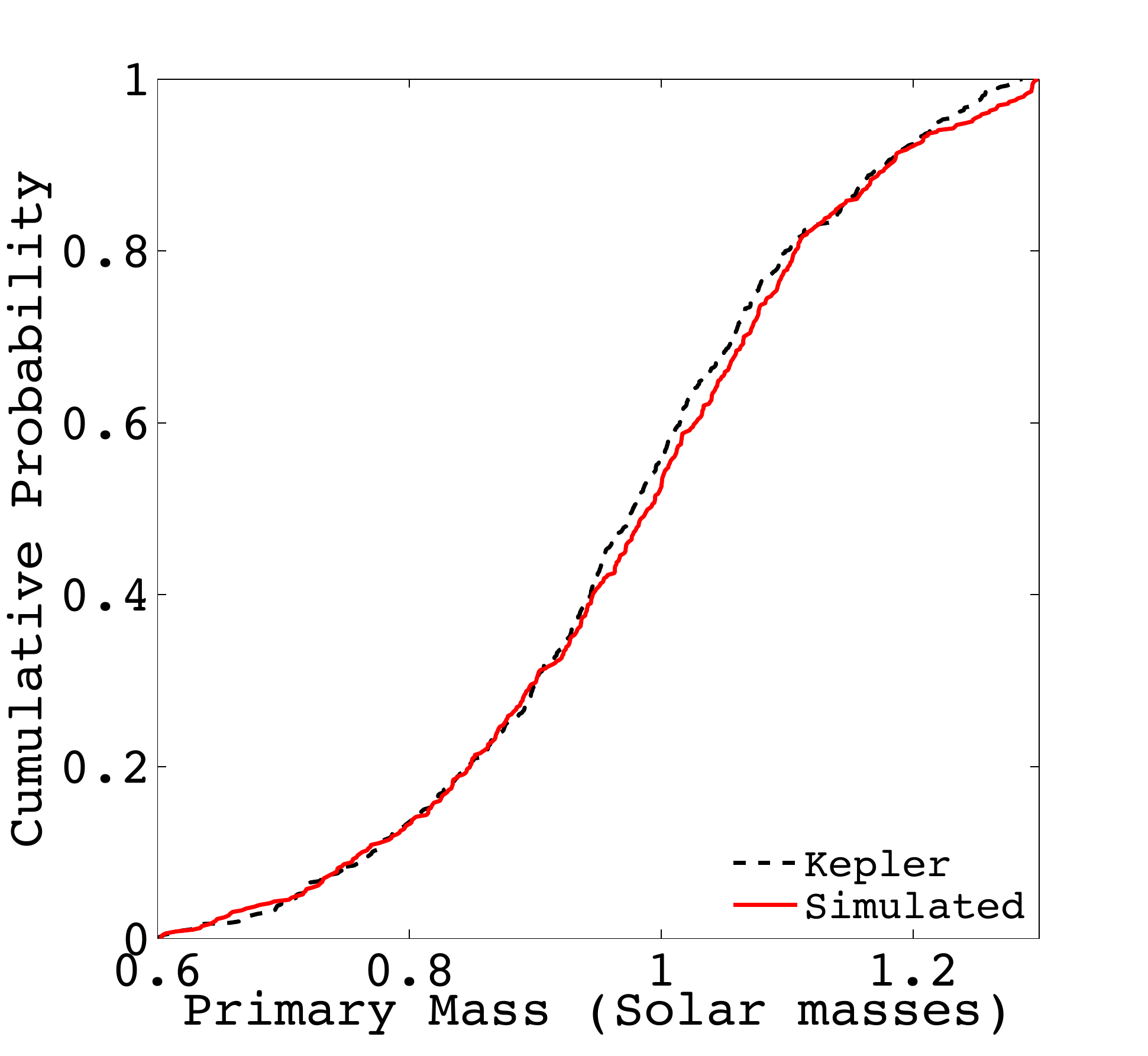}  
		\label{fig:BinaryTests_Mass}  
	\end{subfigure}
	\begin{subfigure}[b]{0.31\textwidth}
		\caption{}
		\includegraphics[width=\textwidth]{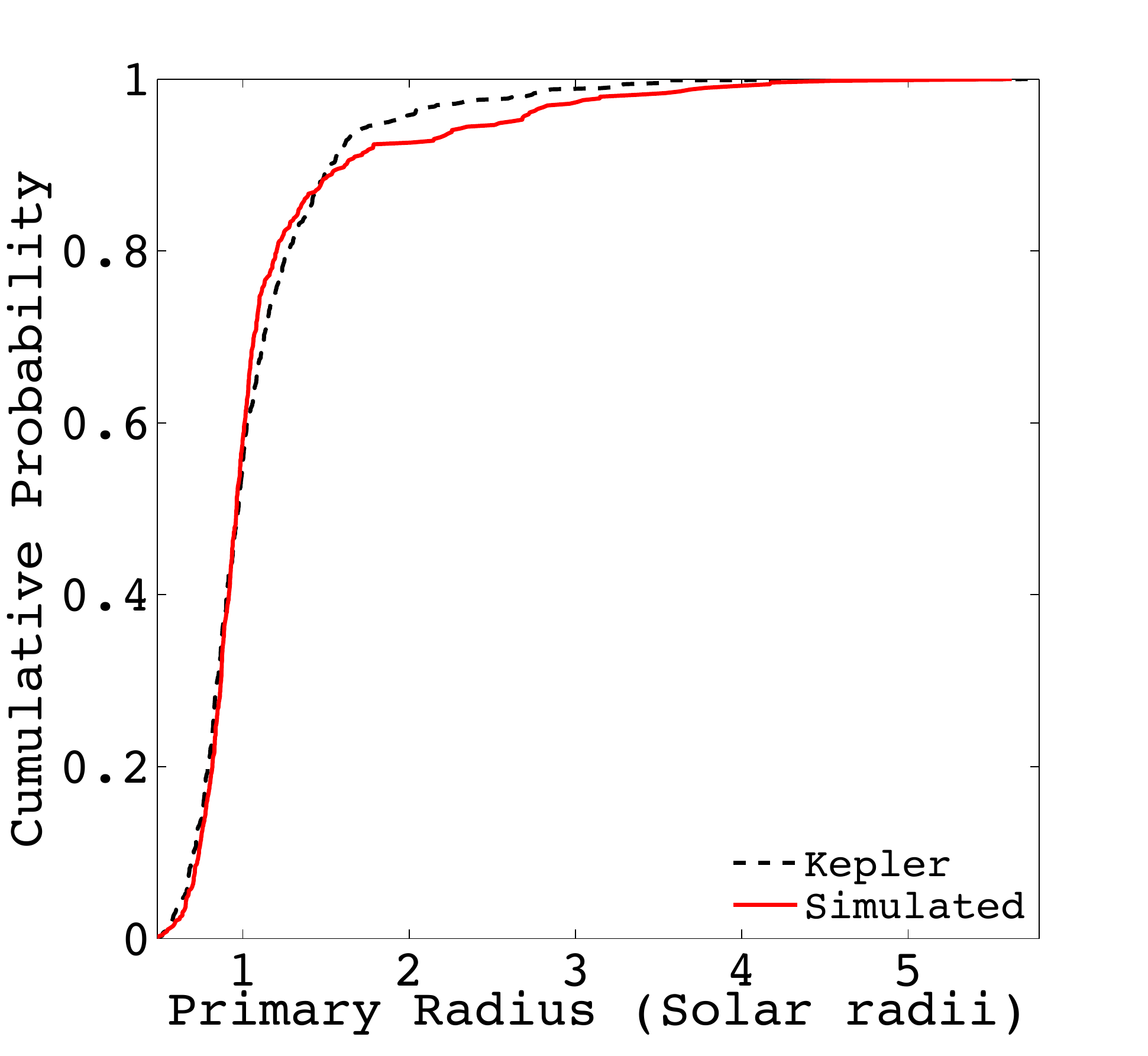}  
		\label{fig:BinaryTests_Radius}  
	\end{subfigure}
	\caption{Cumulative probability functions comparing the simulated eclipsing binary distributions based on \citet{Halbwachs:2003lr} (plain red) and the binaries found in the {\it Kepler} eclipsing binary catalog (dashed black). The comparison is done for binary periods between 5 and 200 days.}
	 \label{fig:BinaryTests}
	 \end{center}  
\end{figure*}

	\citet{Halbwachs:2003lr} conducted a volume-limited radial-velocity survey of binary stars in the solar neighbourhood, building upon the work of \citet{Duquennoy:1991kx}. \citeauthor{Halbwachs:2003lr} estimated that $13.5^{+1.8}_{-1.6}$ \% of main sequence stars exist in multiple systems, with orbital periods ranging from 1 day to 10 years. We have used their observed distribution to draw our binary systems.
Our reasons for choosing a volume-limited radial-velocity survey are that we perceived this sample is well controlled and defined, especially at periods longer than typically found in {\it Kepler}. 
	
	The distinction between the \citet{Halbwachs:2003lr} and {\it Kepler} binaries is that {\it Kepler}'s only contains eclipsing binaries, whilst \citet{Halbwachs:2003lr} observed binaries of any orientation. Consequently, {\it Kepler} contains a strong geometric observing bias towards short periods, whilst \citet{Halbwachs:2003lr} does not. We artificially applied this observing bias to \citet{Halbwachs:2003lr}, to make the two distributions comparable. This is a way to validate our approach, but also to verify for the first time whether the binary star population is similar throughout the Galaxy, by comparing the solar neighbourhood to the {\it Kepler} field. 
	
First, we needed to estimate the size of the {\it Kepler} sample. We searched the {\it Kepler} target catalog and selected all objects having a {\it Data Availability Flag} = 2. This is a flag marking all stars that have been observed and logged. As of May 10, 2013 there were over 200\,000 entries in the {\it Kepler} target catalog, but only 189\,440 had measured effective temperatures ($T_{\rm eff}$), surface gravities ($\log g$) and metallicities ($[Fe/H]$). We used the Torres relation \citep{Torres:2010uq} to estimate masses and radii for each of those entries. We only kept objects whose mass was estimated within: $0.6 < M_\star < 1.3 M_\odot$. This restriction is necessary to coincide with the sample that has been collected by \citeauthor{Halbwachs:2003lr} In addition, later, we will need radii estimates and those are better defined using the Torres relation over the range that we just chose. This leaves us with a list of 149\,526 systems.
Therefore, using \citeauthor{Halbwachs:2003lr}, we expect  20\,186 binaries in {\it Kepler} field, with periods between 1 day and 10 years.

We then assigned a mass and a radius to the primary stars by picking 20\,186 entries randomly within our selection of 149\,526. The secondaries were created using the mass ratio distribution from \citeauthor{Halbwachs:2003lr} Their radii were computed using the same mass--radius relation shown in Eq.~\ref{eq:massradius}. The need to use different methods to assign the primary and secondary radii led to occasional inconsistencies, where the lower-mass secondary star was assigned with a slightly larger radius. In these cases we scaled the primary star radius up to that of the secondary. The binary periods and eccentricities were drawn from distributions presented by \citeauthor{Halbwachs:2003lr}. We repeated the procedure for the upper and lower 1 and 2$\sigma$ errors of the \citeauthor{Halbwachs:2003lr} binary abundance.

The 3D orientation was randomised using the same N-body code as earlier. We counted all eclipsing systems, including those that showed only the primary or secondary eclipse. The simulated eclipsing systems were then compared with those in Villanova's {\it Kepler} eclipsing binary 
catalog\footnote{its beta version can be found on the servers of \href{http://keplerebs.villanova.edu/}{Villanova University} at the following address \href{http://keplerebs.villanova.edu/}{http://keplerebs.villanova.edu/} maintained by Andrej Pr\v sa et al.}, 
that we downloaded on May 10, 2013. This catalog is an up to date version of those presented in \citet{Prsa:2011qy} and \citet{Slawson:2011uq}. It contained 2\,400 eclipsing binaries. By cross-referencing with the {\it Kepler} target catalog, 2\,191 of the eclipsing binaries have measurements for $T_{\rm eff}$, $\log g$ and $[Fe/H]$. We used the Torres relation to calculate the primary star mass and found that 1\,684 binaries have a primary mass falling between 0.6 and 1.3 $M_\odot$.

The simulations cannot be meaningfully compared with the catalog for all period ranges. Due to the nature of {\it Kepler}'s observations and its very steep sensitivity with orbital distance, the eclipsing binary catalog contains a large number of very close and contact binaries, many of which have periods less than a day, whereas \citet{Halbwachs:2003lr} contains no binaries closer than 1 day. At longer periods beyond 200 days, the {\it Kepler} catalog is likely incomplete since the data analysis and binary identification is still ongoing. We think the most accurate comparison is for orbital periods on the range between 5 and 200 days.

Table~\ref{tab:BinaryComp} contains numbers comparing eclipsing binaries in the {\it Kepler} catalog to our simulated systems. Results are presented for several period ranges. Between 5 and 200 days the {\it Kepler} catalog almost matches the upper 95\% confidence interval as predicted by the \citeauthor{Halbwachs:2003lr} distribution but the means differ by about 25\%. There are several possibilities for the discrepancy:
\begin{itemize}
\item The catalog could contain some false positives due to blended background stars, or mis-identification. 
\item Both methods have different biases; if close binaries are part of triples (as observed by \citealt{Tokovinin:2006la}) then the photometry would detect the close pair, while the radial- velocity will be more sensitive to the long period (especially for low mass, synchronously rotating binaries in a triple system). 
\item  \citet{Halbwachs:2003lr}  may have surveyed a slightly binary-deficient region of the Galaxy (or {\it Kepler} a binary-rich region). 
\item Our method for assigning and calculating masses and radii is possibly inaccurate, which would impact the eclipse probabilities (e.g. \citealt{Dressing:2013xy,Batalha:2013lr}). 
\item The mass range we chose does not exactly correspond the spectral type range defined in \citet{Halbwachs:2003lr}. 
\item The period range used for comparison may not be optimal.
\end{itemize}
A combination of the above is conceivable, and the last three options would be the easiest to manipulate in order to match the eclipse numbers. While attempting to make the numbers compatible would prove to be an interesting study, it is beyond what is necessary for our current exercise.

\begin{table}
\caption{Simulated eclipsing binary numbers from \citet{Halbwachs:2003lr} compared with the {\it Kepler} eclipsing binary catalog. } 
\centering 
\begin{tabular}{l | c |ccccc} 
\hline\hline 
Period Range & {\it Kepler} & \multicolumn{5}{c}{Simulated} \\ 
& & $-2\sigma$ & $-1\sigma$ & mean & $+1\sigma$ & $+2\sigma$\\
[0.5ex] 
\hline 
All & 1684 & 721 & 844 & 937 & 1054 & 1208\\
$>$ 1 day & 1089 & 721 & 844 & 937 & 1054 & 1208\\
1 to 200 days & 1053 & 661 & 741 & 812 & 905 & 1046\\
$>$ 5 days & 695 & 449 & 545 & 612 & 708 & 799\\
5 to 200 days & 659 & 389 & 442 & 487 & 559 & 637\\
\hline 
\end{tabular}
\label{tab:BinaryComp}
\end{table}

Further comparisons were made between the two distributions. In Fig.~\ref{fig:BinaryTests} we compare the eclipsing binary data from the simulated and catalog distributions by plotting the cumulative probabilities for the binary period (a), primary mass (b) and primary radius (c). The period and mass distribution match very well over the range 5 to 200 days. This suggests that whilst the raw numbers in Table~\ref{tab:BinaryComp} may contain some discrepancies, the underlying distributions are the same. Our approach compliments the upcoming work of Orosz et al. (in prep), who numerically invert the {\it Kepler} eclipsing binary catalog to determine the underlying distribution.

According to the plot for the primary radius (Fig.~\ref{fig:BinaryTests_Radius}), the {\it Kepler} catalog contains a slightly lower abundance of stars above 1.5 $R_\odot$ than we simulate. The affected population is of order 10\% and does not impact much our simulations (Fig.~\ref{fig:BinaryTests_Period} \& \ref{fig:BinaryTests_Mass}).

As a result of these tests, the binary distribution was considered to be sufficiently realistic to be used in the wider simulations. The lower eclipsing binary counts in the simulated distribution can only act to slightly underestimate predicted planet numbers, something that can be corrected by scaling them up.




\subsection{The planet distribution}\label{sec:planetdist}

The binary distribution (Sec.~\ref{sec:binarydist}) could be created with reasonable confidence since binary stars have  well studied properties. Circumbinary planets, however, have only been conclusively detected within the past few years.
Due to the limited knowledge of how circumbinary planets are distributed, we tested a wide range of possible orbital configurations, constructed from observational and theoretical results.

We dedicate subsections to the two parameters that have the largest impact on transit detections: a planet's inclination with respect to the binary plane, Section~\ref{sec:orbitalincl}, and its orbital period, Sections~\ref{sec:logdist} to \ref{sec:planetdist_stability_limit}.

We restricted our simulations to gas giants, i.e. planets with masses larger than 50 $M_\oplus$ ($\sim 0.15\,M_{\rm Jup}$). This choice is not arbitrary but based on the fact that the orbital distribution of planets with smaller mass at orbital periods larger than 100 days is largely uncertain \citep{Mayor:2011fj,Howard:2010zr,Howard:2012qy} and would be hard to simulate without making many ad-hoc assumptions. Furthermore, large planets cause deep transits that are more easily identified from few events, as expected in the inclined case.

The abundance of planets above $0.15\,M_{\rm Jup}$ orbiting single, solar-like stars, with periods less than 10 years is 13.9$\pm$1.5\% \citep{Mayor:2011fj}. We used this abundance in the construction of our distributions.


\subsubsection{Mutual inclination distributions}\label{sec:orbitalincl}

The distribution in mutual inclination has a large effect on the amount of transits observed (Sec.~\ref{subsec:incl}) which is currently an unknown quantity. We built a number of distributions that are described by five mathematical functions: uniform, Gaussian, Rayleigh, isotropic, and coplanar. For the uniform, Gaussian and Rayleigh distributions, respectively, the upper boundary, standard deviation and Rayleigh parameter, received the values: 1, 5, and 20$^\circ$. 

Some of these descriptions are supported by observational and theoretical studies. \citet{Foucart:2013ys} theorise that circumbinary planets should be very close to coplanar. Planet-planet scattering in single star systems was tested in a number of numerical experiments by \citet{Chatterjee:2008uq}; the resulting inclination distribution for the inner planet (the one we would be most sensitive to detect) closely matches a 20$^\circ$ Gaussian distribution centred on zero. Hot Jupiters orbiting stars with mass $> 1.2 M_\odot$ and younger than 2.5 Gyr have spin--orbit angles compatible with an isotropic distribution \citep{Triaud:2011fk}. Finally, objects of the Solar System follow a 1$^\circ$ Rayleigh distribution relative to the invariant plane (e.g. \citealt{Lissauer:2011rt,Clemence:1955ly}).

We also tested a hybrid distribution that was constructed by combining two descriptions. This is to account that two migration mechanisms can be responsible for shaping the orbital parameters of gas giants \citep{Dawson:2013yq}. Disc-driven migration and planet-planet scattering can both send planet within the {\it snow line} but both together may be required to understand the existence of {\it hot Jupiters}, notably, their relation to metallicity \citep{Dawson:2013yq} and the spread of their orbital obliquities \citep{Triaud:2010fr,Winn:2010rr,Triaud:2011qy,Albrecht:2012lp}. We therefore created a new distribution from the combination of a 1$^\circ$ Rayleigh with a 20$^\circ$ Gaussian, represented with equal fraction, that we called {\it hybrid}.

\subsubsection{A featureless distribution of planets}\label{sec:logdist}

We started with the simplest possible scenario: a uniform $\log_{10}$ planet distribution. We assigned planets on 13.9\% of the synthetic binaries created in Section~\ref{sec:binarydist}. From the flat $\log_{10}$ distribution we drew the planet period between 2.1 days and 8.6 years, where these bounds came from radial-velocity surveys (\href{http://exoplanets.org}{exoplanets.org}). The planet eccentricities, were also taken from the same surveys.

We removed any planets that were placed on dynamically unstable orbits, after computing the stability criterion defined in \citet{Holman:1999lr}. This significantly reduced the simulated occurrence rate from 13.9\% to 3.1\%

For single stars, transit probability is proportional to $1/P_{\rm p}^{3/2}$, leading to an easily correctable observing bias towards short period planets. The circumbinary geometry is more complex because there are two periods to consider: $P_{\rm bin}$ and $P_{\rm p}$. Thanks to our choice of a featureless period distribution, we can investigate the geometric observing biases of circumbinary transiting planets.


\begin{table*}
\caption{Predicted transiting circumbinary planets drawn from a uniform $\log_{10}$ distribution} 
\centering 
\begin{tabular}{l | rrr|rrr|rrr|rrr}
\hline\hline 
Inclination Dist. &\multicolumn{3}{c}{NEBs} &\multicolumn{3}{|c}{EBs consecutive} &\multicolumn{3}{|c}{EBs sparse} & \multicolumn{3}{|c}{EBs total} \\ 
[0.5ex] 
& Min & Mean & Max & Min & Mean & Max & Min & Mean & Max & Min & Mean & Max \\
\hline 
Coplanar & 0 & 0.00 & 0 & 0 & 0.95 & 3 & 0 & 1.00 & 3 & 0 & 1.95 & 5 \\
Uniform 1$^\circ$ & 0 & 0.00 & 0 & 0 & 1.60 & 5 & 0 & 0.70 & 2 & 0 & 2.30 & 5 \\
Uniform 5$^\circ$ & 0 & 0.45 & 2 & 0 & 1.05 & 4 & 0 & 1.90 & 5 & 0 & 2.95 & 8 \\
Uniform 20$^\circ$ & 2 & 4.30 & 8 & 0 & 0.65 & 3 & 1 & 2.05 & 7 & 1 & 2.70 & 8 \\
Gaussian 1$^\circ$ & 0 & 0.00 & 0 & 0 & 1.15 & 5 & 0 & 0.85 & 4 & 0 & 2.00 & 7 \\
Gaussian 5$^\circ$ & 0 & 0.85 & 2 & 0 & 0.95 & 3 & 0 & 2.55 & 6 & 1 & 3.50 & 9 \\
Gaussian 20$^\circ$ & 2 & 6.75 & 14 & 0 & 0.55 & 3 & 0 & 2.65 & 7 & 1 & 3.20 & 9 \\
Rayleigh 1$^\circ$ & 0 & 0.00 & 0 & 0 & 1.80 & 4 & 0 & 1.65 & 4 & 1 & 3.45 & 6 \\
Rayleigh 5$^\circ$ & 0 & 3.40 & 6 & 0 & 0.70 & 2 & 0 & 2.55 & 7 & 0 & 3.25 & 8 \\
Rayleigh 20$^\circ$ & 4 & 9.65 & 14 & 0 & 0.15 & 1 & 0 & 2.15 & 8 & 0 & 2.30 & 8 \\
Hybrid & 1 & 4.00 & 7 & 0 & 0.60 & 4 & 0 & 2.10 & 4 & 0 & 2.70 & 6 \\
Isotropic & 8 & 16.70 & 28 & 0 & 0.10 & 1 & 0 & 2.30 & 5 & 0 & 2.40 & 5 \\

\hline 
\multicolumn{13}{l}{\footnotesize{Note: effective fraction of systems with gas giants:  2.5\%  (for $M_{\rm p}>0.15M_{\rm Jup}$, $P_{\rm p}<10$ years, and $P_{\rm bin}>5$ days).}}
\end{tabular}
\label{tab:ResultsLOG_5daycut}
\end{table*}

\begin{table*}
\caption{Predicted transiting circumbinary planets drawn from radial-velocity surveys} 
\centering 
\begin{tabular}{l | rrr|rrr|rrr|rrr}
\hline\hline 
Inclination Dist. &\multicolumn{3}{c}{NEBs} &\multicolumn{3}{|c}{EBs consecutive} &\multicolumn{3}{|c}{EBs sparse} & \multicolumn{3}{|c}{EBs total} \\ 
[0.5ex] 
& Min & Mean & Max & Min & Mean & Max & Min & Mean & Max & Min & Mean & Max \\
\hline 
Coplanar & 0 & 0.00 & 0 & 0 & 1.05 & 4 & 0 & 1.60 & 6 & 0 & 2.65 & 8 \\
Uniform 1$^\circ$ & 0 & 0.05 & 1 & 0 & 1.05 & 5 & 0 & 2.35 & 5 & 0 & 3.40 & 8 \\
Uniform 5$^\circ$ & 0 & 0.80 & 4 & 0 & 0.90 & 2 & 0 & 2.65 & 6 & 1 & 3.55 & 7 \\
Uniform 20$^\circ$ & 2 & 5.50 & 8 & 0 & 0.60 & 3 & 0 & 2.80 & 6 & 0 & 3.40 & 7 \\
Gaussian 1$^\circ$ & 0 & 0.05 & 1 & 0 & 1.15 & 4 & 0 & 2.50 & 6 & 1 & 3.65 & 8 \\
Gaussian 5$^\circ$ & 0 & 1.55 & 5 & 0 & 1.10 & 5 & 0 & 3.10 & 7 & 1 & 4.20 & 10 \\
Gaussian 20$^\circ$ & 4 & 7.65 & 14 & 0 & 0.40 & 2 & 0 & 2.45 & 5 & 0 & 2.85 & 5 \\
Rayleigh 1$^\circ$ & 0 & 0.35 & 2 & 0 & 1.05 & 4 & 0 & 2.75 & 7 & 1 & 3.80 & 9 \\
Rayleigh 5$^\circ$ & 0 & 3.15 & 8 & 0 & 0.70 & 4 & 0 & 2.75 & 6 & 0 & 3.45 & 7 \\
Rayleigh 20$^\circ$ & 6 & 12.05 & 20 & 0 & 0.10 & 1 & 0 & 2.25 & 6 & 0 & 2.35 & 6 \\
Hybrid & 0 & 3.75 & 10 & 0 & 0.60 & 5 & 0 & 2.55 & 7 & 0 & 3.15 & 9 \\
Isotropic & 10 & 18.15 & 27 & 0 & 0.00 & 0 & 0 & 1.90 & 5 & 0 & 1.90 & 5 \\

\hline 
\multicolumn{13}{l}{\footnotesize{Note: effective fraction of systems with gas giants:  4.5\%  (for $M_{\rm p}>0.15M_{\rm Jup}$, $P_{\rm p}<10$ years, and $P_{\rm bin}>5$ days).}}
\end{tabular}
\label{tab:ResultsRV_5daycut}
\end{table*}

\begin{table*}
\caption{Predicted transiting circumbinary planets drawn from a distribution of planets with a stability limit pile-up} 
\centering 
\begin{tabular}{l | rrr|rrr|rrr|rrr}
\hline\hline 
Inclination Dist. &\multicolumn{3}{c}{NEBs} &\multicolumn{3}{|c}{EBs consecutive} &\multicolumn{3}{|c}{EBs sparse} & \multicolumn{3}{|c}{EBs total} \\ 
[0.5ex] 
& Min & Mean & Max & Min & Mean & Max & Min & Mean & Max & Min & Mean & Max \\
\hline 
Coplanar & 0 & 0.00 & 0 & 0 & 2.10 & 4 & 0 & 2.25 & 5 & 1 & 4.35 & 8 \\
Uniform 1$^\circ$ & 0 & 0.05 & 1 & 0 & 2.10 & 5 & 0 & 2.25 & 7 & 1 & 4.35 & 9 \\
Uniform 5$^\circ$ & 0 & 1.55 & 4 & 0 & 2.00 & 6 & 0 & 3.90 & 7 & 2 & 5.90 & 13 \\
Uniform 20$^\circ$ & 5 & 11.80 & 17 & 0 & 0.80 & 3 & 3 & 5.40 & 10 & 3 & 6.20 & 11 \\
Gaussian 1$^\circ$ & 0 & 0.30 & 1 & 0 & 1.70 & 5 & 1 & 2.50 & 7 & 1 & 4.20 & 9 \\
Gaussian 5$^\circ$ & 1 & 3.70 & 7 & 0 & 1.85 & 4 & 0 & 3.80 & 9 & 0 & 5.65 & 11 \\
Gaussian 20$^\circ$ & 9 & 15.65 & 25 & 0 & 0.75 & 3 & 0 & 3.60 & 7 & 0 & 4.35 & 7 \\
Rayleigh 1$^\circ$ & 0 & 0.50 & 2 & 0 & 2.05 & 5 & 0 & 3.20 & 6 & 1 & 5.25 & 10 \\
Rayleigh 5$^\circ$ & 2 & 6.50 & 10 & 0 & 1.75 & 5 & 2 & 4.80 & 8 & 4 & 6.55 & 12 \\
Rayleigh 20$^\circ$ & 12 & 24.65 & 36 & 0 & 0.30 & 2 & 1 & 4.00 & 10 & 2 & 4.30 & 11 \\
Hybrid & 5 & 9.10 & 13 & 0 & 1.35 & 5 & 1 & 4.20 & 8 & 1 & 5.55 & 10 \\
Isotropic & 29 & 33.80 & 46 & 0 & 0.15 & 1 & 0 & 2.55 & 5 & 0 & 2.70 & 5 \\

\hline 
\multicolumn{13}{l}{\footnotesize{Note: effective fraction of systems with gas giants:  9.0\%  (for $M_{\rm p}>0.15M_{\rm Jup}$, $P_{\rm p}<10$ years, and $P_{\rm bin}>5$ days).}}
\end{tabular}
\label{tab:ResultsSave_5daycut}
\end{table*}

\begin{figure*}  
\begin{center}  
	\begin{subfigure}[b]{0.33\textwidth}
		\caption{Uniform $\log_{10}$ period distribution}
		\includegraphics[width=\textwidth]{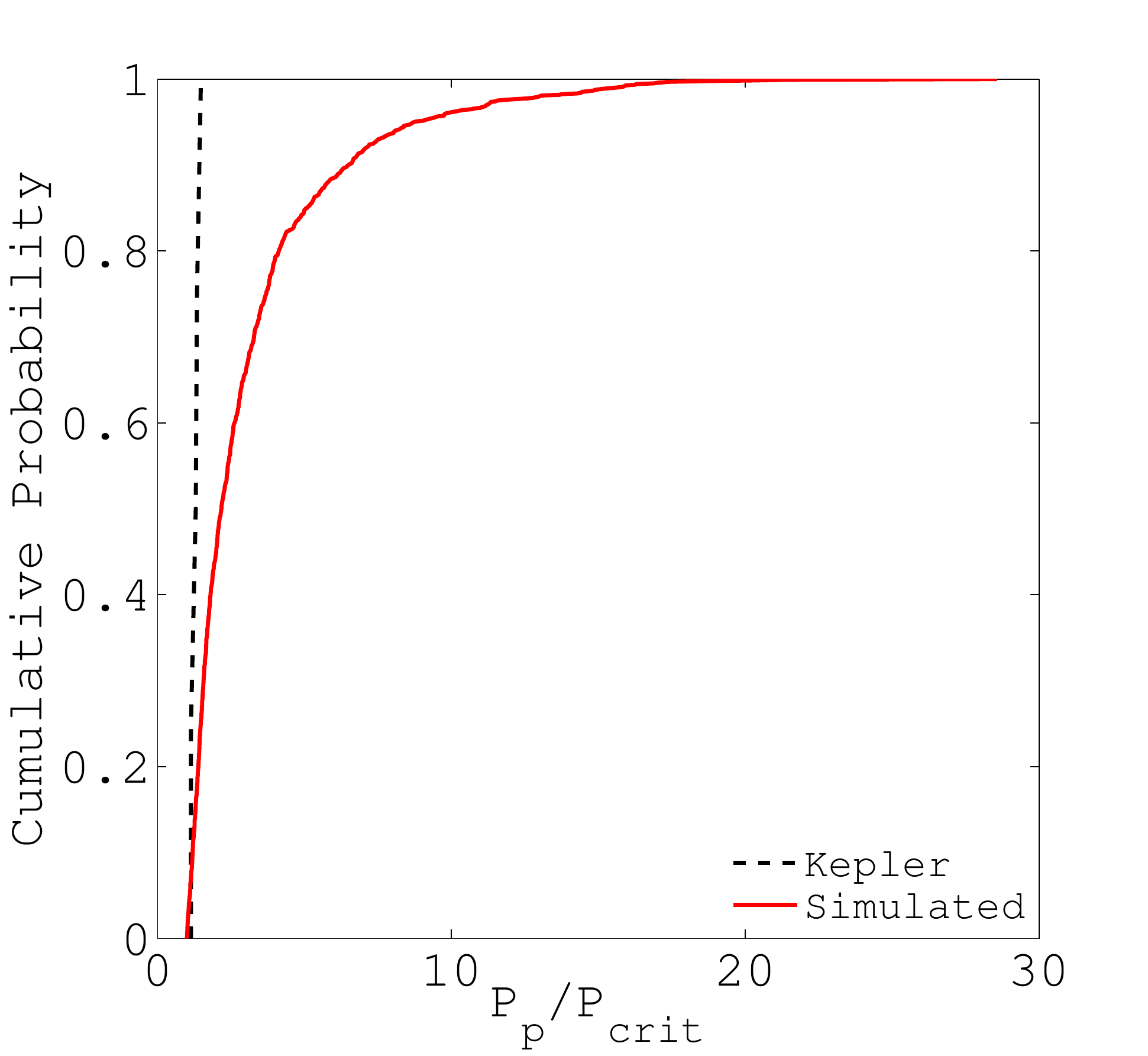}  
		\label{fig:StabilityLimitTest_log}  
	\end{subfigure}
	\begin{subfigure}[b]{0.33\textwidth}
		\caption{RV period distribution}
		\includegraphics[width=\textwidth]{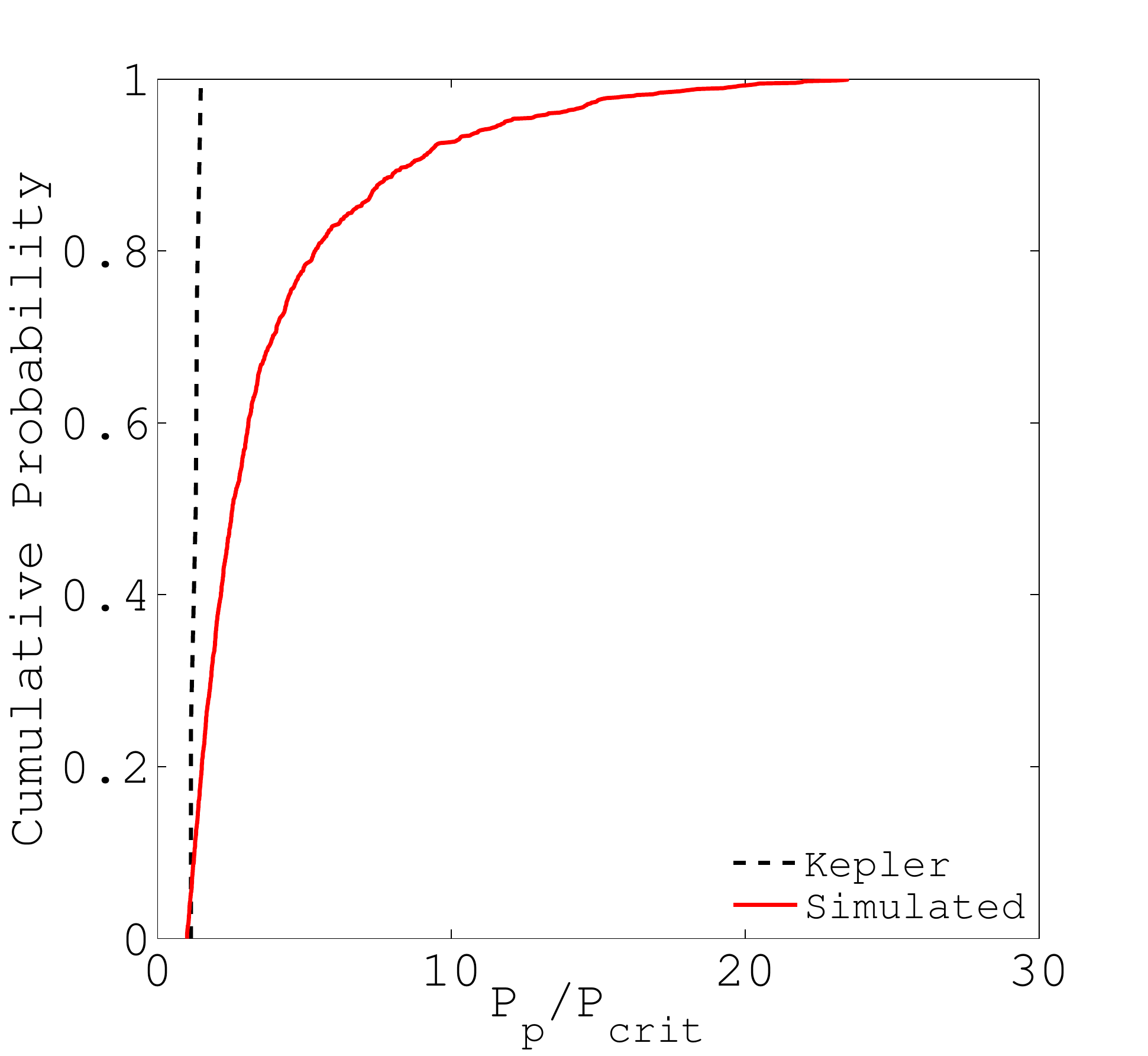}  
		\label{fig:StabilityLimitTest_rv}  
	\end{subfigure}
	\begin{subfigure}[b]{0.33\textwidth}
		\caption{Pile-up at stability limit}
		\includegraphics[width=\textwidth]{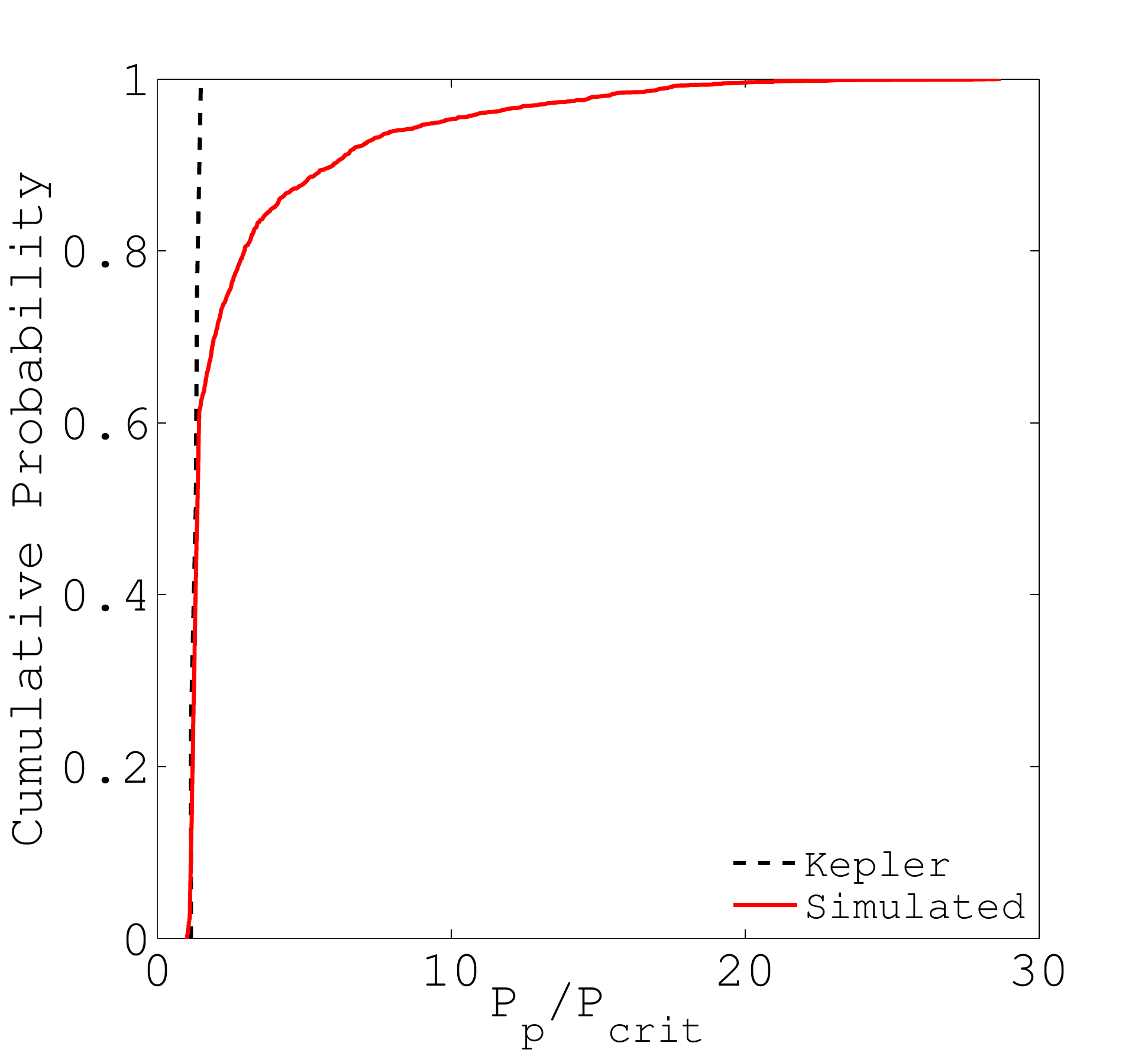}  
		\label{fig:StabilityLimitTest_save}  
	\end{subfigure}
\caption{Cumulative probability functions of the ratio $P_{\rm p}/P_{\rm crit}$ for the simulations (red, solid) compared with the four largest {\it Kepler} circumbinary systems: Kepler-16, -34, -35 and -64 (black, dashed).}
\label{fig:StabilityLimitTest}  
\end{center}  
\end{figure*}

\subsubsection{Using planets from radial-velocity surveys}\label{sec:planetdist_rv}

Our second set of simulations relies on the period distribution of radial-velocity-detected planets. While we may or may not expect that the circumbinary planet population should occupy a different orbital distribution from the circumstellar distribution, the current distribution is the only one that we know for sure has some basis in reality. Testing it can reveal similarities and  distinctions between the populations orbiting single and binary stars.

Radial-velocity surveys have yielded planets with an orbital separation distribution that contains a number of structures, of which the most spectacular is an important and sudden rise in the number of objects beyond $\sim$ 1~AU separation, dubbed the {\it 1~AU bump}. As it turns out, the current {\it Kepler} circumbinary detections straddle on either side of this bump.

We used \href{http://exoplanets.org}{exoplanets.org} \citep{Wright:2011fj} to compile a list of planets, detected by the radial-velocity technique, with masses in the range $M_{\rm p} > 50 M_\oplus$. We only kept planets orbiting stars with masses between $0.6 < M_\star < 1.3 M_\odot$, to be consistent with the synthetic binary distribution. We also removed periods longer than 10 years.  We then probabilistically assigned planets to the binaries according to the 13.9\% abundance and drew the period and eccentricity from the observed radial-velocity distributions.

After removing planets due to the \citet{Holman:1999lr} criterion, the occurrence rate is effectively reduced from 13.9\% to 5.5\%. The occurrence rate of gas giants around single stars reflects the efficiency of planet formation. Assuming this efficiency is the same in circumbinary proto-systems nevertheless leads to anticipating 60\% fewer planets, most of them being lost within the instability region.

\subsubsection{ Accumulating planets near the stability limit}\label{sec:planetdist_stability_limit}

Our third and final period distribution was crafted to match the currently observed location of circumbinary planets. We take here the hypothesis that the accumulation of {\it Kepler}-discovered planets near the stability limit has a physical origin\footnote{We can remind here that HD\,202206c, discovered thanks to the radial-velocity technique, is also near the stability limit \citep{Correia:2005lr}.}. Some scenarios, including a halted disc-driven migration at the disc's inner cavity, are discussed in Section~\ref{sec:formation}.

In the previously simulated set-up of Section~\ref{sec:planetdist_rv} we removed any planet that was placed within the instability region. We created a new set-up, similar to Section~\ref{sec:planetdist_rv}, with the exception that we {\it rescue} a fraction of otherwise lost systems. To do so, we reassigned 50\% of the planets placed within instability, to a new, stable, near-circular orbit with a period randomly selected such that $P_{\rm p}$/$P_{\rm crit}$ is between 1.1 and 1.43. These bounds correspond to the RMS of $P_{\rm p}$/$P_{\rm crit}$ for the observed innermost circumbinary planets (Table \ref{tab:KeplerDiscoveries}). Rescuing those planets simulates a sort of stopping mechanism that would prevent some but not all planet loss. 

In this ``pile-up" scenario, the circumbinary planet occurrence rate fell from 13.9\% to 9.5\%.

\subsection{Simulation set-up}

Two binary populations were created that simulate the contents of the {\it Kepler} field, composed of 21\,553 and 18\,275 systems. For each binary distribution ten sets of planets were created. This was reproduced for all three planet period distributions. This means our results are issued from the combination of 20 different sets of circumbinary systems for each planet period distribution (i.e. 60 sets in total). For each of these 60 sets, the mutual inclination between the binary and planetary orbits was simulated using the 12 misalignment distributions described in Section~\ref{sec:orbitalincl} and presented in tables in the appendix.

Finally, the 3D orientation of the circumbinary system in space was randomised. Using the N-body code, these systems were then simulated over a 4 year timespan. All stellar eclipses and planetary transits were recorded.

\section{Results}\label{sec:results}

\subsection{Simulated planet abundances}\label{sec:simulated_planet_abundances}

In Sections~\ref{sec:logdist} to \ref{sec:planetdist_stability_limit} the effective planet abundances before the start of our simulations were 3.4\%, 5.5\% and 9.7\%  for the $\log_{10}$, radial-velocity and pile-up periods distributions. Our final numbers are affected by two additional factors: First, after running the N-body simulations a small number of unstable systems were discovered and discarded from our statistics. We missed them earlier because the stability criterion of \citet{Holman:1999lr}  does not include the planet's eccentricity. Second, we removed all binaries with periods below 5 days. Our justification for this is that no planets were found transiting {\it Kepler}'s very short period binaries, where the a-priori probability that they would is largest. We expand on this in Section~\ref{sec:short_period_binaries}. The final fraction of systems with gas giants are:

\begin{itemize}
\item uniform $\log_{10}$ distribution: 2.5\%;
\item radial-velocity distribution: 4.5\%;
\item physical stability limit pile-up distribution: 9.0\%.
\end{itemize}

\subsection{Predicted transiting planets}\label{sec:predicted_planet_numbers}

In Tables \ref{tab:ResultsLOG_5daycut} to \ref{tab:ResultsSave_5daycut} we show the predicted transiting planet numbers for each of our three period distributions. We give the mean number that we find across the 20 runs (including two binary populations) that compose each table entry. We also provide the minimum and maximum numbers of transiting systems across any of those runs. Results are separated in the three sub-classes that we defined in Section~\ref{sec:types_of_transits}. We further combine the ``EBs consecutive" and ``EBs sparse" systems in a broader bin enclosing all planets transiting eclipsing binaries\footnote{The minimum and maximum numbers correspond to the extrema in one particular simulation. This is why those values are always larger than in either of the two sub-classes, but less or equal than their sum.}. 

The most interesting result is that in most scenarios, and regardless of the period distribution we used, planetary transits are expected in non-eclipsing binary systems. Furthermore, it is striking to see that they may be much more numerous than in the eclipsing case. This bodes well for the search of inclined circumbinary planets.

The number of planets expected to transit non-eclipsing binaries increases with the extent in the mutual inclination distribution. The amount of transits in the ``EBs consecutive" category generally decreases with the inclination spread, but the maximum does not necessarily correspond to the coplanar simulation. The amount of transits in the ``EBs sparse" category does not have a monotonic relationship with the inclination spread. All of these behaviours were announced in Section~\ref{subsec:incl}.

Current examination of the {\it Kepler} data is focusing solely on the ``EBs consecutive" category. However, we note that a planet occurrence rate cannot be securely estimated from just those: there is a degeneracy between the fraction of systems with planets and the underlying distribution in mutual inclinations. 

The widest distribution in planetary orbital inclination will produce some systems that are near coplanarity and that have a significant chance to transit and produce consecutive events. Their existence does not say anything about all the other systems that could not be detected in transit. This means that a given number of detection is equally consistent between 1) a rare population of mostly coplanar circumbinary gas giants and 2) a population that is widely distributed in inclination and that has a generally higher planet frequency.


Only a minimum occurrence rate can be inferred from just considering planets transiting eclipsing binary in a semi-regular fashion. Understanding and searching for planets transiting non-eclipsing binaries is required to help break this degeneracy.

\subsection{Multi-transiting planets}

Knowing that a planet transited once and that the system is a non-eclipsing binary (from spectroscopic follow-up for example), can be sufficient to know that the planet is on an inclined orbit. However we would be left without much additional information about the orbit occupied by the planet. Because of the binary motion, the transit width, that can be used to obtained a vague idea of the orbital parameters on single stars (e.g. \citealt{Dawson:2012lr}), becomes a degenerate parameter for binaries. Having two transits would help in many respects, from the identification of the system to its characterisation. Exploring our synthetic population, we estimated the fraction of systems that experience two transits, compared to the total number of circumbinary transiting planets. Results are given in Table \ref{tab:MultiTransit}. Of order 50\% of systems that have transited once, will transit at least a second time\footnote{We denote by ``N/A'' the fact that no planets transiting non-eclipsing binaries were predicted for the $\log_{10}$ planet distribution with 1$^\circ$ Rayleigh mutual inclinations. The transit probability is not zero for such planets, but it is sufficiently small such that it is rounded to zero due to statistical noise.}. 

\subsection{Planets around very short period binaries}\label{sec:short_period_binaries}

The current lack of planets around short-period binaries ($P_{\rm bin}<5$ days) is considered surprising for two reasons: First, the median binary period in the {\it Kepler} eclipsing binary catalog is 2.9 days (as of May 10, 2013). Second, for a given ratio of planet and binary periods, the transit probability is maximised for a very close binary (Sec.~\ref{subsec:incl}). The inclusion of all binaries with periods under 5 days causes a significant increase in the amount of expected transiting planets. The number of transiters on eclipsing binaries more than doubled the results described in Section~\ref{sec:predicted_planet_numbers}, whilst there was a $\sim50\%$ increase in planets transiting non-eclipsing binaries. This result reaffirms the common thought that these planets should be numerously detected, if they existed.

\section{Comparisons with {\it Kepler}}\label{sec:comparisons_with_kepler}

The {\it Kepler} circumbinary gas giants all fall in the ``EBs consecutive" category. We can use our predictions and see if they reproduce existing discoveries. This comparison is able to distinguish between the three orbital distribution we investigated, to find that one is more likely to be true. This allowed us to reach a minimum occurrence rate compatible with current detection levels.

\begin{table}
\caption{The probability of planets transiting multiple times a non-eclipsing binary, given that they transited at least once.} 
\centering 
\begin{tabular}{l | ccc} 
\hline\hline 
Inclination Dist. & RV & pile-up & $\log_{10}$\\ 
\hline 
Gaussian 20$^\circ$ & 0.52 & 0.51 & 0.40\\ 
Rayleigh 1$^\circ$ & 0.14 & 0.40 & N/A\\ 
Hybrid & 0.44 & 0.51 & 0.64\\ 
Isotropic & 0.30 & 0.44 & 0.50\\ 
\hline 
\end{tabular}
\label{tab:MultiTransit}
\end{table}

\begin{table}
\caption{Kolmogorov-Smirnov test scores comparing the ratio $P_{\rm p}/P_{\rm crit}$ for the four {\it Kepler} discoveries and the simulation in Fig.~\ref{fig:Rayleigh_save}.} 
\centering 
\begin{tabular}{l | c} 
\hline\hline 
Period Dist. & K-S score\\ 
\hline 
$\log_{10}$ & 0.0032\\
RV & 0.0009\\
pile-up & 0.3858\\ 
\hline 
\end{tabular}
\label{tab:KS}
\end{table}

\begin{figure}  
\begin{center}  
\includegraphics[width=0.45\textwidth]{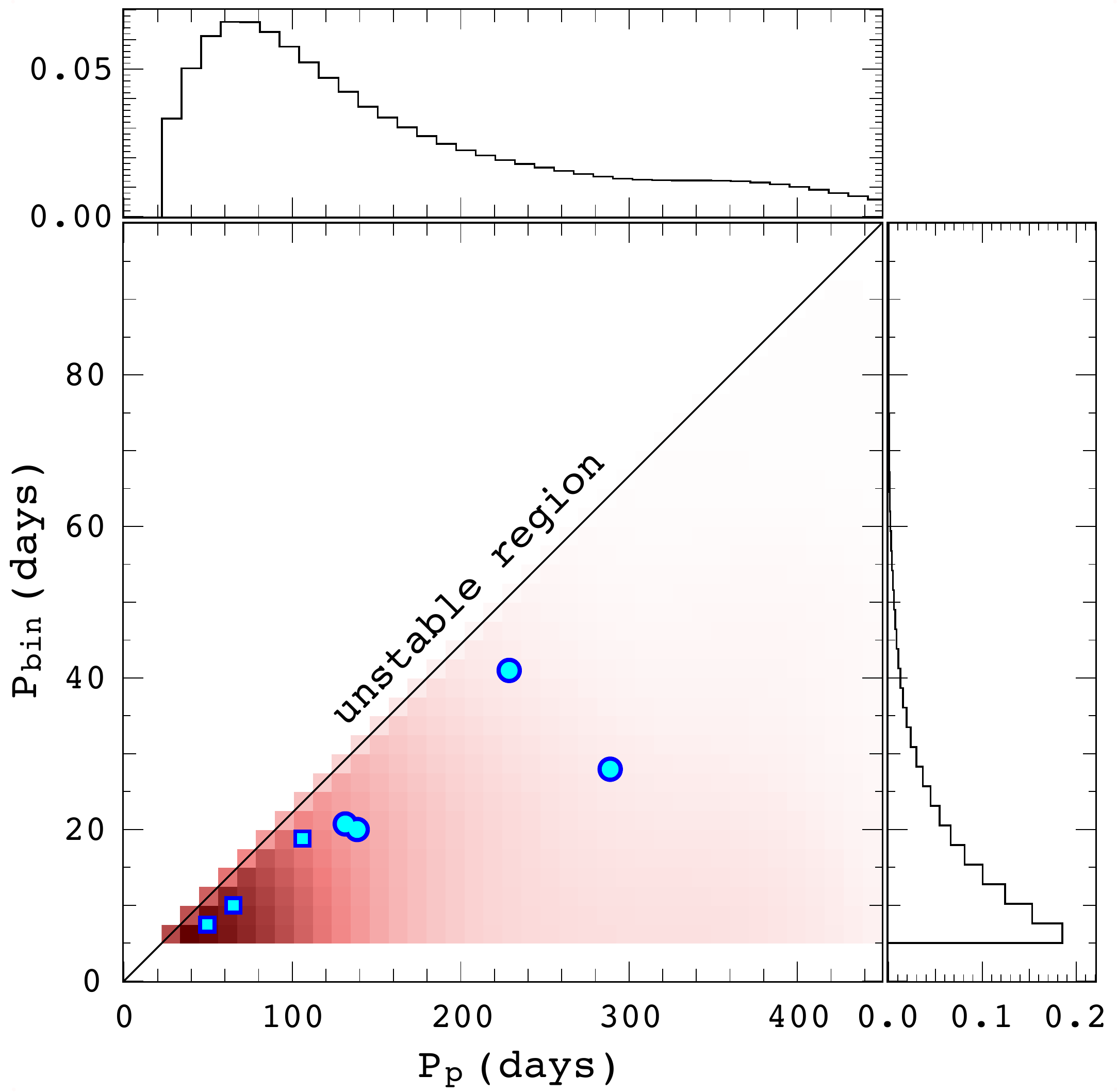}  
\caption{A smoothed 2D probability distribution function of planets consecutively transiting eclipsing binaries, having a mutual inclination following a $1^\circ$ Rayleigh function. The darker the more probable planets are. Blue dots indicate Kepler-16, -34, -35 and -64. Squares are for the other circumbinary systems which fall outside the bounds of our simulation. Only the innermost circumbinary planets are shown. The black line delineate an approximation of the stability limit $P_{\rm p} = 4.5 P_{\rm bin}$. Side histograms display the marginalised probabilities.}
\label{fig:Rayleigh_save}
\end{center}  
\end{figure}

\subsection{On the planet distribution}\label{sec:planet_distribution}

We chose the $1^\circ$ Rayleigh mutual inclination distribution to compare our three proposed period distributions to {\it Kepler}'s discoveries. 
Our choice is based on the premises that 1) strict coplanarity is ruled out by the {\it Kepler} systems and that 2) the $1^\circ$ Rayleigh distribution produces a relatively high number of transiting planets in the ``EBs consecutive" category, which is the only type of system that can be compared with {\it Kepler}.

The empirical data and our simulations are compared to each others using the ratio $P_{\rm p}/P_{\rm crit}$. To improve our statistics, we increased the number of simulations from 20 (used for Sec.~\ref{sec:results}) to 1\,000. Like in Section~\ref{sec:results}, only binaries with periods longer than 5 days were included. Their respective cumulative probability distributions are displayed in Fig. \ref{fig:StabilityLimitTest}.
The empirical cumulative distribution of $P_{\rm p}/P_{\rm crit}$, only includes the four {\it Kepler} discoveries that closest match our tested mass range: Kepler-16b, -34b, -35b and -64b. A Kolmogorov-Smirnov (K-S) test was used to compare the simulated and observed systems; we test the relative shapes of the distributions and not the planet's numbers. Results of the K-S test are in Table~\ref{tab:KS}.

\subsubsection{The flat $\log_{10}$ period distribution}

A uniform distribution in period is a poor match to the {\it Kepler} observations (Fig.~\ref{fig:StabilityLimitTest_log}). This implies the observed distribution of circumbinary gas giants is not featureless, but has some structures. 
Notably, observing biases appear to be insufficient to explain the stability limit pile-up in the {\it Kepler} results. This points to a physical origin for this over-density. Some possible explanations are discussed in Section~\ref{sec:formation}.

\subsubsection{The radial-velocity period distribution}

The radial-velocity distribution of planets has an even poorer match to the {\it Kepler} data (Fig.~\ref{fig:StabilityLimitTest_rv}). We can draw from this that the circumbinary gas giant planet period distribution is inconsistent with the distribution of gas giants orbiting single stars. This is not surprising, since planet formation around single and binary stars is likely different enough to result in distinct distributions of planets. However, this had not been empirically demonstrated before. 

\subsubsection{The pile-up period distribution}

The planet period distribution with a pre-imposed stability limit pile-up is by far the closest match to the {\it Kepler} data (Fig.~\ref{fig:StabilityLimitTest_save}). Indeed, it is 120 times more likely to be true than the uniform, $\log_{10}$ distribution, and is more than 400 times better compared to the distribution of planets based on radial-velocity results. This comparison to empirical data leads us to conclude that  the observed accumulation of gas giants near the stability limit is more likely to have a physical origin than from an observing bias.

Despite its higher K-S test score, this distribution remains only marginally compatible with the known systems: our simulation predicts that we should find a higher fraction of transiting planets far from the stability limit than has been observed. This suggests that our mechanism, rescuing 50\% of planets to populate the pile-up may not be sufficient, that the accumulation is likely to be more pronounced that what we set it to be. 

Fig.~\ref{fig:Rayleigh_save} illustrates our results. It is a 2D histogram of the planet and binary periods for systems in the ``EBs consecutive" category only. Darker tones represent a higher probability. On the top and right are the marginalised probabilities. Like previous results shown, results for $P_{\rm bin}<5$ days are excluded. We delineated the approximate stability limit boundary at $P_{\rm p}/P_{\rm bin} = 4.5$. The innermost planets in each of the {\it Kepler} systems are overplotted, and approximately follow the relation $P_{\rm p}/P_{\rm bin} \approx 6$.

Transits are more numerous when the planet and binary periods are short, as expected, but this is for two unrelated reasons: 
\begin{itemize}
\item shorter period binaries are more likely to eclipse;
\item shorter period planets are more likely to transit. 
\end{itemize}
The circumbinary geometry does not make us preferentially observe planets at any special ratio of binary and planet periods.

\subsection{On the planet occurrence}\label{sec:planet_occurrence}

Strictly within the bounds of our synthetic population ($P_{\rm bin} > 5~$d; $0.6 < M_{\star} < 1.3$; $M_{\rm p} > 0.15 M_{\rm Jup}$) \footnote{Here $M_{\star}$ only refers to the mass of the primary star.}, {\it Kepler} has found two systems: Kepler-16 and -34.

In Section~\ref{sec:planet_distribution}, we demonstrated that the planet distribution with a pre-imposed stability limit pile-up is by far the closest match to the {\it Kepler} observations. From now on, we will only consider this distribution. The $1^\circ$ Rayleigh inclination distribution predicts a most likely number of discoveries in the ``EBs consecutive" category at 2.05 systems (Table~\ref{tab:ResultsSave_5daycut}). This is evidently compatible with {\it Kepler's} two discoveries. 

From these results, we can affirm that {\it Kepler}'s observations are compatible with a ~9\% abundance of circumbinary gas giants, whose distribution in period has a physical pile-up. 

It is likely that this abundance is an underestimate, for the following reasons:
\begin{itemize}
\item Our simulated eclipsing binary numbers are on the low side.
\item The {\it Kepler} discoveries are likely not exhaustive (e.g. \citealt{Kostov:2014qy}).
\item Two more systems (Kepler-35 and -64) lie just outside the range of our synthetic population, that if included would inflate our number. 
\item There is a degeneracy between abundance and mutual inclination spread, described earlier in Section~\ref{sec:predicted_planet_numbers}, that can only lead to an increase in the occurrence rate.
\end{itemize}

Based on physical parameters of three discovered systems, \citet{Welsh:2012lr} had inferred early estimates of the occurrence rate between 1\% and 10\%. Recently, \citet{Armstrong:2014yq} who used an independent method of analysing the {\it Kepler} light-curves to constrain the abundance, found a compatible estimate. It will also be interesting to see how our values compares with future studies.

\section{Discussion}\label{sec:discuss}

We conceptually and numerically explored what an inclined circumbinary planet implies for its probability to transit. Using this knowledge we constructed a series of simulations that built a synthetic population of circumbinary systems. We tested three planetary period distributions, on 12 different mutual inclination distributions for two binary populations. The reason we computed so many different alternatives was to tabulate enough results that can then be compared in the eventual detection of planets transiting non-eclipsing binaries. The period distributions we tested were either mathematical or derived from observational evidence. The purpose of these choices was to make the work as model-independent as possible.  It would, however, be interesting to repeat this work based on a theoretical population synthesis, similar to those created for single stars (\citealt{Benz:2014lr}, and references there-in).

The comparison of our distributions to empirical data collected by the {\it Kepler} spacecraft indicates that there is a physical pile-up of planets close to the stability limit as defined by \citet{Holman:1999lr}. The number of detections implies that at least 9\% of binaries host a circumbinary gas giant with an orbital period shorter than 10 years.

Once it will be estimated that the {\it Kepler} data has been exhaustively analysed, around both eclipsing and non-eclipsing binaries, ratios between inclined and coplanar systems can be computed and compared to our results. Whether inclined planets are found or not would be interesting regardless, as it would constrain some formation and evolution models and place a limit on the maximum occurrence rate.

\subsection{Implications about planet formation and evolution in circumbinary systems}\label{sec:formation}

Stellar formation produces binaries with separations usually larger than 10 AU whose separation can then decrease (e.g. \citealt{Bate:2012rt}). This orbital rearrangement could have been detrimental to the production of planets. Yet, as we have shown, reaching {\it Kepler}'s number of detections requires effective occurrence rates close to those found for single stars. From this, we can conclude that gas giant formation is probably as likely to take place in circumbinary discs as it does in circumstellar environments in opposition to some theoretical expectations (e.g. \citealt{Meschiari:2012qy}). This may however exclude systems involving binaries on periods shorter than five days whose likely, late dynamical origin \citep{Mazeh:1979eu,Tokovinin:2006la,Fabrycky:2007pd} would probably lead to the disruption of any planetary system that would have formed earlier.

If gas giant formation rates are similar in binaries and single stars, the location of formation, or the planets' subsequent orbital evolution appears different. Despite low numbers of detections, we are able to demonstrate that the observed over-density of objects found near the stability limit is difficult to explain by an observing bias selecting only the shortest possible planet period for a given binary period. We conclude that there is more likely a physical pile-up, which is reminiscent of the hot Jupiter pile-up whose origin remains hotly debated \citep{Kuchner:2002uq,Eisner:2005qy,Ford:2006lr,Matsumura:2010ve,Guillochon:2011fk,Dawson:2013yq,Plavchan:2013ll}. There is however one important distinction: if the over-density of hot Jupiters can be explained by a special orbit towards which highly eccentric planets circularise to \citep{Ford:2006lr,Matsumura:2010ve,Guillochon:2011fk}, the necessary tidal friction is absent in the circumbinary case. 

\citet{Meschiari:2012uq} studied planetesimal accretion in a circumbinary disc. His results imply that planet formation is likely inhibited between 1.75 and 4 AUs around Kepler-16 meaning that its planet either has migrated (e.g. \citealt{Goldreich:1979fk,Goldreich:1980lr,Lin:1996yq,Ward:1997kx,Rasio:1996ly,Baruteau:2013lr,Davies:2013qy}) or has formed in-situ. In-situ formation would probably reproduce fairly well the pile-up that we observe, however gas giant formation inner to the snow line requires specific conditions \citep{Bodenheimer:2000vn} and is not expected to take place in circumbinary discs \citep{Paardekooper:2012kx}. 

At this point it would seem more likely that the {\it Kepler} discoveries have formed beyond 4~AU and migrated-in. Disc-driven migration was first studied for circumbinary planets by \citet{Nelson:2003lr}, which, with more recent work \citep{Pierens:2007fj,Pierens:2008uq,Pierens:2013kx},  shows that planets can migrate inwards before being halted near the inner cavity of the disc. The cavity's location is close to the stability limit for planetary orbits. The wtork of these authors is consistent with the pile-up observed by {\it Kepler}, but is unable to accurately fit the circumbinary orbits of specific systems. The independent study of \citet{Kley:2014rt}, using models containing more sophisticated thermodynamics and boundary conditions, successfully reproduces the observed orbit of Kepler-38b, near the stability limit.

Gas giants could also migrate via an exchange of angular momentum with other planetary companions. From this postulate it is natural to expect that inclined planets should exist since planet-planet scattering has been shown an important mechanism to explain the eccentricity and inclination of gas giants \citep{Juric:2008qy,Chatterjee:2008uq,Nagasawa:2008gf,Matsumura:2010ul}. The inclination distribution may be enhanced by the fact that circumbinary discs are more likely to be inclined with respect to the inner binary at large separations than close-in \citep{Lodato:2013uq,Plavchan:2013lr}.


Instead of involving one orbital distribution, a proper expectation of circumbinary planets may involve two. For instance the presence of a pile-up at the stability limit could be linked to the coplanarity of the system. In this case there would be no reason to expect that inclined planets would also show a pile-up.

Undoubtedly more detections are needed to better understand planet formation and orbital evolution in the circumbinary context.  


\subsection{Towards the detection of planets transiting non-eclipsing binaries}\label{sec:detecting_planets}

We showed that only a minimum occurrence rate can be extracted from only considering systems with semi-regular transits in a search of the {\it Kepler} data. To place a limit on the maximum occurrence rate, we must find if any, and how many systems transit non-eclipsing binaries. Here, we explore potential search strategies to be applied in the search for these as-of-yet non-considered planets.


One strategy is to first search for non-eclipsing binaries, which can be spotted from a variety of sources. Reconnaissance spectra of the {\it Kepler} field may already show that a number of targets are double line binaries. Non-eclipsing binaries can also be identified thanks to their photometric variability that combines ellipsoidal, reflection and Doppler beaming effects (e.g. \citealt{Faigler:2011fk,Shporer:2011qy}). This method is sensitive to {\it Kepler} binaries for periods as large as 20 days, and has recently been improved to include eccentric orbits (Mazeh, priv. com.). In the case of high eccentricities, binaries can exhibit a {\it heartbeat-like} photometric signature; over 100 of these have already been noted in the {\it Kepler} data. Once a binary is confirmed, a search of the light curve can be examined for transit like signatures. The quality of {\it Kepler}'s photometry implies that the transit of a Jovian planet (the objects we made predictions for) can be remarked visually without the need for a sophisticated algorithm.

Alternatively, one could search the {\it Kepler} data without any a-priori binary knowledge, looking for either single or infrequent transit-like signatures. The standard {\it Kepler} pipeline is not sensitive to these as it requires quasi-periodicity. This type of search may be appropriate for the Planet Hunters team (\href{http://www.planethunters.org}{planethunters.org}; \citealt{Fischer:2012lr}), which has already had success finding a circumbinary planet \citep{Schwamb:2013kx}. Algorithms with the ability to identify single transit events would tremendously help the search. Although they would also likely pick-up many false astrophysical positives \citep{Brown:2003qy} and data artefacts.
This risk can be reduced by forcing that at least two events must be present in the data; we have shown that of order 50\% of systems will do so. 
After finding some candidates in the photometry, a spectroscopic follow-up would reveal easily most single or double-line binaries (e.g. \citealt{Santerne:2012rt,Diaz:2013fr}) fairly inexpensively. The width and depth of a circumbinary planet are often different, depending on the orbital phase of the star that is being crossed. A transit whose shape implies a mean stellar density \citep{Seager:2003qy} compatible with either of the two components would become an interesting candidate for additional follow-up. Other usual techniques, such as searching for potential blends using adaptive optical systems, or using statistical algorithms like \texttt{Blender} \citep{Fressin:2013uq} can help remove many false positives. 

If enough events are captured and follow-up constrains the physical parameters of the system sufficiently, it could become possible to predict when next transit events would happen. A hit or a miss can refine the parameters to impressive precision as illustrated in Fig.~\ref{fig:Chromosomes}. With predictive capacity, transits can be recorded, for example, to study the atmospheres of fairly cool gas giants (e.g. \citealt{Seager:2010kx}). 

The above strategies can be immediately applied to the {\it Kepler} telescope since the data is already freely available. They can also be applied to future photometric surveys. Transits of non-eclipsing binaries may take on a larger importance in NASA's upcoming {\it TESS} mission. The short observing timespan of {\it TESS} (one month for most of the sky, one year at the ecliptic poles, Ricker priv. com.) means that the detection strategy of circumbinary planets will have to evolve from that of {\it Kepler}'s, since getting multiple transits of circumbinary planets will be rare, even when coplanar. Sparse and single transits will need to be identified in the light curves. Confirmation would almost certainly require follow-up spectroscopy, which will be made easier by {\it TESS}'s focus on bright targets across both hemispheres. Identifying non-eclipsing binaries within the field will become also possible thanks to ESA's {\it GAIA} satellite or even prior to launch from dedicated ground-based surveys. It is possible that inclined circumbinary planets will provide the bulk of transiting cold Jupiters in the {\it TESS} program. Additionally, the time-dependent transit probabilities outlined in Section~\ref{sec:transit_prob} support an extended {\it TESS} mission, which is already under consideration. The proposed {\it Kepler} follow-up mission {\it K2} is similar to {\it TESS} but on a smaller scale.

In addition to the interest of detecting more circumbinary planets, there is the appeal of knowing whether misaligned systems can exist in orbiting binaries like they have been observed around single stars. Several techniques have been developed to measure the {\it spin--orbit} angle, spectroscopically, via the Rossiter-McLaughlin effect \citep{Holt:1893fk,Rossiter:1924qy,McLaughlin:1924uq,Queloz:2000rt,Gaudi:2007vn,Boue:2013fk} and the {\it Doppler shadow} \citep{Collier-Cameron:2010lk}, statistically \citep{Schlaufman:2010fk}, photometrically thanks to gravity darkening \citep{Szabo:2011fr} or through spot-crossing events \citep{Nutzman:2011ly,Sanchis-Ojeda:2012ys}, astrometrically \citep{Sahlmann:2011fk}, using asteroseismologic mode splitting \citep{Chaplin:2013zr} or interferometrically \citep{Le-Bouquin:2009mz}. Searching for transits on non-eclipsing binaries would become another technique to detect inclined orbits, by spatially resolving the misalignment.

\section{Conclusions}

We started this paper as an enquiry into whether any circumbinary planet could be caught transiting a non-eclipsing system. After studying and describing how the dynamics and how various physical and orbital parameters affect the probability of transit, we constructed synthetic populations of circumbinary systems. By simulating the orbits over 4 years and checking for transits, we reveal that {\it Kepler} could have caught such planets. 

We urge a change in the protocols behind the search for circumbinary planets. Efforts should not just focus on eclipsing binaries since most binaries do not eclipse. We also find that applying a quasi-periodic transit signal criterion to find and confirm these planets severely restricts the number of objects that can be found, and biases them towards coplanarity.

Our simulations allow us to affirm that the binary mass and period distributions are similar in the solar neighbourhood and within in the {\it Kepler} field. This indicates the formation of binaries might be a universal process, probably linked to the initial mass function (e.g. \citealt{Salpeter:1955fj,Kroupa:2001qy,Chabrier:2003fk,Bate:2012rt}). We also find that it is likely that circumbinary planets are rare for binaries with periods $<$ 5 days lending further support for a dynamical origin for the closest of binaries \citep{Mazeh:1979eu,Fabrycky:2007pd,Tokovinin:2006la}.

Comparing the existing detections with the output of our simulations, we remark that the occurrence rate of circumbinary gas giants may be close to that of single stars. Current models of gas giant formation (e.g. \citealt{Pollack:1996uq,Boss:2000fj,Helled:2013fj}) produce planets far from their star. To a first order it would seem that whether the central object is a singleton or a close pair does not have a large impact on planet formation efficiency. Inclined orbits are frequently found in single star systems and have highlighted the important role dynamical interactions have in shaping  exoplanetary systems. A similar formation rate would suggest similar mutual interactions happening in circumbinary systems. This further supports our claim that there should be a number of planets transiting non-eclipsing binaries waiting to be discovered within the {\it Kepler} timeseries.

The location of the current detections reveals that after formation, circumbinary gas giants have a migration history that is different from single star systems, with planet likely piling-up at orbits near their binary's stability limit. It remains to be seen whether disc-driven migration could place planets on such particular orbits without pushing them onto an unstable orbit, or whether a different mechanism, maybe specific to binaries, would be at work.

\begin{acknowledgements} 

We would like, particularly, to thank Dave Armstrong, Simchon Faigler, Veselin Kostov, Rosemary Mardling, Tsevi Mazeh and St\'ephane Udry, for insightful and important discussions. We  would also like to thank the many people who commented on the concept, or on specific parts of the manuscript. Those people include researchers working on exoplanets at MIT, notably Joshua Winn, as well as colleagues met during the recent second {\it Kepler} Science Conference at NASA's Ames Research Center. 

We also acknowledge the work of two anonymous referees who helped improve the paper with their comments.

D.~V.~Martin is funded by the Swiss National Science Foundation. A.~H.\,M.\,J.~Triaud is a Swiss National Science Foundation fellow under grant numbers PBGEP2-145594 and P300P2-147773. 

We acknowledge frequent visits of the \href{http://exoplanet.eu}{exoplanet.eu} \citep{Schneider:2011lr} and \href{http://exoplanets.org}{exoplanets.org} \citep{Wright:2011fj} websites as well as an extensive use of the  \href{http://adsabs.harvard.edu/abstract_service.html}{ADS} and \href{http://arxiv.org/archive/astro-ph}{arXiv} paper repositories. Finally, our work could not have happened without the important work and dedication of the {\it Kepler} team who compile updated and publicly available candidate lists and stellar catalogs on the \href{https://archive.stsci.edu/}{MAST} repository and at \href{http://keplerebs.villanova.edu/}{Villanova University}.


\end{acknowledgements} 

\bibliographystyle{aa}
\bibliography{../../1Mybib.bib}

\begin{thebibliography}{139}
\expandafter\ifx\csname natexlab\endcsname\relax\def\natexlab#1{#1}\fi

\bibitem[{{Adams} \& {Laughlin}(2003)}]{Adams:2003wd}
{Adams}, F.~C. \& {Laughlin}, G. 2003, \icarus, 163, 290

\bibitem[{{Agol} {et~al.}(2005){Agol}, {Steffen}, {Sari}, \&
  {Clarkson}}]{Agol:2005qy}
{Agol}, E., {Steffen}, J., {Sari}, R., \& {Clarkson}, W. 2005, \mnras, 359, 567

\bibitem[{{Albrecht} {et~al.}(2009){Albrecht}, {Reffert}, {Snellen}, \&
  {Winn}}]{Albrecht:2009fy}
{Albrecht}, S., {Reffert}, S., {Snellen}, I.~A.~G., \& {Winn}, J.~N. 2009,
  \nat, 461, 373

\bibitem[{{Albrecht} {et~al.}(2012){Albrecht}, {Winn}, {Johnson}, {Howard},
  {Marcy}, {Butler}, {Arriagada}, {Crane}, {Shectman}, {Thompson}, {Hirano},
  {Bakos}, \& {Hartman}}]{Albrecht:2012lp}
{Albrecht}, S., {Winn}, J.~N., {Johnson}, J.~A., {et~al.} 2012, \apj, 757, 18

\bibitem[{{Albrecht} {et~al.}(2014){Albrecht}, {Winn}, {Torres}, {Fabrycky},
  {Setiawan}, {Gillon}, {Jehin}, {Triaud}, {Queloz}, {Snellen}, \&
  {Eggleton}}]{Albrecht:2014lr}
{Albrecht}, S., {Winn}, J.~N., {Torres}, G., {et~al.} 2014, \apj, 785, 83

\bibitem[{{Armstrong} {et~al.}(2013){Armstrong}, {Martin}, {Brown}, {Faedi},
  {G{\'o}mez Maqueo Chew}, {Mardling}, {Pollacco}, {Triaud}, \&
  {Udry}}]{Armstrong:2013rt}
{Armstrong}, D., {Martin}, D.~V., {Brown}, G., {et~al.} 2013, \mnras, 434, 3047

\bibitem[{{Armstrong} {et~al.}(2014){Armstrong}, {Osborn}, {Brown}, {Faedi},
  {G{\'o}mez Maqueo Chew}, {Martin}, {Pollacco}, \& {Udry}}]{Armstrong:2014yq}
{Armstrong}, D.~J., {Osborn}, H., {Brown}, D., {et~al.} 2014, ArXiv e-prints

\bibitem[{{Artymowicz} \& {Lubow}(1996)}]{Artymowicz:1996fk}
{Artymowicz}, P. \& {Lubow}, S.~H. 1996, \apjl, 467, L77

\bibitem[{{Arvo}({1992})}]{Arvo:1992lr}
{Arvo}, J. {1992}, in {Graphics Gems III}, ed. {David Kirk} ({Academic Press}),
  {117--120}

\bibitem[{{Barnes} {et~al.}(2013){Barnes}, {van Eyken}, {Jackson}, {Ciardi}, \&
  {Fortney}}]{Barnes:2013lv}
{Barnes}, J.~W., {van Eyken}, J.~C., {Jackson}, B.~K., {Ciardi}, D.~R., \&
  {Fortney}, J.~J. 2013, \apj, 774, 53

\bibitem[{{Baruteau} {et~al.}(2013){Baruteau}, {Crida}, {Paardekooper},
  {Masset}, {Guilet}, {Bitsch}, {Nelson}, {Kley}, \&
  {Papaloizou}}]{Baruteau:2013lr}
{Baruteau}, C., {Crida}, A., {Paardekooper}, S.-J., {et~al.} 2013, ArXiv
  e-prints

\bibitem[{{Batalha} {et~al.}(2013){Batalha}, {Rowe}, {Bryson}, {Barclay},
  {Burke}, {Caldwell}, {Christiansen}, {Mullally}, {Thompson}, {Brown},
  {Dupree}, {Fabrycky}, {Ford}, {Fortney}, {Gilliland}, {Isaacson}, {Latham},
  {Marcy}, {Quinn}, {Ragozzine}, {Shporer}, {Borucki}, {Ciardi}, {Gautier},
  {Haas}, {Jenkins}, {Koch}, {Lissauer}, {Rapin}, {Basri}, {Boss}, {Buchhave},
  {Carter}, {Charbonneau}, {Christensen-Dalsgaard}, {Clarke}, {Cochran},
  {Demory}, {Desert}, {Devore}, {Doyle}, {Esquerdo}, {Everett}, {Fressin},
  {Geary}, {Girouard}, {Gould}, {Hall}, {Holman}, {Howard}, {Howell},
  {Ibrahim}, {Kinemuchi}, {Kjeldsen}, {Klaus}, {Li}, {Lucas}, {Meibom},
  {Morris}, {Pr{\v s}a}, {Quintana}, {Sanderfer}, {Sasselov}, {Seader},
  {Smith}, {Steffen}, {Still}, {Stumpe}, {Tarter}, {Tenenbaum}, {Torres},
  {Twicken}, {Uddin}, {Van Cleve}, {Walkowicz}, \& {Welsh}}]{Batalha:2013lr}
{Batalha}, N.~M., {Rowe}, J.~F., {Bryson}, S.~T., {et~al.} 2013, \apjs, 204, 24

\bibitem[{{Bate}(2012)}]{Bate:2012rt}
{Bate}, M.~R. 2012, \mnras, 419, 3115

\bibitem[{{Benz} {et~al.}(2014){Benz}, {Ida}, {Alibert}, {Lin}, \&
  {Mordasini}}]{Benz:2014lr}
{Benz}, W., {Ida}, S., {Alibert}, Y., {Lin}, D.~N.~C., \& {Mordasini}, C. 2014,
  ArXiv e-prints

\bibitem[{{Beuermann} {et~al.}(2010){Beuermann}, {Hessman}, {Dreizler},
  {Marsh}, {Parsons}, {Winget}, {Miller}, {Schreiber}, {Kley}, {Dhillon},
  {Littlefair}, {Copperwheat}, \& {Hermes}}]{Beuermann:2010fk}
{Beuermann}, K., {Hessman}, F.~V., {Dreizler}, S., {et~al.} 2010, \aap, 521,
  L60

\bibitem[{{Bodenheimer} {et~al.}(2000){Bodenheimer}, {Hubickyj}, \&
  {Lissauer}}]{Bodenheimer:2000vn}
{Bodenheimer}, P., {Hubickyj}, O., \& {Lissauer}, J.~J. 2000, \icarus, 143, 2

\bibitem[{{Boss}(2000)}]{Boss:2000fj}
{Boss}, A.~P. 2000, \apjl, 536, L101

\bibitem[{{Bou{\'e}} {et~al.}(2013){Bou{\'e}}, {Montalto}, {Boisse}, {Oshagh},
  \& {Santos}}]{Boue:2013fk}
{Bou{\'e}}, G., {Montalto}, M., {Boisse}, I., {Oshagh}, M., \& {Santos}, N.~C.
  2013, \aap, 550, A53

\bibitem[{{Brown} {et~al.}(2012){Brown}, {Cameron}, {Anderson}, {Enoch},
  {Hellier}, {Maxted}, {Miller}, {Pollacco}, {Queloz}, {Simpson}, {Smalley},
  {Triaud}, {Boisse}, {Bouchy}, {Gillon}, \& {H{\'e}brard}}]{Brown:2012lr}
{Brown}, D.~J.~A., {Cameron}, A.~C., {Anderson}, D.~R., {et~al.} 2012, \mnras,
  423, 1503

\bibitem[{{Brown}(2003)}]{Brown:2003qy}
{Brown}, T.~M. 2003, \apjl, 593, L125

\bibitem[{{Capelo} {et~al.}(2012){Capelo}, {Herbst}, {Leggett}, {Hamilton}, \&
  {Johnson}}]{Capelo:2012cr}
{Capelo}, H.~L., {Herbst}, W., {Leggett}, S.~K., {Hamilton}, C.~M., \&
  {Johnson}, J.~A. 2012, \apjl, 757, L18

\bibitem[{{Chabrier}(2003)}]{Chabrier:2003fk}
{Chabrier}, G. 2003, \pasp, 115, 763

\bibitem[{{Chaplin} {et~al.}(2013){Chaplin}, {Sanchis-Ojeda}, {Campante},
  {Handberg}, {Stello}, {Winn}, {Basu}, {Christensen-Dalsgaard}, {Davies},
  {Metcalfe}, {Buchhave}, {Fischer}, {Bedding}, {Cochran}, {Elsworth},
  {Gilliland}, {Hekker}, {Huber}, {Isaacson}, {Karoff}, {Kawaler}, {Kjeldsen},
  {Latham}, {Lund}, {Lundkvist}, {Marcy}, {Miglio}, {Barclay}, \&
  {Lissauer}}]{Chaplin:2013zr}
{Chaplin}, W.~J., {Sanchis-Ojeda}, R., {Campante}, T.~L., {et~al.} 2013, \apj,
  766, 101

\bibitem[{{Chatterjee} {et~al.}(2008){Chatterjee}, {Ford}, {Matsumura}, \&
  {Rasio}}]{Chatterjee:2008uq}
{Chatterjee}, S., {Ford}, E.~B., {Matsumura}, S., \& {Rasio}, F.~A. 2008, \apj,
  686, 580

\bibitem[{{Chiang} \& {Murray-Clay}(2004)}]{Chiang:2004fk}
{Chiang}, E.~I. \& {Murray-Clay}, R.~A. 2004, \apj, 607, 913

\bibitem[{{Clemence} \& {Brouwer}(1955)}]{Clemence:1955ly}
{Clemence}, G.~M. \& {Brouwer}, D. 1955, \aj, 60, 118

\bibitem[{{Collier Cameron} {et~al.}(2010){Collier Cameron}, {Bruce}, {Miller},
  {Triaud}, \& {Queloz}}]{Collier-Cameron:2010lk}
{Collier Cameron}, A., {Bruce}, V.~A., {Miller}, G.~R.~M., {Triaud},
  A.~H.~M.~J., \& {Queloz}, D. 2010, \mnras, 403, 151

\bibitem[{{Correia} {et~al.}(2005){Correia}, {Udry}, {Mayor}, {Laskar}, {Naef},
  {Pepe}, {Queloz}, \& {Santos}}]{Correia:2005lr}
{Correia}, A.~C.~M., {Udry}, S., {Mayor}, M., {et~al.} 2005, \aap, 440, 751

\bibitem[{{Davies} {et~al.}(2013){Davies}, {Adams}, {Armitage}, {Chambers},
  {Ford}, {Morbidelli}, {Raymond}, \& {Veras}}]{Davies:2013qy}
{Davies}, M.~B., {Adams}, F.~C., {Armitage}, P., {et~al.} 2013, ArXiv e-prints

\bibitem[{{Dawson} \& {Johnson}(2012)}]{Dawson:2012lr}
{Dawson}, R.~I. \& {Johnson}, J.~A. 2012, \apj, 756, 122

\bibitem[{{Dawson} \& {Murray-Clay}(2013)}]{Dawson:2013yq}
{Dawson}, R.~I. \& {Murray-Clay}, R.~A. 2013, \apjl, 767, L24

\bibitem[{{D{\'{\i}}az} {et~al.}(2013){D{\'{\i}}az}, {Damiani}, {Deleuil},
  {Almenara}, {Moutou}, {Barros}, {Bonomo}, {Bouchy}, {Bruno}, {H{\'e}brard},
  {Montagnier}, \& {Santerne}}]{Diaz:2013fr}
{D{\'{\i}}az}, R.~F., {Damiani}, C., {Deleuil}, M., {et~al.} 2013, \aap, 551,
  L9

\bibitem[{{Doolin} \& {Blundell}(2011)}]{Doolin:2011lr}
{Doolin}, S. \& {Blundell}, K.~M. 2011, \mnras, 418, 2656

\bibitem[{{Doyle} {et~al.}(2011){Doyle}, {Carter}, {Fabrycky}, {Slawson},
  {Howell}, {Winn}, {Orosz}, {Pr\v{s}a}, {Welsh}, {Quinn}, {Latham}, {Torres},
  {Buchhave}, {Marcy}, {Fortney}, {Shporer}, {Ford}, {Lissauer}, {Ragozzine},
  {Rucker}, {Batalha}, {Jenkins}, {Borucki}, {Koch}, {Middour}, {Hall},
  {McCauliff}, {Fanelli}, {Quintana}, {Holman}, {Caldwell}, {Still},
  {Stefanik}, {Brown}, {Esquerdo}, {Tang}, {Furesz}, {Geary}, {Berlind},
  {Calkins}, {Short}, {Steffen}, {Sasselov}, {Dunham}, {Cochran}, {Boss},
  {Haas}, {Buzasi}, \& {Fischer}}]{Doyle:2011vn}
{Doyle}, L.~R., {Carter}, J.~A., {Fabrycky}, D.~C., {et~al.} 2011, Science,
  333, 1602

\bibitem[{{Dressing} \& {Charbonneau}(2013)}]{Dressing:2013xy}
{Dressing}, C.~D. \& {Charbonneau}, D. 2013, \apj, 767, 95

\bibitem[{{Dumusque} {et~al.}(2012){Dumusque}, {Pepe}, {Lovis},
  {S{\'e}gransan}, {Sahlmann}, {Benz}, {Bouchy}, {Mayor}, {Queloz}, {Santos},
  \& {Udry}}]{Dumusque:2012lr}
{Dumusque}, X., {Pepe}, F., {Lovis}, C., {et~al.} 2012, \nat, 491, 207

\bibitem[{{Duquennoy} \& {Mayor}(1991)}]{Duquennoy:1991kx}
{Duquennoy}, A. \& {Mayor}, M. 1991, \aap, 248, 485

\bibitem[{{Dvorak}(1986)}]{Dvorak:1986fk}
{Dvorak}, R. 1986, \aap, 167, 379

\bibitem[{{Dvorak} {et~al.}(1989){Dvorak}, {Froeschle}, \&
  {Froeschle}}]{Dvorak:1989vn}
{Dvorak}, R., {Froeschle}, C., \& {Froeschle}, C. 1989, \aap, 226, 335

\bibitem[{{Eisner} {et~al.}(2005){Eisner}, {Hillenbrand}, {White}, {Akeson}, \&
  {Sargent}}]{Eisner:2005qy}
{Eisner}, J.~A., {Hillenbrand}, L.~A., {White}, R.~J., {Akeson}, R.~L., \&
  {Sargent}, A.~I. 2005, \apj, 623, 952

\bibitem[{{Fabrycky} \& {Tremaine}(2007)}]{Fabrycky:2007pd}
{Fabrycky}, D. \& {Tremaine}, S. 2007, \apj, 669, 1298

\bibitem[{{Faigler} \& {Mazeh}(2011)}]{Faigler:2011fk}
{Faigler}, S. \& {Mazeh}, T. 2011, \mnras, 415, 3921

\bibitem[{{Farago} \& {Laskar}(2010)}]{Farago:2010fj}
{Farago}, F. \& {Laskar}, J. 2010, \mnras, 401, 1189

\bibitem[{{Figueira} {et~al.}(2012){Figueira}, {Marmier}, {Bou{\'e}}, {Lovis},
  {Santos}, {Montalto}, {Udry}, {Pepe}, \& {Mayor}}]{Figueira:2012lr}
{Figueira}, P., {Marmier}, M., {Bou{\'e}}, G., {et~al.} 2012, \aap, 541, A139

\bibitem[{{Fischer} {et~al.}(2012){Fischer}, {Schwamb}, {Schawinski},
  {Lintott}, {Brewer}, {Giguere}, {Lynn}, {Parrish}, {Sartori}, {Simpson},
  {Smith}, {Spronck}, {Batalha}, {Rowe}, {Jenkins}, {Bryson}, {Prsa},
  {Tenenbaum}, {Crepp}, {Morton}, {Howard}, {Beleu}, {Kaplan}, {Vannispen},
  {Sharzer}, {Defouw}, {Hajduk}, {Neal}, {Nemec}, {Schuepbach}, \&
  {Zimmermann}}]{Fischer:2012lr}
{Fischer}, D.~A., {Schwamb}, M.~E., {Schawinski}, K., {et~al.} 2012, \mnras,
  419, 2900

\bibitem[{{Ford} \& {Rasio}(2006)}]{Ford:2006lr}
{Ford}, E.~B. \& {Rasio}, F.~A. 2006, \apjl, 638, L45

\bibitem[{{Foucart} \& {Lai}(2013)}]{Foucart:2013ys}
{Foucart}, F. \& {Lai}, D. 2013, \apj, 764, 106

\bibitem[{{Fressin} {et~al.}(2013){Fressin}, {Torres}, {Charbonneau}, {Bryson},
  {Christiansen}, {Dressing}, {Jenkins}, {Walkowicz}, \&
  {Batalha}}]{Fressin:2013uq}
{Fressin}, F., {Torres}, G., {Charbonneau}, D., {et~al.} 2013, \apj, 766, 81

\bibitem[{{Gaudi} \& {Winn}(2007)}]{Gaudi:2007vn}
{Gaudi}, B.~S. \& {Winn}, J.~N. 2007, \apj, 655, 550

\bibitem[{{Gillon} {et~al.}(2011){Gillon}, {Bonfils}, {Demory}, {Seager},
  {Deming}, \& {Triaud}}]{Gillon:2011lr}
{Gillon}, M., {Bonfils}, X., {Demory}, B.-O., {et~al.} 2011, \aap, 525, A32

\bibitem[{{Goldreich} \& {Tremaine}(1979)}]{Goldreich:1979fk}
{Goldreich}, P. \& {Tremaine}, S. 1979, \apj, 233, 857

\bibitem[{{Goldreich} \& {Tremaine}(1980)}]{Goldreich:1980lr}
{Goldreich}, P. \& {Tremaine}, S. 1980, \apj, 241, 425

\bibitem[{{Greaves} {et~al.}(2014){Greaves}, {Kennedy}, {Thureau}, {Eiroa},
  {Marshall}, {Maldonado}, {Matthews}, {Olofsson}, {Barlow},
  {Moro-Mart{\'{\i}}n}, {Sibthorpe}, {Absil}, {Ardila}, {Booth},
  {Broekhoven-Fiene}, {Brown}, {Cameron}, {del Burgo}, {Di Francesco},
  {Eisl{\"o}ffel}, {Duch{\^e}ne}, {Ertel}, {Holland}, {Horner}, {Kalas},
  {Kavelaars}, {Lestrade}, {Vican}, {Wilner}, {Wolf}, \&
  {Wyatt}}]{Greaves:2014vn}
{Greaves}, J.~S., {Kennedy}, G.~M., {Thureau}, N., {et~al.} 2014, \mnras, 438,
  L31

\bibitem[{{Guillochon} {et~al.}(2011){Guillochon}, {Ramirez-Ruiz}, \&
  {Lin}}]{Guillochon:2011fk}
{Guillochon}, J., {Ramirez-Ruiz}, E., \& {Lin}, D. 2011, \apj, 732, 74

\bibitem[{{Halbwachs} {et~al.}(2003){Halbwachs}, {Mayor}, {Udry}, \&
  {Arenou}}]{Halbwachs:2003lr}
{Halbwachs}, J.~L., {Mayor}, M., {Udry}, S., \& {Arenou}, F. 2003, \aap, 397,
  159

\bibitem[{{H{\'e}brard} {et~al.}(2008){H{\'e}brard}, {Bouchy}, {Pont},
  {Loeillet}, {Rabus}, {Bonfils}, {Moutou}, {Boisse}, {Delfosse}, {Desort},
  {Eggenberger}, {Ehrenreich}, {Forveille}, {Lagrange}, {Lovis}, {Mayor},
  {Pepe}, {Perrier}, {Queloz}, {Santos}, {S{\'e}gransan}, {Udry}, \&
  {Vidal-Madjar}}]{Hebrard:2008mz}
{H{\'e}brard}, G., {Bouchy}, F., {Pont}, F., {et~al.} 2008, \aap, 488, 763

\bibitem[{{Helled} {et~al.}(2013){Helled}, {Bodenheimer}, {Podolak}, {Boley},
  {Meru}, {Nayakshin}, {Fortney}, {Mayer}, {Alibert}, \&
  {Boss}}]{Helled:2013fj}
{Helled}, R., {Bodenheimer}, P., {Podolak}, M., {et~al.} 2013, ArXiv e-prints

\bibitem[{{Holman} {et~al.}(2010){Holman}, {Fabrycky}, {Ragozzine}, {Ford},
  {Steffen}, {Welsh}, {Lissauer}, {Latham}, {Marcy}, {Walkowicz}, {Batalha},
  {Jenkins}, {Rowe}, {Cochran}, {Fressin}, {Torres}, {Buchhave}, {Sasselov},
  {Borucki}, {Koch}, {Basri}, {Brown}, {Caldwell}, {Charbonneau}, {Dunham},
  {Gautier}, {Geary}, {Gilliland}, {Haas}, {Howell}, {Ciardi}, {Endl},
  {Fischer}, {F{\"u}r{\'e}sz}, {Hartman}, {Isaacson}, {Johnson}, {MacQueen},
  {Moorhead}, {Morehead}, \& {Orosz}}]{Holman:2010lr}
{Holman}, M.~J., {Fabrycky}, D.~C., {Ragozzine}, D., {et~al.} 2010, Science,
  Volume 330, Issue 6000, pp.~51- (2010)., 330, 51

\bibitem[{{Holman} \& {Murray}(2005)}]{Holman:2005fk}
{Holman}, M.~J. \& {Murray}, N.~W. 2005, Science, 307, 1288

\bibitem[{{Holman} \& {Wiegert}(1999)}]{Holman:1999lr}
{Holman}, M.~J. \& {Wiegert}, P.~A. 1999, \aj, 117, 621

\bibitem[{{Holt}(1893)}]{Holt:1893fk}
{Holt}, J.~R. 1893, Astro-Physics, XII, 646

\bibitem[{{Howard} {et~al.}(2012){Howard}, {Marcy}, {Bryson}, {Jenkins},
  {Rowe}, {Batalha}, {Borucki}, {Koch}, {Dunham}, {Gautier}, {Van Cleve},
  {Cochran}, {Latham}, {Lissauer}, {Torres}, {Brown}, {Gilliland}, {Buchhave},
  {Caldwell}, {Christensen-Dalsgaard}, {Ciardi}, {Fressin}, {Haas}, {Howell},
  {Kjeldsen}, {Seager}, {Rogers}, {Sasselov}, {Steffen}, {Basri},
  {Charbonneau}, {Christiansen}, {Clarke}, {Dupree}, {Fabrycky}, {Fischer},
  {Ford}, {Fortney}, {Tarter}, {Girouard}, {Holman}, {Johnson}, {Klaus},
  {Machalek}, {Moorhead}, {Morehead}, {Ragozzine}, {Tenenbaum}, {Twicken},
  {Quinn}, {Isaacson}, {Shporer}, {Lucas}, {Walkowicz}, {Welsh}, {Boss},
  {Devore}, {Gould}, {Smith}, {Morris}, {Prsa}, {Morton}, {Still}, {Thompson},
  {Mullally}, {Endl}, \& {MacQueen}}]{Howard:2012qy}
{Howard}, A.~W., {Marcy}, G.~W., {Bryson}, S.~T., {et~al.} 2012, \apjs, 201, 15

\bibitem[{{Howard} {et~al.}(2010){Howard}, {Marcy}, {Johnson}, {Fischer},
  {Wright}, {Isaacson}, {Valenti}, {Anderson}, {Lin}, \& {Ida}}]{Howard:2010zr}
{Howard}, A.~W., {Marcy}, G.~W., {Johnson}, J.~A., {et~al.} 2010, Science, 330,
  653

\bibitem[{{Juri{\'c}} \& {Tremaine}(2008)}]{Juric:2008qy}
{Juri{\'c}}, M. \& {Tremaine}, S. 2008, \apj, 686, 603

\bibitem[{{Kennedy} {et~al.}(2012{\natexlab{a}}){Kennedy}, {Wyatt},
  {Sibthorpe}, {Duch{\^e}ne}, {Kalas}, {Matthews}, {Greaves}, {Su}, \&
  {Fitzgerald}}]{Kennedy:2012zr}
{Kennedy}, G.~M., {Wyatt}, M.~C., {Sibthorpe}, B., {et~al.} 2012{\natexlab{a}},
  \mnras, 421, 2264

\bibitem[{{Kennedy} {et~al.}(2012{\natexlab{b}}){Kennedy}, {Wyatt},
  {Sibthorpe}, {Phillips}, {Matthews}, \& {Greaves}}]{Kennedy:2012lr}
{Kennedy}, G.~M., {Wyatt}, M.~C., {Sibthorpe}, B., {et~al.} 2012{\natexlab{b}},
  \mnras, 426, 2115

\bibitem[{{Kippenhahn} \& {Weigert}(1994)}]{Kippenhahn:1994fj}
{Kippenhahn}, R. \& {Weigert}, A. 1994, {Stellar Structure and Evolution}

\bibitem[{{Kley} \& {Haghighipour}(2014)}]{Kley:2014rt}
{Kley}, W. \& {Haghighipour}, N. 2014, \aap, 564, A72

\bibitem[{{Kostov} {et~al.}(2014){Kostov}, {McCullough}, {Carter}, {Deleuil},
  {D{\'{\i}}az}, {Fabrycky}, {H{\'e}brard}, {Hinse}, {Mazeh}, {Orosz},
  {Tsvetanov}, \& {Welsh}}]{Kostov:2014qy}
{Kostov}, V.~B., {McCullough}, P.~R., {Carter}, J.~A., {et~al.} 2014, \apj,
  784, 14

\bibitem[{{Kostov} {et~al.}(2013){Kostov}, {McCullough}, {Hinse}, {Tsvetanov},
  {H{\'e}brard}, {D{\'{\i}}az}, {Deleuil}, \& {Valenti}}]{Kostov:2013lr}
{Kostov}, V.~B., {McCullough}, P.~R., {Hinse}, T.~C., {et~al.} 2013, \apj, 770,
  52

\bibitem[{{Kratter} \& {Perets}(2012)}]{Kratter:2012lr}
{Kratter}, K.~M. \& {Perets}, H.~B. 2012, \apj, 753, 91

\bibitem[{{Kroupa}(2001)}]{Kroupa:2001qy}
{Kroupa}, P. 2001, \mnras, 322, 231

\bibitem[{{Kuchner} \& {Lecar}(2002)}]{Kuchner:2002uq}
{Kuchner}, M.~J. \& {Lecar}, M. 2002, \apjl, 574, L87

\bibitem[{{Le Bouquin} {et~al.}(2009){Le Bouquin}, {Absil}, {Benisty}, {Massi},
  {M{\'e}rand}, \& {Stefl}}]{Le-Bouquin:2009mz}
{Le Bouquin}, J.-B., {Absil}, O., {Benisty}, M., {et~al.} 2009, \aap, 498, L41

\bibitem[{{Leung} \& {Hoi Lee}(2013)}]{Leung:2013fk}
{Leung}, G.~C.~K. \& {Hoi Lee}, M. 2013, \apj, 763, 107

\bibitem[{{Lin} {et~al.}(1996){Lin}, {Bodenheimer}, \&
  {Richardson}}]{Lin:1996yq}
{Lin}, D.~N.~C., {Bodenheimer}, P., \& {Richardson}, D.~C. 1996, \nat, 380, 606

\bibitem[{{Lissauer} {et~al.}(2011){Lissauer}, {Ragozzine}, {Fabrycky},
  {Steffen}, {Ford}, {Jenkins}, {Shporer}, {Holman}, {Rowe}, {Quintana},
  {Batalha}, {Borucki}, {Bryson}, {Caldwell}, {Carter}, {Ciardi}, {Dunham},
  {Fortney}, {Gautier}, {Howell}, {Koch}, {Latham}, {Marcy}, {Morehead}, \&
  {Sasselov}}]{Lissauer:2011rt}
{Lissauer}, J.~J., {Ragozzine}, D., {Fabrycky}, D.~C., {et~al.} 2011, \apjs,
  197, 8

\bibitem[{{Lodato} \& {Facchini}(2013)}]{Lodato:2013uq}
{Lodato}, G. \& {Facchini}, S. 2013, \mnras, 433, 2157

\bibitem[{{Matsumura} {et~al.}(2010{\natexlab{a}}){Matsumura}, {Peale}, \&
  {Rasio}}]{Matsumura:2010ve}
{Matsumura}, S., {Peale}, S.~J., \& {Rasio}, F.~A. 2010{\natexlab{a}}, \apj,
  725, 1995

\bibitem[{{Matsumura} {et~al.}(2010{\natexlab{b}}){Matsumura}, {Thommes},
  {Chatterjee}, \& {Rasio}}]{Matsumura:2010ul}
{Matsumura}, S., {Thommes}, E.~W., {Chatterjee}, S., \& {Rasio}, F.~A.
  2010{\natexlab{b}}, \apj, 714, 194

\bibitem[{{Mayor} {et~al.}(2011){Mayor}, {Marmier}, {Lovis}, {Udry},
  {S{\'e}gransan}, {Pepe}, {Benz}, {Bertaux}, {Bouchy}, {Dumusque}, {Lo Curto},
  {Mordasini}, {Queloz}, \& {Santos}}]{Mayor:2011fj}
{Mayor}, M., {Marmier}, M., {Lovis}, C., {et~al.} 2011, eprint arXiv:1109.2497

\bibitem[{{Mayor} \& {Queloz}(1995)}]{Mayor:1995uq}
{Mayor}, M. \& {Queloz}, D. 1995, \nat, 378, 355

\bibitem[{{Mazeh} \& {Shaham}(1979)}]{Mazeh:1979eu}
{Mazeh}, T. \& {Shaham}, J. 1979, \aap, 77, 145

\bibitem[{{McLaughlin}(1924)}]{McLaughlin:1924uq}
{McLaughlin}, D.~B. 1924, \apj, 60, 22

\bibitem[{{Meschiari}(2012{\natexlab{a}})}]{Meschiari:2012uq}
{Meschiari}, S. 2012{\natexlab{a}}, \apj, 752, 71

\bibitem[{{Meschiari}(2012{\natexlab{b}})}]{Meschiari:2012qy}
{Meschiari}, S. 2012{\natexlab{b}}, \apjl, 761, L7

\bibitem[{{Miralda-Escud{\'e}}(2002)}]{Miralda-Escude:2002uq}
{Miralda-Escud{\'e}}, J. 2002, \apj, 564, 1019

\bibitem[{{Moeckel} \& {Veras}(2012)}]{Moeckel:2012lr}
{Moeckel}, N. \& {Veras}, D. 2012, ArXiv e-prints

\bibitem[{{Morais} \& {Giuppone}(2012)}]{Morais:2012qy}
{Morais}, M.~H.~M. \& {Giuppone}, C.~A. 2012, \mnras, 424, 52

\bibitem[{{Mustill} {et~al.}(2013){Mustill}, {Marshall}, {Villaver}, {Veras},
  {Davis}, {Horner}, \& {Wittenmyer}}]{Mustill:2013lr}
{Mustill}, A.~J., {Marshall}, J.~P., {Villaver}, E., {et~al.} 2013, \mnras,
  436, 2515

\bibitem[{{Nagasawa} {et~al.}(2008){Nagasawa}, {Ida}, \&
  {Bessho}}]{Nagasawa:2008gf}
{Nagasawa}, M., {Ida}, S., \& {Bessho}, T. 2008, \apj, 678, 498

\bibitem[{{Nelson}(2003)}]{Nelson:2003lr}
{Nelson}, R.~P. 2003, \mnras, 345, 233

\bibitem[{{Nutzman} {et~al.}(2011){Nutzman}, {Fabrycky}, \&
  {Fortney}}]{Nutzman:2011ly}
{Nutzman}, P.~A., {Fabrycky}, D.~C., \& {Fortney}, J.~J. 2011, \apjl, 740, L10

\bibitem[{{Orosz} {et~al.}(2012{\natexlab{a}}){Orosz}, {Welsh}, {Carter},
  {Brugamyer}, {Buchhave}, {Cochran}, {Endl}, {Ford}, {MacQueen}, {Short},
  {Torres}, {Windmiller}, {Agol}, {Barclay}, {Caldwell}, {Clarke}, {Doyle},
  {Fabrycky}, {Geary}, {Haghighipour}, {Holman}, {Ibrahim}, {Jenkins},
  {Kinemuchi}, {Li}, {Lissauer}, {Pr{\v s}a}, {Ragozzine}, {Shporer}, {Still},
  \& {Wade}}]{Orosz:2012yq}
{Orosz}, J.~A., {Welsh}, W.~F., {Carter}, J.~A., {et~al.} 2012{\natexlab{a}},
  \apj, 758, 87

\bibitem[{{Orosz} {et~al.}(2012{\natexlab{b}}){Orosz}, {Welsh}, {Carter},
  {Fabrycky}, {Cochran}, {Endl}, {Ford}, {Haghighipour}, {MacQueen}, {Mazeh},
  {Sanchis-Ojeda}, {Short}, {Torres}, {Agol}, {Buchhave}, {Doyle}, {Isaacson},
  {Lissauer}, {Marcy}, {Shporer}, {Windmiller}, {Barclay}, {Boss}, {Clarke},
  {Fortney}, {Geary}, {Holman}, {Huber}, {Jenkins}, {Kinemuchi}, {Kruse},
  {Ragozzine}, {Sasselov}, {Still}, {Tenenbaum}, {Uddin}, {Winn}, {Koch}, \&
  {Borucki}}]{Orosz:2012lr}
{Orosz}, J.~A., {Welsh}, W.~F., {Carter}, J.~A., {et~al.} 2012{\natexlab{b}},
  Science, 337, 1511

\bibitem[{{Paardekooper} {et~al.}(2012){Paardekooper}, {Leinhardt},
  {Th{\'e}bault}, \& {Baruteau}}]{Paardekooper:2012kx}
{Paardekooper}, S.-J., {Leinhardt}, Z.~M., {Th{\'e}bault}, P., \& {Baruteau},
  C. 2012, \apjl, 754, L16

\bibitem[{{Pierens} \& {Nelson}(2007)}]{Pierens:2007fj}
{Pierens}, A. \& {Nelson}, R.~P. 2007, \aap, 472, 993

\bibitem[{{Pierens} \& {Nelson}(2008)}]{Pierens:2008uq}
{Pierens}, A. \& {Nelson}, R.~P. 2008, \aap, 483, 633

\bibitem[{{Pierens} \& {Nelson}(2013)}]{Pierens:2013kx}
{Pierens}, A. \& {Nelson}, R.~P. 2013, \aap, 556, A134

\bibitem[{{Pilat-Lohinger} {et~al.}(2003){Pilat-Lohinger}, {Funk}, \&
  {Dvorak}}]{Pilat-Lohinger:2003qy}
{Pilat-Lohinger}, E., {Funk}, B., \& {Dvorak}, R. 2003, \aap, 400, 1085

\bibitem[{{Plavchan} \& {Bilinski}(2013)}]{Plavchan:2013ll}
{Plavchan}, P. \& {Bilinski}, C. 2013, \apj, 769, 86

\bibitem[{{Plavchan} {et~al.}(2008){Plavchan}, {Gee}, {Stapelfeldt}, \&
  {Becker}}]{Plavchan:2008fk}
{Plavchan}, P., {Gee}, A.~H., {Stapelfeldt}, K., \& {Becker}, A. 2008, \apjl,
  684, L37

\bibitem[{{Plavchan} {et~al.}(2013){Plavchan}, {G{\"u}th}, {Laohakunakorn}, \&
  {Parks}}]{Plavchan:2013lr}
{Plavchan}, P., {G{\"u}th}, T., {Laohakunakorn}, N., \& {Parks}, J.~R. 2013,
  \aap, 554, A110

\bibitem[{{Pollack} {et~al.}(1996){Pollack}, {Hubickyj}, {Bodenheimer},
  {Lissauer}, {Podolak}, \& {Greenzweig}}]{Pollack:1996uq}
{Pollack}, J.~B., {Hubickyj}, O., {Bodenheimer}, P., {et~al.} 1996, \icarus,
  124, 62

\bibitem[{{Pr{\v s}a} {et~al.}(2011){Pr{\v s}a}, {Batalha}, {Slawson}, {Doyle},
  {Welsh}, {Orosz}, {Seager}, {Rucker}, {Mjaseth}, {Engle}, {Conroy},
  {Jenkins}, {Caldwell}, {Koch}, \& {Borucki}}]{Prsa:2011qy}
{Pr{\v s}a}, A., {Batalha}, N., {Slawson}, R.~W., {et~al.} 2011, \aj, 141, 83

\bibitem[{{Queloz} {et~al.}(2000){Queloz}, {Eggenberger}, {Mayor}, {Perrier},
  {Beuzit}, {Naef}, {Sivan}, \& {Udry}}]{Queloz:2000rt}
{Queloz}, D., {Eggenberger}, A., {Mayor}, M., {et~al.} 2000, \aap, 359, L13

\bibitem[{{Raghavan} {et~al.}(2010){Raghavan}, {McAlister}, {Henry}, {Latham},
  {Marcy}, {Mason}, {Gies}, {White}, \& {ten Brummelaar}}]{Raghavan:2010lr}
{Raghavan}, D., {McAlister}, H.~A., {Henry}, T.~J., {et~al.} 2010, \apjs, 190,
  1

\bibitem[{{Rasio} \& {Ford}(1996)}]{Rasio:1996ly}
{Rasio}, F.~A. \& {Ford}, E.~B. 1996, Science, 274, 954

\bibitem[{{Rossiter}(1924)}]{Rossiter:1924qy}
{Rossiter}, R.~A. 1924, \apj, 60, 15

\bibitem[{{Sahlmann} {et~al.}(2011){Sahlmann}, {Lovis}, {Queloz}, \&
  {S{\'e}gransan}}]{Sahlmann:2011fk}
{Sahlmann}, J., {Lovis}, C., {Queloz}, D., \& {S{\'e}gransan}, D. 2011, \aap,
  528, L8

\bibitem[{{Salpeter}(1955)}]{Salpeter:1955fj}
{Salpeter}, E.~E. 1955, \apj, 121, 161

\bibitem[{{Sanchis-Ojeda} {et~al.}(2012){Sanchis-Ojeda}, {Fabrycky}, {Winn},
  {Barclay}, {Clarke}, {Ford}, {Fortney}, {Geary}, {Holman}, {Howard},
  {Jenkins}, {Koch}, {Lissauer}, {Marcy}, {Mullally}, {Ragozzine}, {Seader},
  {Still}, \& {Thompson}}]{Sanchis-Ojeda:2012ys}
{Sanchis-Ojeda}, R., {Fabrycky}, D.~C., {Winn}, J.~N., {et~al.} 2012, \nat,
  487, 449

\bibitem[{{Santerne} {et~al.}(2012){Santerne}, {D{\'{\i}}az}, {Moutou},
  {Bouchy}, {H{\'e}brard}, {Almenara}, {Bonomo}, {Deleuil}, \&
  {Santos}}]{Santerne:2012rt}
{Santerne}, A., {D{\'{\i}}az}, R.~F., {Moutou}, C., {et~al.} 2012, \aap, 545,
  A76

\bibitem[{{Schlaufman}(2010)}]{Schlaufman:2010fk}
{Schlaufman}, K.~C. 2010, \apj, 719, 602

\bibitem[{{Schneider}(1994)}]{Schneider:1994lr}
{Schneider}, J. 1994, \planss, 42, 539

\bibitem[{{Schneider} {et~al.}(2011){Schneider}, {Dedieu}, {Le Sidaner},
  {Savalle}, \& {Zolotukhin}}]{Schneider:2011lr}
{Schneider}, J., {Dedieu}, C., {Le Sidaner}, P., {Savalle}, R., \&
  {Zolotukhin}, I. 2011, \aap, 532, A79

\bibitem[{{Schwamb} {et~al.}(2013){Schwamb}, {Orosz}, {Carter}, {Welsh},
  {Fischer}, {Torres}, {Howard}, {Crepp}, {Keel}, {Lintott}, {Kaib}, {Terrell},
  {Gagliano}, {Jek}, {Parrish}, {Smith}, {Lynn}, {Simpson}, {Giguere}, \&
  {Schawinski}}]{Schwamb:2013kx}
{Schwamb}, M.~E., {Orosz}, J.~A., {Carter}, J.~A., {et~al.} 2013, \apj, 768,
  127

\bibitem[{{Seager} \& {Deming}(2010)}]{Seager:2010kx}
{Seager}, S. \& {Deming}, D. 2010, \araa, 48, 631

\bibitem[{{Seager} \& {Mall{\'e}n-Ornelas}(2003)}]{Seager:2003qy}
{Seager}, S. \& {Mall{\'e}n-Ornelas}, G. 2003, \apj, 585, 1038

\bibitem[{{Shporer} {et~al.}(2011){Shporer}, {Jenkins}, {Rowe}, {Sanderfer},
  {Seader}, {Smith}, {Still}, {Thompson}, {Twicken}, \&
  {Welsh}}]{Shporer:2011qy}
{Shporer}, A., {Jenkins}, J.~M., {Rowe}, J.~F., {et~al.} 2011, \aj, 142, 195

\bibitem[{{Slawson} {et~al.}(2011){Slawson}, {Pr{\v s}a}, {Welsh}, {Orosz},
  {Rucker}, {Batalha}, {Doyle}, {Engle}, {Conroy}, {Coughlin}, {Gregg},
  {Fetherolf}, {Short}, {Windmiller}, {Fabrycky}, {Howell}, {Jenkins}, {Uddin},
  {Mullally}, {Seader}, {Thompson}, {Sanderfer}, {Borucki}, \&
  {Koch}}]{Slawson:2011uq}
{Slawson}, R.~W., {Pr{\v s}a}, A., {Welsh}, W.~F., {et~al.} 2011, \aj, 142, 160

\bibitem[{{Szab{\'o}} {et~al.}(2011){Szab{\'o}}, {Szab{\'o}}, {Benk{\H o}},
  {Lehmann}, {Mez{\H o}}, {Simon}, {K{\H o}v{\'a}ri}, {Hodos{\'a}n},
  {Reg{\'a}ly}, \& {Kiss}}]{Szabo:2011fr}
{Szab{\'o}}, G.~M., {Szab{\'o}}, R., {Benk{\H o}}, J.~M., {et~al.} 2011, \apjl,
  736, L4

\bibitem[{{Terquem} {et~al.}(1999){Terquem}, {Eisl{\"o}ffel}, {Papaloizou}, \&
  {Nelson}}]{Terquem:1999pd}
{Terquem}, C., {Eisl{\"o}ffel}, J., {Papaloizou}, J.~C.~B., \& {Nelson}, R.~P.
  1999, \apjl, 512, L131

\bibitem[{{Tokovinin} {et~al.}(2006){Tokovinin}, {Thomas}, {Sterzik}, \&
  {Udry}}]{Tokovinin:2006la}
{Tokovinin}, A., {Thomas}, S., {Sterzik}, M., \& {Udry}, S. 2006, \aap, 450,
  681

\bibitem[{{Torres} {et~al.}(2010){Torres}, {Andersen}, \&
  {Gim{\'e}nez}}]{Torres:2010uq}
{Torres}, G., {Andersen}, J., \& {Gim{\'e}nez}, A. 2010, \aapr, 18, 67

\bibitem[{{Triaud}(2011{\natexlab{a}})}]{Triaud:2011qy}
{Triaud}, A.~H.~M.~J. 2011{\natexlab{a}}, PhD thesis, Observatoire Astronomique
  de l'Universite de Geneve, http://archive-ouverte.unige.ch/unige:18065

\bibitem[{{Triaud}(2011{\natexlab{b}})}]{Triaud:2011fk}
{Triaud}, A.~H.~M.~J. 2011{\natexlab{b}}, \aap, 534, L6

\bibitem[{{Triaud} {et~al.}(2010){Triaud}, {Collier Cameron}, {Queloz},
  {Anderson}, {Gillon}, {Hebb}, {Hellier}, {Loeillet}, {Maxted}, {Mayor},
  {Pepe}, {Pollacco}, {S{\'e}gransan}, {Smalley}, {Udry}, {West}, \&
  {Wheatley}}]{Triaud:2010fr}
{Triaud}, A.~H.~M.~J., {Collier Cameron}, A., {Queloz}, D., {et~al.} 2010,
  \aap, 524, A25

\bibitem[{{Ward}(1997)}]{Ward:1997kx}
{Ward}, W.~R. 1997, \icarus, 126, 261

\bibitem[{{Watson} {et~al.}(2011){Watson}, {Littlefair}, {Diamond}, {Collier
  Cameron}, {Fitzsimmons}, {Simpson}, {Moulds}, \& {Pollacco}}]{Watson:2011mz}
{Watson}, C.~A., {Littlefair}, S.~P., {Diamond}, C., {et~al.} 2011, \mnras,
  413, L71

\bibitem[{{Welsh} {et~al.}(2013){Welsh}, {Orosz}, {Carter}, \&
  {Fabrycky}}]{Welsh:2013lr}
{Welsh}, W.~F., {Orosz}, J.~A., {Carter}, J.~A., \& {Fabrycky}, D.~C. 2013,
  ArXiv e-prints

\bibitem[{{Welsh} {et~al.}(2012){Welsh}, {Orosz}, {Carter}, {Fabrycky}, {Ford},
  {Lissauer}, {Pr{\v s}a}, {Quinn}, {Ragozzine}, {Short}, {Torres}, {Winn},
  {Doyle}, {Barclay}, {Batalha}, {Bloemen}, {Brugamyer}, {Buchhave},
  {Caldwell}, {Caldwell}, {Christiansen}, {Ciardi}, {Cochran}, {Endl},
  {Fortney}, {Gautier}, {Gilliland}, {Haas}, {Hall}, {Holman}, {Howard},
  {Howell}, {Isaacson}, {Jenkins}, {Klaus}, {Latham}, {Li}, {Marcy}, {Mazeh},
  {Quintana}, {Robertson}, {Shporer}, {Steffen}, {Windmiller}, {Koch}, \&
  {Borucki}}]{Welsh:2012lr}
{Welsh}, W.~F., {Orosz}, J.~A., {Carter}, J.~A., {et~al.} 2012, \nat, 481, 475

\bibitem[{{Winn}(2010)}]{Winn:2010lz}
{Winn}, J.~N. 2010, {Exoplanet Transits and Occultations}, ed. S.~{Seager},
  55--77

\bibitem[{{Winn} {et~al.}(2010){Winn}, {Fabrycky}, {Albrecht}, \&
  {Johnson}}]{Winn:2010rr}
{Winn}, J.~N., {Fabrycky}, D., {Albrecht}, S., \& {Johnson}, J.~A. 2010, \apjl,
  718, L145

\bibitem[{{Winn} {et~al.}(2006){Winn}, {Hamilton}, {Herbst}, {Hoffman},
  {Holman}, {Johnson}, \& {Kuchner}}]{Winn:2006qy}
{Winn}, J.~N., {Hamilton}, C.~M., {Herbst}, W.~J., {et~al.} 2006, \apj, 644,
  510

\bibitem[{{Winn} {et~al.}(2004){Winn}, {Holman}, {Johnson}, {Stanek}, \&
  {Garnavich}}]{Winn:2004lr}
{Winn}, J.~N., {Holman}, M.~J., {Johnson}, J.~A., {Stanek}, K.~Z., \&
  {Garnavich}, P.~M. 2004, \apjl, 603, L45

\bibitem[{{Winn} {et~al.}(2009){Winn}, {Johnson}, {Albrecht}, {Howard},
  {Marcy}, {Crossfield}, \& {Holman}}]{Winn:2009lr}
{Winn}, J.~N., {Johnson}, J.~A., {Albrecht}, S., {et~al.} 2009, \apjl, 703, L99

\bibitem[{{Wright} {et~al.}(2011){Wright}, {Fakhouri}, {Marcy}, {Han}, {Feng},
  {Johnson}, {Howard}, {Fischer}, {Valenti}, {Anderson}, \&
  {Piskunov}}]{Wright:2011fj}
{Wright}, J.~T., {Fakhouri}, O., {Marcy}, G.~W., {et~al.} 2011, \pasp, 123, 412

\bibitem[{{Zhou} \& {Huang}(2013)}]{Zhou:2013fk}
{Zhou}, G. \& {Huang}, C.~X. 2013, \apjl, 776, L35

\end{thebibliography}

\end{document}